\documentclass[twocolumn,preprintnumbers,amsmath,amssymb]{revtex4-1}

\usepackage{graphicx}
\newcommand{\commentoutA}[1]{}

\begin{document}

\preprint{LA-UR-17-23007}

\title{Next generation extended Lagrangian first principles molecular dynamics}

\author{Anders M. N. Niklasson}
\email{amn@lanl.gov, anders.niklasson@gmail.com}
\affiliation{Theoretical Division, Los Alamos National Laboratory, Los Alamos, New Mexico 87545}

\date{\today}

\begin{abstract}
Extended Lagrangian Born-Oppenheimer molecular dynamics 
[Phys. Rev. Lett., ${\bf 100}$, 123004 (2008)] 
is formulated 
for general Hohenberg-Kohn density functional theory and compared
to the extended Lagrangian framework of
first principles molecular dynamics by Car and Parrinello 
[Phys. Rev. Lett. ${\bf 55}$, 2471 (1985)].
It is shown how extended Lagrangian Born-Oppenheimer molecular dynamics
overcomes several shortcomings of regular, direct Born-Oppenheimer molecular dynamics,
while improving or maintaining important features of Car-Parrinello simulations.
The accuracy of the electronic degrees of freedom in extended Lagrangian 
Born-Oppenheimer molecular dynamics, with respect to the exact Born-Oppenheimer 
solution, is of second order in the size of the integration time step and
of fourth order in the potential energy surface.
Improved stability over recent formulations of extended Lagrangian 
Born-Oppenheimer molecular dynamics is achieved
by generalizing the theory to finite temperature ensembles, using fractional
occupation numbers in the calculation of the inner-product kernel of the 
extended harmonic oscillator that appears as a preconditioner in the electronic 
equations of motion.  Materials systems that normally 
exhibit slow self-consistent field convergence can be simulated
using integration time steps of the same order as in direct Born-Oppenheimer
molecular dynamics, but without the requirement of an iterative, 
non-linear electronic ground state optimization prior to the force evaluations
and without a systematic drift in the total energy.
In combination with proposed low-rank and on-the-fly updates of the kernel, 
this formulation provides an efficient and general framework for
quantum based Born-Oppenheimer molecular dynamics simulations.
\end{abstract}

\maketitle

\section{Introduction}

Quantum based molecular dynamics, using classical molecular trajectories, but with 
the interatomic forces calculated on-the-fly from a quantum mechanical 
description of the underlying electronic structure, offers an almost 
universal simulation engine for computational materials science, chemistry and molecular biology.
Similar to classical molecular dynamics 
\cite{MAllen90,MKarplus90,vanGusteren90,MTuckerman00,DFrenkel02,MKarplus02,KYSanbonmatsu07,MOrozco14,JRPerilla15}, quantum based 
molecular dynamics can be used to analyze, predict and manipulate a broad range of systems 
and phenomena \cite{DMarx00,MTuckerman02,BKirchner12,JHutter12}. However,
since quantum based molecular dynamics 
includes an explicit quantum mechanical description of the electrons,
it can be used also for materials that are difficult or impossible to simulate
with classical force field methods. Quantum based molecular dynamics therefore
provides a more general approach that can handle
phenomena such as electronic excitations
in photo voltaics, quantum size effects in nanodevices, spin-polarization in magnetic
materials, quantum response properties such as the conductivity or the polarizability,
thermal excitations in warm dense matter, as well as 
bond formation and dissociation with the associated charge transfer 
in chemical reactions.

While some early applications of quantum based molecular dynamics 
simulations date back to the mid 70's \cite{MKarplus73,CLeforestier78,AWarshel75}, 
it was only after the introduction of the ingenious unified extended Lagrangian 
approach to first principles electronic structure theory
by Car and Parrinello \cite{RCar85,DRemler90,GPastore91,FBornemann98,DMarx00,JHutter12} that stable 
and efficient molecular dynamics simulations with predictive quantum accuracy 
became practically feasible for a broad range of materials systems.
Their pioneering contribution unraveled major developments
of first principles molecular dynamics simulation tools and
opened the door to a number of new applications in materials science, 
chemistry and molecular biology \cite{DRemler90,DMarx00,PCarloni02,MTuckerman02,HSenn09,BKirchner12,JHutter12}.
In contrast to the straightforward direct Born-Oppenheimer approach, where the interatomic 
forces are calculated with an optimized relaxed electronic ground state in each time step
\cite{WHeitler27,MBorn27,RBarnett93,GKresse93,DMarx00}, Car-Parrinello 
molecular dynamics avoids the ground state relaxation. Instead, the electronic
degrees of freedom are included as dynamical variables that 
approximately follow the electronic ground state. This formulation avoids
problems arising from the non-linear iterative ground state optimization 
that is necessary in a direct Born-Oppenheimer simulation, 
such as a high computational overhead or non-conservative forces and a 
systematic drift in the total energy \cite{DRemler90,PPulay04,ANiklasson06,TDKuhne07}. 

Extended Lagrangian Car-Parrinello molecular dynamics has been adapted
to multiple research fields. For example, the plane-wave basis functions
that were used to represent the electronic degrees of freedom in the original 
formulation, have been modified to include localized atomic-orbitals \cite{BHartke92}, 
the charge density 
\cite{FLambert06}, or the atomic-orbital density matrix \cite{HBSchlegel01,SIyengar01,JMHerbert04}. 
This allows Car-Parrinello molecular dynamics to be applied to a broad range of
problems including, for example, warm dense matter \cite{FLambert06},
quantum chemistry simulations using correlated wavefunction methods \cite{JLi16},
and to extended systems using linear scaling electronic structure theory \cite{SGoedecker99,DBowler12,HBSchlegel01,SIyengar01,JMHerbert04}.
Car-Parrinello based techniques are often referred to simply as "the extended Lagrangian approach".
This is unfortunate, since there are a number
of different extended Lagrangian approaches.
Probably the first extended Lagrangian formulations in molecular dynamics 
were introduced to include the effects of thermostats or barostats in classical 
force field simulations \cite{HCAndersen80,MParrinello80,SNose84}
-- and recently, a new extended Lagrangian formulation 
of first-principles Born-Oppenheimer molecular dynamics was proposed
by the author and his co-workers in a series of papers 
\cite{ANiklasson08,ANiklasson09,AOdell09,PSteneteg10,GZheng11,MCawkwell12,
PSouvatzis14,ANiklasson14,BAradi15,EMartinez15,VVitale17,AAlbaugh17}.
This formulation is the main focus of this article. 

The extended Lagrangian formulation of first-principles Born-Oppenheimer molecular dynamics 
differs from the Car-Parrinello method, although some interesting connections have
been discussed \cite{BKirchner12,JHutter12,LLin14}. 
Both methods provide general theoretical frameworks that are applicable to
a broad class of materials problems and electronic structure methods.
However, extended Lagrangian Born-Oppenheimer molecular dynamics
is a higher-order method 
and it is based on a different Lagrangian that avoids the orthonormalization
(or the idempotency) constraints as well as the problem of choosing
material-dependent fictitious electron mass parameters.

Extended Lagrangian Born-Oppenheimer
molecular dynamics \cite{ANiklasson06,ANiklasson08} was originally developed 
to overcome shortcomings of direct
Born-Oppenheimer molecular dynamics, such as the 
irreversible and unphysical evolution of the electronic degrees of freedom, which is causing a 
systematic drift in the total energy and temperature \cite{DRemler90,PPulay04}.
The original idea was based on a time-reversible extrapolation of the electronic 
degrees of freedom from previous time steps followed by a fast, approximate update 
of the Born-Oppenheimer potential energy surface and the forces \cite{ANiklasson06}. 
This {\em ad hoc} approach, which has many similarities with, for example, the Fock-dynamics scheme 
by Pulay and Fogarasi \cite{PPulay04,JMHerbert05} as well as the technique by K\"{u}hne {\em et. al} \cite{TDKuhne07,JKolafa03}, 
was thereafter generalized and formulated in terms of an extended Lagrangian framework in 
the spirit of Car-Parrinello molecular dynamics \cite{ANiklasson08}.
This framework represents a new generation extended Lagrangian first principles molecular
dynamics that overcomes shortcomings of direct Born-Oppenheimer molecular
dynamics while maintaining important benefits of Car's and Parrinello's formulation. 

First I present extended Lagrangian Born-Oppenheimer molecular dynamics 
in terms of Hohenberg-Kohn density functional theory and discuss its
relation to Car-Parrinello molecular dynamics.
Thereafter, I formulate a generalized extended Lagrangian Born-Oppenheimer molecular dynamics
for thermal (or ensemble) Kohn-Sham density functional theory with fractional occupation numbers.  
I then focus on the calculation and approximation of the inner-product kernel 
that appears in the extended harmonic oscillator of the Lagrangian. I show how this kernel
can be constructed on-the-fly through a sequence of low-rank updates. 
At the end I demonstrate the framework with examples using self-consistent-charge 
density functional based tight-binding (SCC-DFTB) theory, before I outline
remaining challenges and give a summary and conclusions.

\section{Next generation extended Lagrangian first principles molecular dynamics}

Extended Lagrangian Born-Oppenheimer molecular dynamics was first formulated 
in Ref.\ \cite{ANiklasson08}, and then developed and adapted to a number of electronic structure codes 
at different levels of theory and choices of dynamical variables for the electronic degrees of freedom 
\cite{ANiklasson09,AOdell09,PSteneteg10,GZheng11,ANiklasson11,MCawkwell12,PSouvatzis14,ANiklasson14,BAradi15,MArita14,LLin14,KNomura15,AAlbaugh15,ANiklasson16,VVitale17,AAlbaugh17}.
Recently the framework was introduced in a more general form, including a kernel 
that defines the inner product norm of the extended harmonic oscillator 
\cite{ANiklasson14}. This kernel appears similar to a preconditioner 
in the equations of motion of the extended electronic degrees of freedom, which
can be used to improve the stability and the accuracy of a simulation.
In early formulations of extended Lagrangian Born-Oppenheimer molecular dynamics
a few self-consistent field iterations were sometimes required prior to the force evaluations.
With the generalized framework, which is based on an underlying shadow Hamiltonian
dynamics, the self-consistent field procedure can be avoided completely.
In this section I will briefly review this formulation in a more rigorous form
based on Hohenberg's and Kohn's non-degenerate ground state density functional theory.
Extensions beyond Hohenberg-Kohn density functional theory 
should be straightforward, e.g.\ within Levy's and Lieb's 
formulations \cite{MLevy79,ELieb83}, but this will not be discussed here. 
The purpose is to present a clear and fairly brief formulation of 
extended Lagrangian Born-Oppenheimer molecular dynamics.

\subsection{Density functional theory for Born-Oppenheimer molecular dynamics}

In Hohenberg's and Kohn's original density functional theory \cite{hohen,RParr89,RMDreizler90}
the electronic ground state  density, $\rho_{\rm min}({\bf r})$,
of a non-degenerate electron system in an external potential, $v({\bf R, r})$, that moves with the
nuclear coordinates, ${\bf R} = \{R_I\}$, is given through a constrained functional
minimization over all $v$-representable densities, $\rho({\bf r}) \in {\cal V}$, such that
\begin{equation}\label{rho_min}
\rho_{\rm min}({\bf r}) = \arg \min_{\rho \in {\cal V}} \left\{ F[\rho] + \int v({\bf R, r}) \rho({\bf r}) d{\bf r}\right\},
\end{equation}
where $F[\rho]$ is a universal functional of the electron density.
The Born-Oppenheimer potential energy surface is then given by
\begin{equation}\label{U_min}
U_{\rm BO}({\bf R}) = F[\rho_{\rm min}] + \int v({\bf R, r}) \rho_{\rm min}({\bf r}) d{\bf r} + V_{nn}({\bf R}),
\end{equation}
including a nuclear-nuclear (or ion-ion) interaction potential, $V_{nn}({\bf R})$.
A Born-Oppenheimer molecular dynamics based on density functional theory 
can then be defined through the Lagrangian
\begin{equation}\label{BOMD}
{\cal L}_{\rm BO}({\bf R},{\bf {\dot R}}) = \frac{1}{2}\sum_I M_I {\dot R}_I^2 - U_{\rm BO}({\bf R}),
\end{equation}
where $M_I$ are the nuclear masses and the dots denote time derivatives.
The Euler-Lagrange's equations, 
\begin{equation}
\frac{d}{dt} \left( \frac{\partial {\cal L}_{\rm BO}}{\partial {\dot R}_I} \right) = \frac{\partial {\cal L}_{\rm BO}}{\partial {R}_I},
\end{equation}
then gives us the equations of motion,
\begin{equation}\label{BOMD_eq}
M_I {\ddot R_I} = -\frac{\partial U_{\rm BO}({\bf R})}{\partial R_I}.
\end{equation}
The corresponding constant of motion is the Born-Oppenheimer Hamiltonian
\begin{equation}\label{BOH}
{\cal H}_{\rm BO} = \frac{1}{2}\sum_I M_I {\dot R}_I^2 + U_{\rm BO}({\bf R}).
\end{equation}

\subsection{Shadow Born-Oppenheimer potential energy surface}

The main cost associated with density functional based Born-Oppenheimer molecular dynamics
is the non-linear ground state minimization in Eq.\ (\ref{rho_min}), which is required prior 
to each force evaluation in Eq.\ (\ref{BOMD_eq}).
The ground state optimization is typically given through an iterative
self-consistent-field optimization procedure \cite{CGBroyden65,DGAnderson65,PPulay80,GPSrivastava84,GPKerker81,DDJohnson88} 
or with a constrained direct energy functional minimization 
using, for example, a non-linear conjugate gradient method \cite{RFletcher70,IStich89,JHutter94,VWeber08a}. 
In practice these techniques 
are always approximate and incomplete. This incompleteness, combined with initial guesses 
to the optimization that are extrapolated from previous time steps, which is used to reduce 
the computational overhead, leads to errors in the forces and a systematic drift in the
total energy or an unphysical statistical temperature distribution in canonical simulations
\cite{DRemler90,PPulay04,EMartinez15}. These problems all arise from the
non-linearity of the universal density functional $F[\rho]$.
To avoid the expensive, yet still approximate, iterative ground state optimization, we can replace
the universal functional $F[\rho]$ with an approximate expression $F^{(1)}[\rho,n]$, 
given from a linearization of $F[\rho]$ around some approximate, constant 
ground state density $n({\bf r})$, i.e. where
\begin{equation}\label{rho_1_min}
F[\rho] \approx F^{(1)}[\rho,n ] = F[n] + \int \left. \frac{\delta F[\rho]}{\delta \rho ({\bf r})}
 \right \vert_{\rho = n}  
\left( \rho ({\bf r}) - n({\bf r})\right) d{\bf r}.
\end{equation}
This gives us the corresponding $n$-dependent linearized density functional,
\begin{equation}
E^{(1)}[\rho,n] = F^{(1)}[\rho,n] + \int v({\bf R},{\bf r}) \rho({\bf r}) d{\bf r}.
\end{equation}
We now approximate the exact ground state density, $\rho_{\rm min}({\bf r})$, 
by the constrained minimization
\begin{equation}\label{rho1_min}
\rho^{(1)}[n]({\bf r}) = \arg \min_{\rho \in {\cal V}} \left\{ F^{(1)}[\rho,n] + \int v({\bf R, r}) \rho({\bf r}) d{\bf r}\right\},
\end{equation}
or more generally by a variationally stationary solution, $\rho^{(1)}_{\rm min}[n]({\bf r})$,
of the linearized functional with respect to the constrained density $\rho({\bf r}) \in {\cal V}$, 
i.e. by solving 
\begin{equation}\label{Rho_stat}
\left. \frac{\delta E^{(1)}[\rho,n]}{\delta \rho}\right \vert_{\rho \in {\cal V}} = 0.
\end{equation}
The approximate, but still {\em variationally optimized}, ground state density $\rho^{(1)}_{\rm min}[n]({\bf r})$ 
defines the $n$-dependent, approximate Born-Oppenheimer potential energy surface,
\begin{equation}\label{LinFunc}
U^{(1)}_{\rm BO}[n]({\bf R}) = F^{(1)}[\rho^{(1)}_{\rm min},n] 
+ \int v({\bf R},{\bf r}) \rho^{(1)}_{\rm min}({\bf r}) d{\bf r} + V_{nn}({\bf R}).
\end{equation}
Since the solution of Eq.\ (\ref{Rho_stat}) is given for a linearized functional, 
the iterative self-consistent-field optimization procedure is
avoided and the $n$-dependent ground state density, $\rho_{\rm min}^{(1)}[n]({\bf r})$, 
can be calculated directly in a single optimization step at only a fraction of the cost of the original problem.

The idea is to use $U^{(1)}_{\rm BO}[n]({\bf R})$ as an approximate {\em shadow potential} from which
exact forces can be calculated with a low computational cost. Such a backward analysis approach 
is frequently used in classical molecular dynamics. Instead of calculating approximate forces
for an underlying exact Hamiltonian, exact forces are calculated for an 
approximate {\em shadow Hamiltonian} that closely follows the exact Hamiltonian. A large class of 
symplectic or geometric integration schemes 
can be constructed in this way, e.g.\ the well-known velocity Verlet or leap-frog scheme
\cite{JPChannel90,McLachlan92,BJLeimkuhler04,RDEngle05}.
As I will describe below, 
the approximate Born-Oppenheimer potential energy surface, $U^{(1)}_{\rm BO}[n]({\bf R})$,
makes this shadow Hamiltonian approach possible also for non-linear self-consistent-field theory.
However, the general application of this technique has some limitations.
In particular, the conditions under which the construction of $U^{(1)}_{\rm BO}[n]({\bf R})$, 
Eqs.\ (\ref{rho_1_min})-(\ref{LinFunc}), 
works as an accurate approximation of the exact ground state potential, 
$U_{\rm BO}({\bf R})$ in Eq.\ (\ref{U_min}), 
are not well understood for general Hohenberg-Kohn theory and some caution is therefore necessary. 
In fact, it is straightforward to
chose hypothetical functional expressions and constraints under which a construction
like the one in Eqs.\ (\ref{rho_1_min})-(\ref{LinFunc}) both 
works and fails as an accurate approximation.
Here we have to assume that the right (although unknown) conditions 
are met and that the construction gives an approximation where 
$U^{(1)}_{\rm BO}[n]({\bf R}) \approx U_{\rm BO}({\bf R})$, 
with the size of the (unsigned) energy difference between
$U^{(1)}_{\rm BO}[n]({\bf R})$ and the exact Born-Oppenheimer surface $U_{\rm BO}({\bf R})$ scaling as $\Delta \rho^2$, 
where $\Delta \rho$
is the size in the difference $(\rho_{\rm min}-n)$ or the residual $(\rho_{\rm min}^{(1)}[n] - n)$.
We further assume that
\begin{equation}\label{dUdn}
\partial U^{(1)}_{\rm BO}[n]({\bf R})/\partial n \sim \rho_{\rm min}^{(1)}[n] - n.
\end{equation}
So far, the motivations for these assumptions are given mainly by model examples
and {\em a posteriori} numerical analysis. 
However, in Kohn-Sham density functional 
theory, $U^{(1)}_{\rm BO}[n]({\bf R})$
is directly related to the Harris-Foulkes functional \cite{JHarris85,MFoulkes89}, which
has the desired properties, and in the related Hartree-Fock theory the corresponding assumptions (with
the density replaced by a density matrix) also hold.
For general Hohenberg-Kohn density functional theory,
any formal proof of the validity and accuracy of the approximation
based on the variationally optimized constrained solution 
using the linearized universal functional approximation is left for the future. 

The approximation of $U_{\rm BO}({\bf R})$ 
presented above
is not the only option. Other approaches following the same basic idea 
that provides a simplified shadow potential energy surface for non-linear self-consistent 
field theory are also possible. This is of particular interest to methods that are different
from Kohn-Sham implementations of density functional theory, e.g. orbtial-free 
Thomas-Fermi density functional theory, 
excited state dynamics, methods for strongly correlated electrons, and extensions 
to polarizable force field methods \cite{AAlbaugh15,VVitale17,AAlbaugh17}.
Shadow potential methods for non-linear self-consistent field theory represents an interesting
and potentially fruitful area of research.

\subsection{Extended Lagrangian}

A straightforward application of the shadow potential energy surface $U^{(1)}_{\rm BO}[n]({\bf R})$
in the molecular dynamics scheme in Eq.\ (\ref{BOMD}) would not work. As the molecular trajectories evolve, 
the accuracy of the linearized expression around a constant density $n({\bf r})$ would become progressively worse
and if we update $n({\bf r})$ as a function of the nuclear positions ${\bf R}$ we would get
additional force terms depending on $\partial n/\partial R_I$ that are difficult to calculate.
A solution to these problems is given by including $n({\bf r})$ as a dynamical variable field, which follows 
the ground state density through a harmonic oscillator. We can do this with the extended Lagrangian,
\begin{equation}\label{XL}\begin{array}{l}
{\displaystyle {\cal L}_{\rm XBO}({\bf R},{\bf {\dot R}}, n, {\dot n}) = \frac{1}{2}\sum_I M_I {\dot R_I}^2 - U^{(1)}_{\rm BO}[n]({\bf R}) }\\
~~ \\
{\displaystyle ~~ + \frac{\mu}{2} \int ({\dot n}({\bf r}))^2 d{\bf r}}\\
~~\\
{\displaystyle ~~~~ - \frac{\mu \omega^2}{2} \iiint \left(\rho_{\rm min}^{(1)}[n]({\bf r})-n({\bf r}) \right) K({\bf r'},{\bf r})}\\
~~ \\
{\displaystyle ~~~~~~ \times  K({\bf r'},{\bf r''})\left(\rho_{\rm min}^{(1)}[n]({\bf r''})-n({\bf r''})\right)d{\bf r} d{\bf r'}d{\bf r''}},
\end{array}
\end{equation}
where the kernel $K({\bf r'},{\bf r})$ is defined as the inverse Jacobian of the residual function, 
$ \rho_{\rm min}^{(1)}[n]({\bf r})-n({\bf r})$,
i.e. such that
\begin{equation}\label{Kernel_1}
{\displaystyle \int K({\bf r},{\bf r'}) \left(\frac{\delta \rho_{\rm min}^{(1)}[n]({\bf r'})}{\delta n({\bf r''})}
- \frac{\delta n({\bf r'})}{\delta n({\bf r''})}\right) d{\bf r'} = \delta({\bf r-r''})}.
\end{equation}
In the extended Lagrangian, $\mu$ is a fictitious mass parameter and $\omega$ 
is the frequency of the extended harmonic oscillator.

\subsection{Equations of motion from a classical (Born-Oppenheimer-like) adiabatic approximation}

To formulate a quantum based molecular dynamics scheme 
we need the equations of motion corresponding to the postulated Lagrangian in Eq.\ (\ref{XL})
in a simple and practically feasible form.
We can achieve this by deriving the Euler-Lagrange equations of motion in 
a classical adiabatic limit, where we assume that the oscillator frequency $\omega$ 
is high compared to the fastest nuclear degrees of freedom $\Omega$,
i.e. when $\omega \gg \Omega$.  This is similar to the quantum adiabatic limit of the Born-Oppenheimer approximation 
in electronic structure theory, which also imposes a decoupling between the nuclear and the electronic
degrees of freedom. The classical adiabatic separation is applied to
the extended Lagrangian assuming the approximation $\omega^2/\Omega^2 \rightarrow \infty$.
We perform this classical adiabatic limit in the derivation of the equations of motion by letting
$\omega \rightarrow \infty$ and the mass parameter $\mu \rightarrow 0$,
while $\mu \omega \rightarrow {\rm constant}~ k$, under the assumptions that 
$K$ and $\delta K/\delta n$ are bounded and that 
\begin{equation}
\delta U^{(1)}[n]/\delta n \sim (\rho_{\rm min}^{(1)}[n]-n) \sim 1/\omega^{2}.
\end{equation}
In this limit \cite{ANiklasson14} we get the equations of motion
\begin{equation}\label{EqMot}\begin{array}{l}
{\displaystyle M_I {\ddot R}_I = \left.-\frac{\partial U^{(1)}_{\rm BO}[n]({\bf R})}{\partial R_I}\right\vert_{n}},\\
~~\\
{\displaystyle {\ddot n}({\bf r}) = -\omega^2 \int K({\bf r},{\bf r'})\left(\rho_{\rm min}^{(1)}[n]({\bf r'})-n({\bf r'})\right)d{\bf r'}}.
\end{array}
\end{equation}
No iterative ground state optimization is necessary prior to the force evaluations, since the
calculation of $\rho_{\rm min}^{(1)}[n]({\bf r})$ in Eq.\ (\ref{Rho_stat}), which defines $U^{(1)}_{\rm BO}[n]({\bf R})$, 
is given directly in a single step thanks to the
linearized form of the universal functional, $F^{(1)}[n,\rho]$. Both the nuclear and the electronic degrees of freedom
can be integrated with accurate time-reversible or symplectic integration schemes.
The nuclear force evaluation in Eq.\ (\ref{EqMot}) is calculated with the partial derivatives 
under constant density $n({\bf r})$, since $n({\bf r})$ occurs as a dynamical variable,
and all $\delta  U^{(1)}_{\rm BO}[n]/\delta \rho$ terms vanish because of Eq.\ (\ref{Rho_stat}).
In this way we do not rely on the Hellmann-Feynman theorem for the exact ground state and no additional
adjustment terms to the forces are needed despite an incomplete ground state optimization with respect
to $U_{\rm BO}({\bf R})$. The kernel $K({\bf r},{\bf r'})$, which was introduced as an inner-product
norm of the extended harmonic oscillator in Eq.\ (\ref{XL}), appears like a preconditioner in the equation
of motion for the extended electronic degrees of freedom and makes $n({\bf r})$ behave as if
it oscillates around a close approximation to the exact ground state density $\rho_{\rm min}({\bf r})$.
This can be understood from the fact that the kernel $K$ acts like in a quadratically convergent 
Newton minimization, such that
\begin{equation}\begin{array}{l}\label{EffOfK}
{\displaystyle \int K({\bf r},{\bf r'})\left(\rho_{\rm min}^{(1)}[n]({\bf r'})-n({\bf r'})\right)d{\bf r'} }\\
~~\\
{\displaystyle ~~ = (\rho_{\rm min}({\bf r}) - n({\bf r})) + {\cal O}\left[(\rho_{\rm min}^{(1)}[n] -n)^2\right] }.
\end{array}
\end{equation}

The constant of motion can be derived from Noether's theorem, where
\begin{equation}
{\displaystyle \sum_I \frac{\partial {\cal L}_{\rm XBO}}{\partial {\dot R}_I} 
+ \frac{\partial {\cal L}_{\rm XBO}}{\partial {\dot n}} - {\cal L}_{\rm XBO} = {\rm Constant}}.
\end{equation}
In the adiabatic limit when $\omega \rightarrow \infty$ and $\mu \rightarrow 0$ 
this constant of motion is given by the Born-Oppenheimer shadow Hamiltonian
\begin{equation}\label{ShadowH1}
{\displaystyle {\cal H}_{\rm XBO}^{(1)} = \frac{1}{2} \sum_I M_I {\dot R}_I^2 + U^{(1)}_{\rm BO}[n]({\bf R})},
\end{equation}
which closely follows the exact Born-Oppenheimer constant of motion ${\cal H}_{\rm BO}$ in Eq.\ (\ref{BOH}). 
As long as there is
a clear adiabatic separation between the fastest nuclear degrees of freedom $\Omega$ and $\omega$, we thus expect
a molecular dynamics scheme that closely follows the exact Born-Oppenheimer potential energy surface.
It can be shown that this classical adiabatic separation, assumed in the derivation of the equations of motion, is automatically fulfilled for 
any material system as long as we use integration time steps of the same order as 
in a regular direct Born-Oppenheimer molecular dynamics simulation, in combination with, for example, 
the Verlet algorithm \cite{ANiklasson14}, as will be discussed below.

\subsection{Integrating the equations of motion}

To integrate the equations of motion we first need to define 
the initial boundary conditions at time $t = t_0$ for the electronic degrees of freedom,
\begin{equation} \label{IBC} \begin{array}{l}
{\displaystyle n(t_0) = \rho_{\rm min}(t_0)},\\
~~\\
{\displaystyle {\dot n}(t_0) = 0},
\end{array}
\end{equation}
and the nuclear degrees of freedom,
\begin{equation} \begin{array}{l}
{\displaystyle {\bf R}(t_0) = {\bf R}_0},\\
~~\\
{\displaystyle {\bf {\dot R}}(t_0) = {\bf {\dot R}}_0}.\\
\end{array}
\end{equation}
Here I have dropped the ${\bf r}$ in $n({\bf r},t)$ and included time $t$ for clarity.
The equations of motion can then in principle be integrated with any
standard method used in classical molecular dynamics, e.g. a combined leap-frog and Verlet scheme:
\begin{equation}\label{Integration}\begin{array}{l}
{\displaystyle {\bf {\dot R}}(t + \frac{\delta t}{2})  = {\bf {\dot R}}(t) + \frac{\delta t}{2} {\bf {\ddot R}}(t)}\\
~~\\
{\displaystyle {\bf R}(t + \delta t) = {\bf R}(t) + \delta t {\bf {\ddot R}}(t + \frac{\delta t)}{2}}\\
~~\\
{\displaystyle n(t + \delta t) = 2n(t) - n(t - \delta t) + \delta t^2 {\ddot n}(t)}\\
~~\\
{\displaystyle {\bf {\dot R}}(t + \delta t) = {\bf {\dot R}}(t + \frac{\delta t}{2}) 
+ \frac{\delta t}{2} {\bf {\ddot R}}(t+\delta t)}.\\
\end{array}
\end{equation}
However, there is one important difference to classical molecular dynamics.
In the presence of numerical noise, due to finite numerical precision or an approximate
sparse matrix algebra used to achieve linear scaling complexity \cite{MCawkwell12,MArita14,CNegre16,VVitale17,THirakawa17}, 
the electronic degrees of freedom might start to diverge from the exact ground state density, which is determined by the nuclear
degrees of freedom. This happens as soon
as we use any integration scheme with a perfect time reversibility. In this case, numerical
noise is never removed but is accumulated in the electronic degrees of freedom until
divergence occurs when $n(t)$ and ${\bf R}(t)$ no longer are aligned, i.e.\ 
when $n(t)$ no longer is sufficiently close to the electronic ground state 
for the external potential $v({\bf R,r})$ in Eq.\ (\ref{rho_min}).
Some form of electron dissipation (or {\em synchronization} \cite{LPecora90,LPecora15}) is needed to avoid this breakdown.
We typically use a modified Verlet integration scheme to introduce a weak dissipation without
causing any significant drift in the total energy \cite{ANiklasson09,PSteneteg10,GZheng11}.
In this case the Verlet integration of the electronic degrees of freedom, Eq.\ (\ref{Integration}), is given by
\begin{equation}\label{VRL_Damp}\begin{array}{l}
{\displaystyle n(t + \delta t) = 2n(t) - n(t - \delta t) + \delta t^2 {\ddot n}(t)}\\
~~\\
{\displaystyle ~~~~~~~~~~~~ + \alpha \sum_{k = 0}^{K_{\rm max}} c_k n(t-k \delta t)},\\
\end{array}
\end{equation}
for some optimized set of coefficients, $\alpha$ and $\{c_k\}$ in Tab.\ \ref{Tab_Coef}.
The first three terms on the right hand side correspond to the regular Verlet integration
and the last term is a dissipative force term.
Alternative damping schemes can also be adapted to the integration of wavefunctions \cite{PSteneteg10}
and higher-order symplectic algorithms \cite{AOdell09,AOdell11}.
Recently a more general class of integrators was introduced that was shown to work well
in combination with polarizable force field models \cite{AAlbaugh15,VVitale17,AAlbaugh17}.

\subsection{Natural adiabatic separation}

In the modified Verlet integration scheme, Eqs.\ (\ref{EqMot}) and (\ref{VRL_Damp}) above, the term 
\begin{equation}
\delta t^2 {\ddot n}({\bf r}) = -\delta t^2 \omega^2 \int K({\bf r},{\bf r'})(\rho^{(1)}_{\rm min}[n]({\bf r'}) - n({\bf r'}))d{\bf r'}
\end{equation}
includes a dimensionless variable $\kappa = \delta t^2 \omega^2$. To achieve stability in
the Verlet integration, this constant should to be smaller than two \cite{ANiklasson09}, i.e.\ $\kappa < 2$.
For normal integration time steps in direct Born-Oppenheimer molecular dynamics, 
$\delta t$ is some fraction $T_\Omega/m$ of the fastest period of the nuclear motion, $T_\Omega$.
Typically $m\in[15,20]$, which means that $\omega$ automatically is multiple times higher than 
the corresponding frequency, $\Omega = 2\pi /T_\Omega$,  of the nuclear degrees of freedom \cite{ANiklasson14}. 
This guarantees a natural and {\em system-independent} classical adiabatic separation 
between the nuclear and electronic degrees of 
freedom. This also motivates the separation 
assumed in the derivation of the equations of motion.

\begin{table}[t]
  \centering
  \caption{\protect Optimized $\alpha$ and $\kappa = \delta t^2 \omega^2$ values and the $c_k$ coefficient
    for the dissipative electronic force term in Eq.\ (\ref{VRL_Damp}), see Refs. \cite{ANiklasson09,PSteneteg10,GZheng11}.
  }\label{Tab_Coef}
  \begin{ruledtabular}
  \begin{tabular}{llccccccccc}
    $K_{\rm max}$ & $\kappa$ & \!\!$\alpha$\!\! & \!$c_{0}$ & $c_{1}$ & $c_{2}$ & $c_{3}$ & $c_{4}$ &
$c_{5}$ & $c_{6}$ & $c_{7}$ \\
     \hline
     5  & 1.82 & \!\!0.018\!\! &-6    &   14   &  -8  & -3  &  4  &  -1  &     &     \\
     6  & 1.84 & \!\!0.0055\!\! &-14    &   36  & -27  & -2  &  12 &  -6  & 1   &    \\
     7  & 1.86 & \!\!0.0016\!\! &-36    &   99  & -88  & 11  &  32 & -25  & 8   &  -1  \\
  \end{tabular}
  \end{ruledtabular}
\end{table}

\subsection{A few remarks}

The extended Lagrangian density functional Born-Oppenheimer molecular dynamics scheme, formulated above, 
is based on three key ideas: {\em i)} the first idea is to define an approximate shadow Born-Oppenheimer
potential energy surface based on a constrained minimization (or a stationary solution) of an energy expression with
the universal functional that is linearized around some sufficiently accurate approximation, $n({\bf r})$, 
to the exact ground state density $\rho_{\rm min}({\bf r})$; {\em ii)} the second idea is to 
use this linearized functional and to include the density, $n({\bf r})$, as a dynamical field variable, in an extended Lagrangian
formulation, where $n({\bf r})$ follows the optimized ground state density, $\rho^{(1)}[n]({\bf r})$,
through a generalized harmonic oscillator including an inner product kernel 
$K({\bf r},{\bf r'})$; and {\em iii)} the third idea is to apply an classical (Born-Oppenheimer-like) adiabatic approximation, 
assuming  $\omega/\Omega \rightarrow \infty$, in the derivation of the equations of motion.
The result is a fast and efficient simulation framework that closely follows the dynamics determined
by the exact Born-Oppenheimer potential energy surface.

The theory above was presented in terms of general Hohenberg-Kohn density functional theory
and should thus, in principle, be applicable to a broad range of implementations, 
including orbital-free Thomas-Fermi-Dirac
and Kohn-Sham theory. The electron density, chosen as the dynamical variable for
the electronic degrees of freedom above, can easily be replaced by, for example, the single-particle
wavefunctions \cite{PSteneteg10} or the density matrix \cite{PSouvatzis14,MArita14,VVitale17}. 
Modifications also makes the extended Lagrangian framework above
applicable to classical molecular dynamics schemes with polarizable force fields \cite{AAlbaugh15,KNomura15,AAlbaugh17}.

\section{Comparison to extended Lagrangian Car-Parrinello molecular dynamics}

Car and Parrinello introduced a general a class of extended Lagrangians \cite{RCar85,DMarx00,JHutter12}. 
For Kohn-Sham density functional (or Hartree-Fock) theory the Carr-Parrinello Lagrangian can be represented by
\begin{equation}\begin{array}{l}
{\displaystyle {\cal L}_{\rm CP} = \frac{1}{2}\sum_I M_I{\dot R}_I^2 + \frac{1}{2} \sum_{i} \mu_i \langle {\dot \psi}_i \vert {\dot \psi}_i \rangle }\\
~~\\
{\displaystyle - U({\bf R}, \{\psi_i\}) + \sum_{ij} \lambda_{ij} \left(\langle \psi_i\vert \psi_j \rangle - \delta_{ij}\right)},
\end{array}
\end{equation}
where $U({\bf R}, \{\psi_i\})$ is the potential energy calculated for an orthonormal set of occupied
instantaneous single-particle orbitals $\{\psi_i\}$. The last term includes the orthonormalization constraints 
and $\mu_i$ are fictitious electron mass parameters. The corresponding Euler-Lagrange equations of motion are
\begin{equation}\label{CPMD}\begin{array}{l}
{\displaystyle M_I{\ddot R}_I = \left. \frac{\partial U({\bf R}, \{\psi_i\})}{\partial R_I} \right \vert_{\psi}
+ \sum \lambda_{ij} \frac{\partial}{\partial R_I} \langle \psi_i\vert \psi_j\rangle},\\
~~\\
{\displaystyle \mu_i {\ddot \psi_i} = - H({\bf R},\{\psi_i\}) \psi_i + \sum_{j} \lambda_{ij} \psi_j}.
\end{array}
\end{equation}
$H({\bf R},\{\psi_i\})$ is the effective single-particle Hamiltonian, e.g. the Kohn-Sham Hamiltonian 
(or the Fockian).
The constant of motion is given by the Car-Parinello Hamiltonian
\begin{equation}\label{CPH}
{\cal H}_{\rm CP} = \frac{1}{2} \sum_I M_I {\dot R}_I^2 + U({\bf R}, \{\psi_i\})
+ \frac{1}{2} \sum_{i} \mu_i \langle {\dot \psi}_i \vert {\dot \psi}_i \rangle .
\end{equation}
To simulate a dynamics that approximates Born-Oppenheimer properties, the last electronic kinetic energy term 
in ${\cal H}_{\rm CP}$
needs to be kept small, which is possible if the nuclear and the electronic degrees of freedom are
adiabatically decoupled such that $\langle {\dot \psi}_i \vert {\dot \psi}_i \rangle$ does not
increase because of the nuclear motion.
A decoupling can be achieved by choosing sufficiently small mass parameters $\mu_i$.
Unfortunately, this reduces the largest integration time step $\delta t$ that can be used.
The largest possible size of $\mu_i$ depends of the electronic gap of the system and
the appropriate choice of $\mu_i$ and $\delta t$ may thus require some prior estimates 
and expertise to provide efficient and accurate simulations \cite{DMarx00,PTangney02,PTangney06,JHutter12}.

\subsection{Similarities}

There are several interesting relations between Car-Parrinello and extended Lagrangian Born-Oppenheimer molecular dynamics: (1) both
are extended Lagrangian approaches to first principles molecular dynamics
simulations, providing general, unified frameworks that easily can be adapted to
a number of different problems also beyond first principles molecular dynamics; 
(2) in contrast to regular direct Born-Oppenheimer molecular dynamics, 
no iterative self-consistent-field optimization procedure is needed;
(3) while extended Lagrangian Born-Oppenheimer molecular dynamics 
assumes an classical adiabatic separation in the derivation 
of the equations of motion, Car-Parrinello molecular dynamics enforces the corresponding adiabatic
decoupling between the nuclear and electronic degrees of freedom in the
integration of the equations of motion, e.g. by choosing appropriate
electron mass parameters, $\mu_i$, and integration time step, $\delta t$; 
(4) in both methods electronic degrees
of freedom are used as extended dynamical variables that oscillate around the
exact ground state or some close approximation to the ground state; 
(5) the equations of motion for the nuclear and the electronic
degrees of freedom can both be integrated with time-reversible or symplectic
integration schemes that avoid (or drastically reduce) the problem with
a systematic drift in the total energy; and (6) in the limit when the integration 
time step $\delta t \rightarrow 0$ (and when $\mu \rightarrow 0$ in Car-Parrinello molecular dynamics), 
both methods converge to exact Born-Oppenheimer molecular dynamics. In this way extended Lagrangian
Born-Oppenheimer molecular dynamics maintains several important benefits of Car-Parrinello
molecular dynamics simulations.

\subsection{Differences}

Apart from these similarities there are also several significant differences: (1)
a single full diagonalization is required in each time step of extended Lagrangian Born-Oppenheimer
molecular dynamics, i.e.\ in the ground state optimization 
of the linearized functional expression in Eq.\ (\ref{Rho_stat}), compared to the more simplified expression 
of matrix-vector multiplications, $H\psi_i$, necessary in Car-Parrinello molecular dynamics 
in Eq.\ (\ref{CPMD}); (2) on the other hand, 
the non-linear constraints, e.g.\ of wavefunction orthonormality or the equivalent density matrix idempotency, 
is avoided in extended Lagrangian Born-Oppenheimer molecular dynamics;
(3) the adiabatic decoupling between the nuclear and the electronic dynamical variables is
system independent and automatically fulfilled in extended Lagrangian Born-Oppenheimer
molecular dynamics as long as normal Born-Oppenheimer
integration time steps are used and does not depend on the size of the electronic gap
as in Car-Parrinello molecular dynamics; 
(4) the problem of choosing appropriate mass parameters, $\mu_i$, in Car-Parrinello molecular dynamics
is replaced by the challenge of calculating or approximating the kernel, $K({\bf r},{\bf r'})$, 
in extended Lagrangian Born-Oppenheimer molecular dynamics;
(5) extended Lagrangian Born-Oppenheimer molecular dynamics can be applied both for metals and insulators with integration
time steps, $\delta t$, which are of the same size as in direct Born-Oppenheimer molecular dynamics, 
whereas Car-Parrinello molecular dynamics requires
a shorter integration time step as the electronic gap is reduced, which increases
the computational overhead \cite{DMarx00}; (6) the constant of motion in extended Lagrangian Born-Oppenheimer
molecular dynamics, i.e. the shadow Hamiltonian in Eq.\ (\ref{ShadowH1}), 
follows closely the "exact" total Born-Oppenheimer energy in Eq.\ (\ref{BOH}), whereas the constant of motion 
in Car-Parrinello molecular dynamics, Eq.\ (\ref{CPH}), 
includes an additional kinetic energy term from the electronic degrees of freedom; (7)
while both methods converge to the exact Born-Oppenheimer limit as the integration time step $\delta t \rightarrow 0$,
{\em the Car-Parrinello method is a first-order method} \cite{GPastore91, FBornemann98}, since the
the difference $\Delta_{\rm CP-BO} = |\psi_i - \psi_{i,{\rm BO}}|$
between the dynamical variables of the electronic degrees of freedom, $\{\psi_i\}$, and the exact ground state
solution, $\psi_{i,{\rm BO}}$, scales as $\Delta_{\rm CP-BO} \sim \mu^{1/2} \sim \delta t$, whereas
{\em extended Lagrangian Born-Oppenheimer molecular dynamics can be regarded as a second-order method}, since the
convergence of the corresponding difference scales as 
$\Delta_{\rm XLBO-BO} = \|\rho_{\rm min}-n \| \sim \|\rho^{(1)}_{\rm min}[n] - n\| \sim \omega^{-2} \sim \delta t^2$ 
\cite{PSouvatzis13,ANiklasson14}; and (8) the error in the potential energy surface with respect to the 
exact Born-Oppenheimer potential energy surface for each new configuration during a simulation is therefore of second order ($\delta t^2$)
for Car-Parrinello molecular dynamics, whereas it is of fourth order ($\delta t^4$) for extended Lagrangian Born-Oppenheimer 
molecular dynamics (see Fig.\ \ref{fg_10}).

Despite these differences, extended Lagrangian Born-Oppenheimer molecular dynamics 
can be understood in the spirit of Car-Parrinello molecular dynamics 
and be seen as a next generation (second-order) extended Lagrangian first principles 
molecular dynamics that follows in the footsteps of the seminal 
work by Car and Parrinello three decades ago.

\section{Finite temperature Kohn-Sham formulation of extended Lagrangian Born-Oppenheimer molecular dynamics}

Extended Lagrangian density functional based Born-Oppenheimer molecular dynamics can also be applied to
finite temperature ensembles \cite{ANiklasson11}. Here I will use a thermal Kohn-Sham formulation of 
density functional theory \cite{RParr89,NMermin65} to introduce this generalization, including 
the inner-product kernel for temperature dependent, fractional occupation numbers.

\subsection{Kohn-Sham shadow potential}

In Kohn-Sham density functional theory the universal functional, $F[\rho]$, in Eq.\ (\ref{rho_min}), is 
implemented with
\begin{equation}
F_{\rm KS}[\rho] = -\frac{1}{2}\sum_{i} f_i \langle \psi_i \vert \nabla^2 \vert \psi_i \rangle 
+ \frac{1}{2} \iint \frac{\rho({\bf r}) \rho({\bf r'})}{\vert{\bf r-r'}\vert} d{\bf r}d{\bf r'} + E_{\rm xc}[\rho],
\end{equation}
where the $v$-representable density is given by a sum of orthonormal single-particle orbitals, i.e. with 
\begin{equation}
\rho = \sum_{i} f_i \vert \psi_i \vert^2
\end{equation}
and 
\begin{equation}
\langle \psi_i\vert \psi_j \rangle = \delta_{ij}.
\end{equation} 
Here $f_i \in [0,1]$ are Fermi occupation factors. At zero electronic temperature $f_i = 1$ for occupied state 
and $f_i = 0$ for unoccupied state. Special cases occurring due to spin degrees of freedom are left out for simplicity.
The last term, $E_{\rm xc}[\rho]$, is the exchange correlation
functional, which in practice has to be approximated. Our linearized functional, ${F}^{(1)}_{\rm KS}[\rho,n]$, becomes 
\begin{equation}\begin{array}{l}
{\displaystyle F^{(1)}_{\rm KS}[\rho,n] = -\frac{1}{2}\sum_{i} f_i \langle \psi_i \vert \nabla^2 \vert \psi_i \rangle }\\
~~\\
{\displaystyle ~~ + \frac{1}{2} \iint \frac{( 2\rho({\bf r})-n({\bf r})) n({\bf r'})}{\vert{\bf r-r'}\vert} d{\bf r}d{\bf r'} }\\
~~\\
{\displaystyle ~~~~ + E_{\rm xc}[n] + \int v_{\rm xc}[n]({\bf r})(\rho({\bf r})-n({\bf r}))d{\bf r}},
\end{array}
\end{equation}
where $v_{\rm xc}[n]({\bf r})$ is the functional derivative of $E_{\rm xc}[\rho]$ calculated at $\rho = n$. 
The variationally optimized density, $\rho^{(1)}_{\rm min}[n]({\bf r})$, for the linearized functional, including 
the external potential energy, $\int v({\bf R, r}) \rho({\bf r}) d{\bf r}$,
is given from the constrained optimization, as in Eq.\ (\ref{Rho_stat}), under the conditions
that $\rho = \sum_{i} f_i \vert \psi_i \vert^2$ and $\langle \psi_i\vert \psi_j \rangle = \delta_{ij}$. This
leads to the linearized Kohn-Sham eigenvalue equation 
\begin{equation}
H_{\rm KS}[n]\vert \psi^{(1)}_{{\rm min},i}\rangle  = \epsilon_i \vert \psi^{(1)}_{{\rm min},i}\rangle  ~~~,
\end{equation}
where 
\begin{equation}
\rho^{(1)}_{\rm min}[n]({\bf r}) = \sum_{i} f_i \vert \psi^{(1)}_{{\rm min},i}({\bf r})\vert^2,
\end{equation}
and with the Kohn-Sham Hamiltonian 
\begin{equation}
H_{\rm KS}[n] = -\frac{1}{2} \nabla^2 + \int \frac{n({\bf r'})}{\vert {\bf r-r'}\vert} d{\bf r} +  v_{\rm xc}[n]({\bf r}) + v({\bf R, r}).
\end{equation}
The Kohn-Sham shadow potential energy surface is then given by
\begin{equation}\begin{array}{l}
{\displaystyle U^{(1)}_{\rm KS}[n]({\bf R}) = 
-\frac{1}{2}\sum_{i} f_i \langle \psi^{(1)}_{{\rm min},i} \vert \nabla^2 \vert \psi^{(1)}_{{\rm min},i} \rangle } \\
~~\\
{\displaystyle ~~ + \frac{1}{2} \iint \frac{( 2\rho^{(1)}_{\rm min}[n]({\bf r})-n({\bf r})) n({\bf r'})}{\vert{\bf r-r'}\vert} d{\bf r}d{\bf r'}}\\ 
~~\\
{\displaystyle ~~~~ + E_{\rm xc}[n] + \int v_{\rm xc}[n]({\bf r})(\rho^{(1)}_{\rm min}[n]({\bf r})-n({\bf r}))d{\bf r}}\\
~~\\
{\displaystyle ~~~~~~ + \int v({\bf R},{\bf r}) \rho^{(1)}_{\rm min}({\bf r}) d{\bf r} + V_{nn}({\bf R})}.
\end{array}
\end{equation}
No iterative self-consistent field optimization is necessary in the 
calculation of the linearized Kohn-Sham potential $U^{(1)}_{\rm KS}[n]({\bf R})$.

\subsection{Free energy shadow potential}

At finite electronic temperatures, $T_e > 0$, the occupation factors, ${\bf f} = \{ f_i\}$, are fractional and
determined by the Fermi-Dirac function,
\begin{equation}
{\displaystyle f_i = \left[ e^{\beta(\epsilon_i - \mu_e)} +1 \right]^{-1}},
\end{equation}
where $\mu_e$ is the chemical potential, which is set such that the sum of the $f_i$'s is the number of occupied states, i.e. $\sum_i f_i = N_{\rm occ}$,
and $\beta$ is the inverse temperature, i.e. $\beta = 1/(k_B T_e)$.
In the case of finite electronic temperatures \cite{RParr89} we need to include an electronic entropy contribution, 
$S[{\bf f}]$, in a free energy generalization, $\Omega^{(1)}_{\rm KS}[n]({\bf R})$, 
of the shadow potential, $U^{(1)}_{\rm KS}[n]({\bf R})$, where
\begin{equation}
\Omega^{(1)}_{\rm KS}[n]({\bf R}) = U^{(1)}_{\rm KS}[n]({\bf R}) - T_e S[{\bf f}].
\end{equation}
The entropy function for the Fermi-Dirac distribution is
\begin{equation} \label{S_entropy}
S[{\bf f}] = -k_B \sum_i\left( f_i \ln f_i + (1-f_i) \ln (1-f_i) \right).
\end{equation}
The entropy is needed to obtain variationally correct (and thus conservative) force terms with 
a simple Hellmann-Feynman-like expression \cite{MWeinert92,RWentzcovitch92,ANiklasson08b}. 

\subsection{Extended Lagrangian}

The temperature-dependent extended Lagrangian is a straightforward generalization of Eq.\ (\ref{XL}), i.e.
\begin{equation}\label{XL_FT}\begin{array}{l}
{\displaystyle {\cal L}({\bf R},{\bf {\dot R}}, n, {\dot n}) = \frac{1}{2}\sum_I M_I {\dot R_I}^2 - \Omega^{(1)}_{\rm KS}[n]({\bf R}) }\\
~~ \\
{\displaystyle ~~ + \frac{\mu}{2} \int ({\dot n}({\bf r}))^2 d{\bf r}}\\
~~\\
{\displaystyle ~~~~ - \frac{\mu \omega^2}{2} \iiint \left(\rho_{\rm min}^{(1)}[n]({\bf r})-n({\bf r}) \right) K({\bf r'},{\bf r})}\\
~~ \\
{\displaystyle ~~~~~~ \times  K({\bf r'},{\bf r''})\left(\rho_{\rm min}^{(1)}[n]({\bf r''})-n({\bf r''})\right)d{\bf r} d{\bf r'}d{\bf r''}},
\end{array}
\end{equation}
with the corresponding equations of motion in the adiabatic limit,
\begin{equation}\label{EqMotFT}\begin{array}{l}
{\displaystyle M_I {\ddot R}_I = \left.-\frac{\partial \Omega^{(1)}_{\rm KS}[n]({\bf R})}{\partial R_I}\right\vert_{n}},\\
~~\\
{\displaystyle {\ddot n}({\bf r}) = -\omega^2 \int K({\bf r},{\bf r'})\left(\rho_{\rm min}^{(1)}[n]({\bf r'})-n({\bf r'})\right)d{\bf r'}},
\end{array}
\end{equation}
and a constant of motion
\begin{equation}\label{ShadowH1}
{\displaystyle {\cal F}^{(1)}_{\rm KS} = \frac{1}{2} \sum_I M_I {\dot R}_I^2 + \Omega^{(1)}_{\rm KS}[n]({\bf R})}.
\end{equation}
Apart from replacing with $U^{(1)}_{\rm BO}[n]({\bf R})$ with 
$\Omega^{(1)}_{\rm KS}[n]({\bf R})$ and ${\cal H}_{\rm BO}^{(1)}$ with ${\cal F}^{(1)}_{\rm KS}$
everything in the thermal Kohn-Sham density-functional-based dynamics is thus the same as 
for the zero-temperature Hohenberg-Kohn density functional formulation. 
The same integration technique, e.g.\ the leap-frog and modified Verlet scheme as in Eqs.\ (\ref{Integration}) and (\ref{VRL_Damp}), can also be used.

\subsection{A few remarks} 

The Kohn-Sham shadow potential energy surface 
$U^{(1)}_{\rm KS}[n]({\bf R})$ can be reformulated to the well-known
Harris-Foulkes functional \cite{JHarris85,MFoulkes89}, 
where $n$ becomes an approximate ''input density'' (or overlapping atomic densities) 
and $\rho^{(1)}[n]$ the "output density". However, there is both a conceptual and practical difference between the Kohn-Sham formulation of
$U^{(1)}_{\rm BO}[n]({\bf R})$ and the Harris-Foulkes functional. Instead of an approximate energy expression, providing 
a faster energy convergence in a sequence of input and output densities during a self-consistent-field
optimization (or an approximate functional
using overlapping atomic densities), $U^{(1)}_{\rm BO}[n]({\bf R})$
is here introduced as an {\em exact} shadow potential energy surface that is part of 
a {\em dynamical} extended Lagrangian framework in a classical adiabatic limit (as $\omega \rightarrow \infty$),
which can be used to calculate {\em both energies and forces} with high precision. 
The problem that the Harris-Foulkes functional is not variational with respect to ''the input density'', $n({\bf r})$, 
or that the Hellmann-Feynman theorem is not fulfilled, which prohibits force evaluations
for the Harris-Foulkes functional without additional correction terms, is here never an issue, since $n({\bf r})$ 
occurs as a dynamical field variable within the extended Lagrangian framework. Moreover,
since $n({\bf r})$ closely follows the exact ground state, the accuracy
in the force evaluations will be high such that the constant of motion during
a simulation will follow the exact Born-Oppenheimer solution with an error only of
second order in the deviation of $n({\bf r})$ from the exact ground state density.
In practice, the main error of an extended Lagrangian Born-Oppenheimer simulation, compared 
to the theoretically ``exact'' simulation, will therefore be dominated by the 
local truncation error of the Verlet integration algorithm, caused by the finite size of
the integration time step.

\section{The kernel $K({\bf r}, {\bf r})$}

The definition of the kernel, $K({\bf r},{\bf r'})$ in Eq.\ (\ref{Kernel_1}), does not only lead to a simple form
of the equations of motion, it also has a direct physical implication. The kernel 
makes the dynamical variable density, $n({\bf r})$, oscillate as if it was virtually centered 
around the exact Born-Oppenheimer ground state density, $\rho_{\rm min}({\bf r})$.
This improves both the stability and the accuracy of extended Lagrangian dynamics.

\subsection{Coarse graining}

The kernel $K({\bf r},{\bf r'})$ is given as the inverse of the
Jacobian of the residual function, $\rho^{(1)}_{\rm min}[n]({\bf r}) - n({\bf r})$, i.e. as
\begin{equation}
K({\bf r},{\bf r'}) = \left[\frac{\delta \rho_{\rm min}^{(1)}[n]({\bf r})}{\delta n({\bf r'})} - \delta({\bf r- r'})  \right]^{-1}.
\end{equation}
In a coarse-grained algebraic matrix-vector notation, e.g. where $n$ is a vector of length $N$ 
with the values of the integrated charge densities for each atom or the values of discretized
point charges in space, we can describe the kernel as 
\begin{equation}\label{KK}
K = \left[\left\{\frac{\delta \rho[n]}{\delta n}\right\} - I\right]^{-1},
\end{equation}
where $K$ is a $N \times N$ matrix, $\{{\delta \rho[n]}/{\delta n}\}$ 
is a set of $N$ number of $N \times 1$ column vectors, and $I$ is the $N \times N$ identity matrix.
The corresponding equations of motion for the electronic degrees of freedom, Eq.\ (\ref{EqMot}), 
in this matrix-vector notation is
\begin{equation} \label{n_v}
{\displaystyle {\ddot n} = -\omega^2  K\left(\rho[n]-n\right)}.
\end{equation}
The notation ``min'' and $(1)$ in $\rho^{(1)}_{\rm min}[n]$ have here been dropped for simplicity
and they will not be used in the remaining part of the paper.

\subsection{Quantum response calculations with fractional occupation numbers}

At $T_e = 0$ the charge response, $\{{\delta \rho[n]}/{\delta n}\}$, can be calculated
using regular quantum perturbation theory, e.g.\ Rayleigh-Shr\"{o}dinger perturbation theory. 
For example, if $n$ is a vector of atomic net charges, then we can calculate the
response ${\delta \rho[n]}/{\delta n}$ as follows. 
For a perturbation of an atomic charge $\delta n(R_I)$ at position $R_I$ we first calculate
the response in the Coulomb potential $\delta V_{\rm Coul}({\bf R})/\delta n(R_I)$. 
We can then use $\delta V_{\rm Coul}({\bf R})/\delta n(R_I)$ as a perturbation of the Kohn-Sham Hamiltonian, i.e. 
\begin{equation}
H = H^{(0)}_{\rm KS}[n] + \lambda_I H^{(1)}_I,
\end{equation}
where
\begin{equation}
H^{(1)}_I = \delta V_{\rm Coul}({\bf R})/\delta n(R_I).
\end{equation}
Using (non-degenerate) Rayleigh-Shr\"{o}dinger perturbation theory, we can then calculate the first-order response
in the wavefunctions,
\begin{equation}
\vert \psi^{(1)}_i\rangle = \sum_{j \ne i} \frac{\langle \psi^{(0)}_j\vert H^{(1)}_I\vert \psi^{(0)}_i\rangle}{\epsilon^{(0)}_i - \epsilon^{(0)}_j} \vert \psi^{(0)}_j\rangle,
\end{equation} 
from which the first order response in the charge on all the atoms, i.e. the column vector ${\delta \rho[n]}/{\delta n(R_I)}$, can be calculated.
This is repeated for perturbations on each atom $I$.  Once all the column vectors of the charge response, 
$\{{\delta \rho[n]}/{\delta n(R_I)}\}$, in Eq.\ (\ref{KK}) have been calculated, we subtract the identity matrix $I$ 
and perform the matrix inverse to get the kernel $K$.

At $T_e > 0$ it is in principle possible to generalize Rayleigh-Shr\"{o}dinger perturbation theory to
canonical ensembles including fractional occupation factors \cite{YNishimoto17}. However,
an interesting alternative to Rayleigh-Shr\"{o}dinger-based calculations is provided by density matrix 
perturbation theory \cite{ANiklasson04,VWeber04,VWeber05}, which also enables response
calculations with linear scaling complexity.
Density matrix perturbation theory was recently generalized
to finite temperature ensembles \cite{ANiklasson15}. This canonical density matrix perturbation 
theory allows response calculations also for degenerate problems, including systems with fractional 
occupation numbers such as metals and is thus well suited for thermal Kohn-Sham density functional theory.
A full description of canonical density matrix perturbation theory, 
including a detailed algorithm, is given in Ref.\ \cite{ANiklasson15}.

\subsection{Approximating the kernel}

Since the full kernel $K$ is fairly expensive to calculate, some approximations could be useful.
For example: 1) we may use Broyden or Anderson schemes \cite{CGBroyden65,DGAnderson65} to estimate $K$; 
2) for different classes of materials we could approximate the kernel
from a related, but simplified model system as in the Kerker scheme for metals \cite{GPKerker81};
3) we may calculate the full kernel exactly in the first time step and 
then keep it constant or perform some approximate updates on-the-fly at some predefined (or adaptive) frequency; 
4) we can apply some form of coarse graining where $K$ and the charge response, 
$\{{\delta \rho[n]}/{\delta n}\}$, are calculated, for example, per atom
or per molecule; 5) high accuracy could be used in the short-range response 
with some tail approximation for the long-range behavior; 6) we could use
a low-rank estimate of $K$ that is less expensive to calculate;
7) we can use a fully local scaled delta-function approximation where 
$K({\bf r},{\bf r'}) \approx -c \delta({\bf r-r'})$
for some constant $c \in [0,1]$; or 8) we may simply 
replace the effect of the kernel, Eq.\ (\ref{EffOfK}), by regular self-consistent
field optimization. In the test examples below I will demonstrate 
a few of these approaches,  but first I will present 6), the low-rank approach,
in some detail.

\subsection{Construct $K$ from a sequence of rank-1 updates}

I propose a scheme for how the kernel $K$ can be constructed through a sequence of rank-1 updates,
which relies on the Sherman-Morrison formula \cite{JSherman50} and a sequence
of directional derivatives given by quantum perturbation theory.  If we start with
\begin{equation}\label{KCI}
K_0 = [ J_0 -I ]^{-1},
\end{equation}
where
\begin{equation}\label{JCI}
J_0 = cI
\end{equation}
for some constant $c$, ($ c \ne 1$), then if
${\bf v} = \{v_i\}$ is a set of $N \times 1$ orthonormal perturbation vectors, it can be shown that
for $i = 1,2,\ldots, N$ the sequence of operations:
\begin{equation}\label{K_up}\begin{array}{l}
{\displaystyle u_i = \left. \frac{\delta \rho[n 
+ \lambda v_i]}{\delta \lambda}\right\vert_{\lambda = 0} - J_{i-1} v_i}, \\
~~\\
{\displaystyle J_i = J_{i-1} + u_iv_i^T},\\
~~\\
{\displaystyle K_i = K_{i-1}  - \frac{K_{i-1}u_i v_i^T K_{i-1}}{1+v_i^TK_{i-1}u_i}},
\end{array}
\end{equation}
converges to the exact kernel in Eq.\ (\ref{KK}), such that $K = K_N$.
The derivatives, i.e.\ the directional Gateaux differentials, 
in the first line of Eq.\ (\ref{K_up}) can be calculated
using density matrix perturbation theory, both at zero electronic temperature, $T_e = 0$ K 
\cite{ANiklasson04,VWeber04,VWeber05},
and at finite electronic temperature, $T_e > 0$ K \cite{ANiklasson15}.

The construction of $K$ over a general
set of orthonormalized perturbation vectors ${\bf v} = \{v_i\}$,
gives us the flexibility to chose a few optimal perturbation directions 
in a low-rank approximation of $K$. In fact, the scheme can be used to construct 
an approximate kernel by improving any initial estimate of $K_0$
with a single (or a low) rank update using only one or a few well-chosen 
perturbation vectors in Eq.\ (\ref{K_up}).

\subsection{Estimate $K$ from an approximate $K_0$ with single rank-1 updates}

If we choose to use only a single rank-1 update of an approximate kernel $K_0$,
how do we then chose the perturbation vector $v_1$ in the best possible way?
In the equations of motion, Eq.\ (\ref{n_v}), $K$ acts on the vector $\rho[n] - n$.
Based on the update formula for $K_{i-1}$ in Eq.\ (\ref{K_up}) it is easy to
see that we get a significant update of $K_0$ that optimizes the immediate overlap with respect to $\rho[n] - n$
if we chose $v_1$ such that $v_1^TK_0$ is parallel to $\rho[n] - n$, i.e. if we calculate the rank-1 update
using the normalized perturbation vector
\begin{equation}\label{v_1}
v_1 = \frac{K_0(\rho[n] - n)}{\|K_0(\rho[n] - n)\|}.
\end{equation}
Even when $J_0 = - c I$ and $K_0 = [ J_0 -I ]^{-1}$ and with a single ``rank-1 only update'' 
this choice of $v_1$ and
$K_1$ often gives a good approximation of the exact kernel acting on $\rho[n] - n$.
It can therefore be used as a simple and efficient single rank-1 update approximation of the
kernel $K$ in extended Lagrangian Born-Oppenheimer
molecular dynamics. The low-rank approximation of the kernel is of particular interest for plane-wave, 
finite grid, or wavelet implementations.
In these cases it is not possible to construct the full explicit kernel, since the dimension
of the matrix would be very high. In a molecular dynamics simulation 
the approximate kernel from a previous time step, $K_0(t)$,
can be used as an initial estimate of $K_0 (t+\delta t)$ 
that can be updated on-the-fly by a rank-1 update. In this 
way an accumulated and adaptive approximation of the kernel is build 
up, on-the-fly, during a simulation.

\subsection{A self-consistent-field optimization scheme}

The kernel update procedure, Eqs.\ (\ref{KCI})-(\ref{v_1}), also provides a practical
algorithm for regular self-consistent field optimizations, where
an approximate charge density $\rho_j$ can be updated by
\begin{equation}\label{SCF_K}
\rho_{j+1} = \rho_j - K_j\left(\rho[\rho_j]-\rho_j\right).
\end{equation}
This procedure using approximate, low-rank updated kernels that are estimated directly 
from quantum response theory thus serves as an alternative to, for example, Broyden's 
scheme and Pulay or Anderson mixing \cite{CGBroyden65,DGAnderson65,PPulay80}.
In an iterative self-consistent-field optimization scheme we can sometimes improve 
convergence by orthogonalizing each new perturbation vector $v_j = K_{j-1}(\rho[\rho_j]-\rho_j)$ 
to all previous vectors $v_i$ ($i<j$). 

\section{Examples}

The theory presented in this paper is quite general and should be applicable to
a broad range of electronic structure methods. Here I will demonstrate 
extended Lagrangian Born-Oppenheimer molecular dynamics using self-consistent-charge density functional
based tight-binding (SCC-DFTB) theory \cite{MElstner98,MFinnis98,TFrauenheim00,MCawkwell12,BAradi15}. 
It represents a simplified and highly efficient form of Kohn-Sham density functional
theory that includes all essential complexities that are necessary to demonstrate
key features of extended Lagrangian Born-Oppenheimer dynamics. SCC-DFTB uses an atomistic coarse graining of the
electronic structure with a Mulliken net charge on each atom, $\rho(R_I)$. 
The charges interact through a Coulomb field at long range, which is
screened to a Hubbard $U$ for the on site interactions.
The charge-independent part of the Kohn-Sham Hamiltonian is approximated using 
a Slater-Koster tight-binding Hamiltonian and a non-orthogonal atomic-orbital basis-set representation. 
Remaining parts of the exchange-correlation and Hartree energies are approximated with a repulsive pair potential
and the entropy contribution at finite electronic temperatures is given by the regular
expression, Eq.\ (\ref{S_entropy}), as a function of the orbital occupation factors.
The charges, $\rho$, are calculated self-consistently as in regular 
Kohn-Sham density functional theory. The forces are calculated from the gradients of
the SCC-DFTB total free energy, including basis-set dependent Pulay forces.
The extended dynamical variables for the electronic degrees of freedom in the
extended Lagrangian formulation are denoted by $n$, corresponding to approximate 
net Mulliken charges.

Except for the initial conditions of the first time step, Eq.\ (\ref{IBC}), 
no self-consistent field iterations will be used in the simulations.
Any constant occurring in the iterations, e.g.\ linear mixing coefficients or $c$ in $J_0 = cI$,
are kept constant in each simulation after an initial optimization, 
chosen to achieve fastest convergence in the initial
ground-state calculation within a tolerance of $0.1$. All test systems have periodic boundary
conditions. The modified Verlet integration scheme, 
Eqs.\ (\ref{Integration}) and (\ref{VRL_Damp}), was used in all the molecular dynamics simulations
with $\kappa = 1.82$, in Tab.\ \ref{Tab_Coef}.

The implementation and simulations were performed with an experimental, stand-alone version, of 
the open source software package LATTE \cite{LATTE,ESanville10,MCawkwell12}. As examples I choose
three different test systems. The first example, Fig.\ \ref{fg_1}, which is the more challenging system, is 
a hydroqionone radical solvated in water. The initial self-consistent field optimization is  hard to converge to a physically
meaningful ground state without using fractional occupation numbers, i.e.\ thermal density functional theory.
The second test system is liquid water, which represents a ``normal'' simulation, and serves as a standard
for comparison. In the initial, self-consistent ground state optimization, Eq.\ (\ref{IBC}), I also make
a comparison to the Anderson/Pulay acceleration scheme \cite{DGAnderson65,PPulay80,PPulay82} with 
an implementation as in Ref.\ \cite{ASBanerjee16} but without any restarts. The third test system
is liquid nitromethane (7 molecules, 49 atoms), which is used to demonstrate the scaling
of the accuracy in the forces, the electron density, and the free energy shadow potential surface.

\subsection{Hydroquinone radical in water}

\begin{figure}
\includegraphics[scale=0.3]{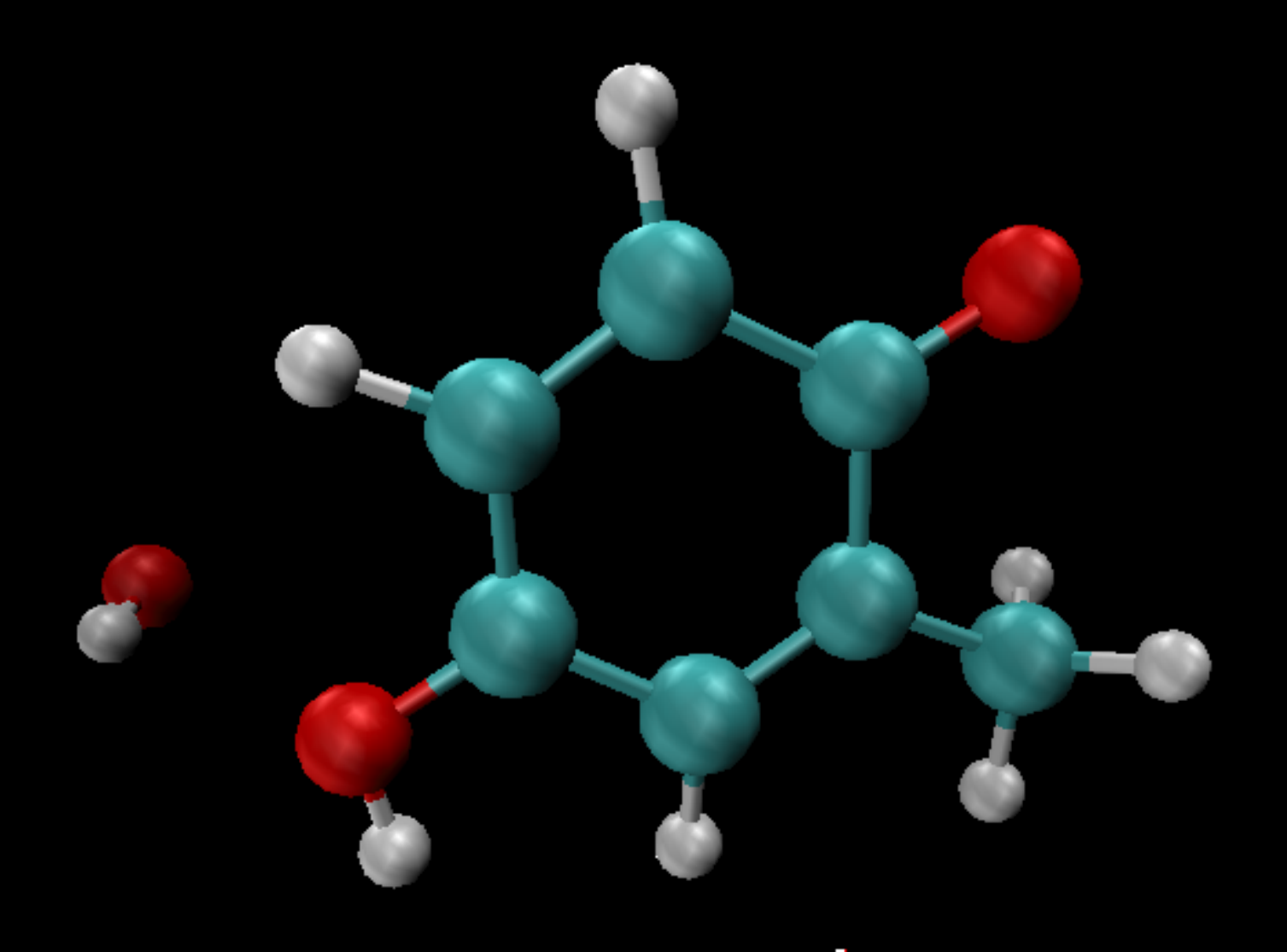}
\caption{\label{fg_1} The hydroquionone radical (C$_7$H$_6$OOH + OH)
that has been solvated in water (surrounding water is not shown).}
\end{figure}

\begin{figure}
\includegraphics[scale=0.3]{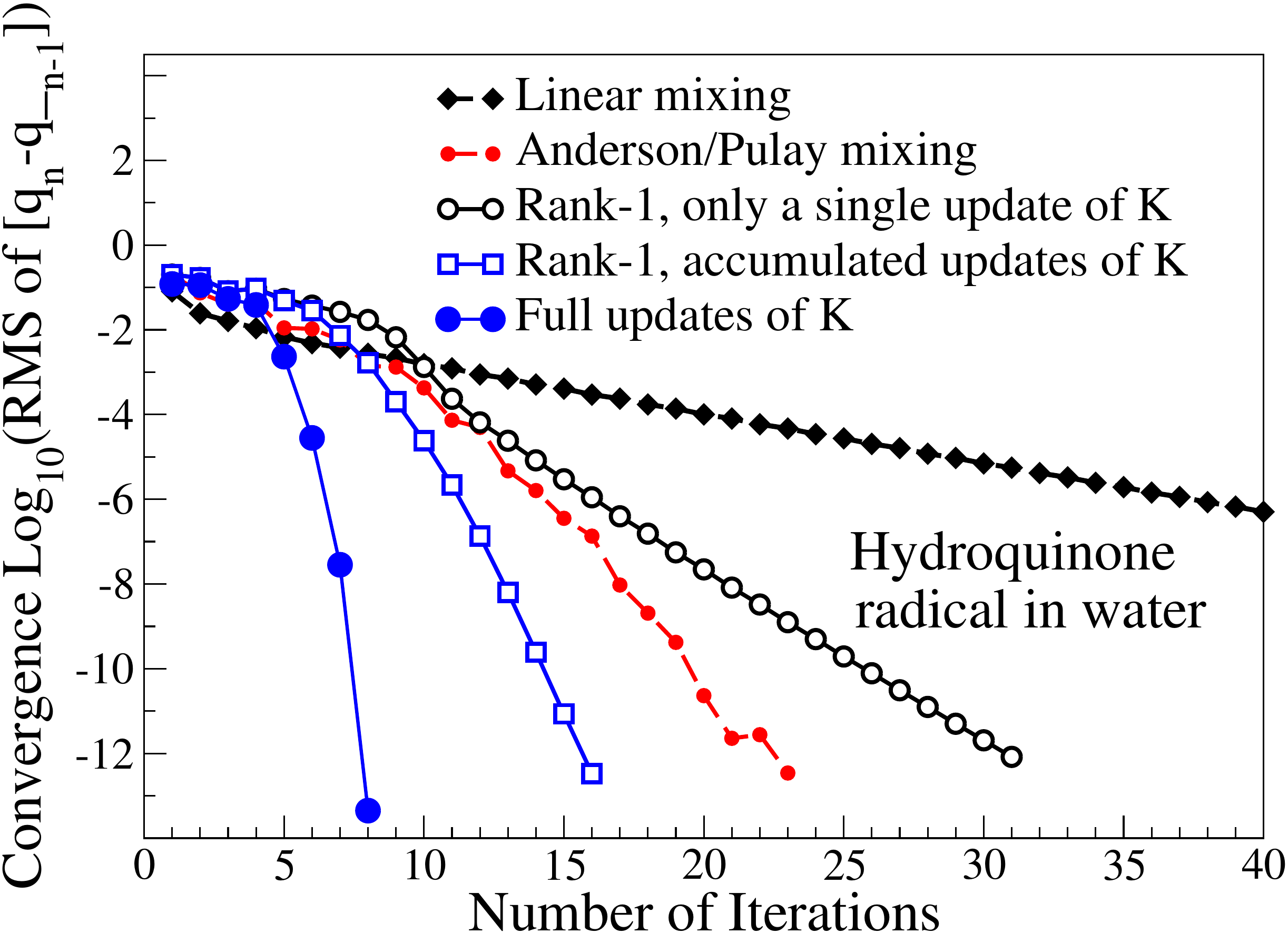}
\caption{\label{fg_2} Convergence or the root mean square (RMS) of the difference in charge between iterations of 
the self-consistent-field optimization 
for a hydroquinone radical in water (with a total of 99 atoms and the electronic temperature $T_e = 3,000$ K) 
using simple linear mixing, 
Anderson/Pulay mixing, rank-1 updates of the kernel $K$ or full updates of the entire $K$ in each
iterations. In practice the Anderson/Pulay scheme is the fastest.}
\end{figure}

\begin{figure}
\includegraphics[scale=0.3]{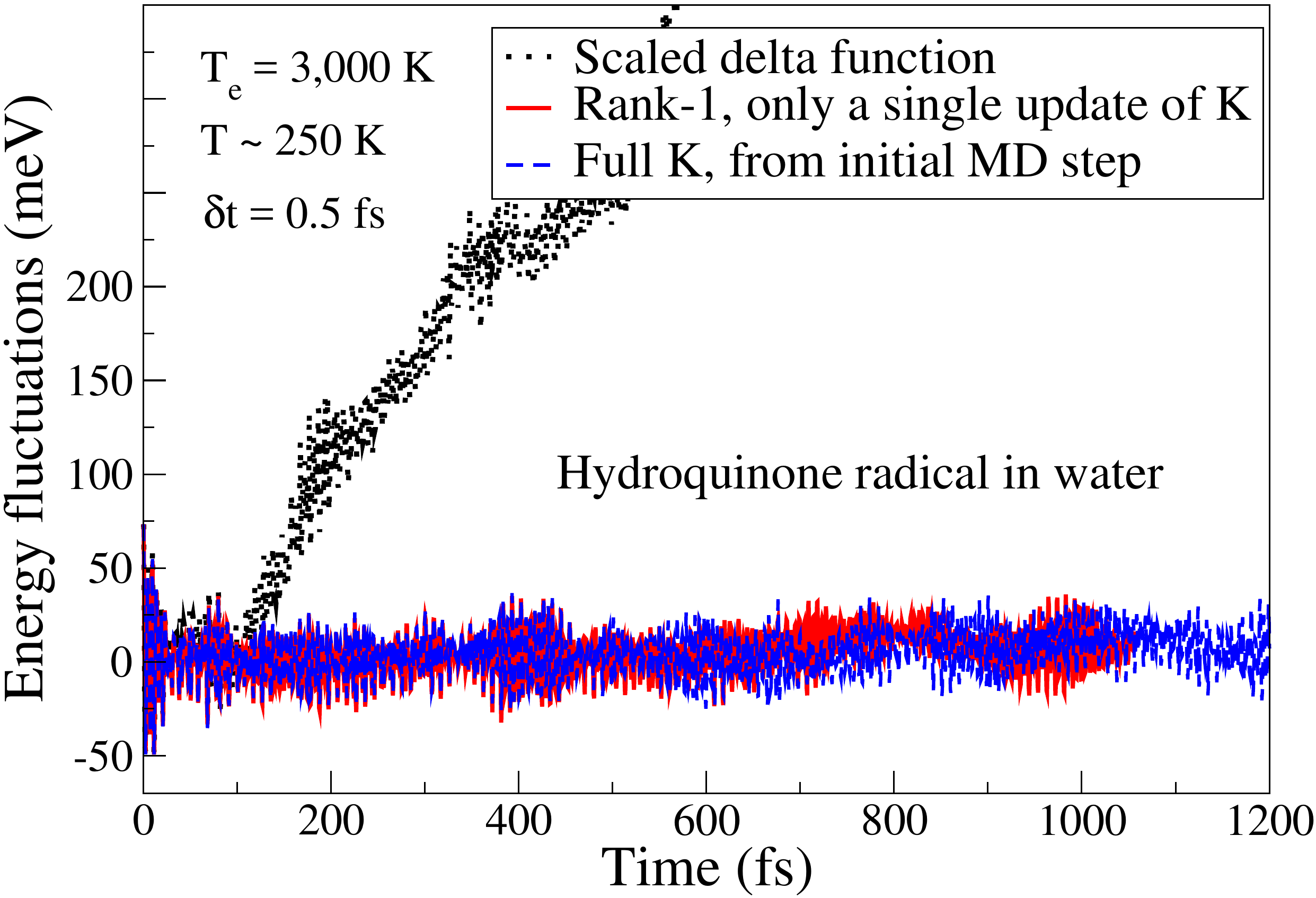}
\caption{\label{fg_3} The fluctuations in the total free energy (potential + entropy + kinetic, 
corresponding to the constant of motion Eq.\ (\ref{ShadowH1}) ) of an extended Lagrangian
Born-Oppenheimer molecular dynamics simulation of a hydroquinone radical in water 
(99 atoms, with a statistical temperature for the molecular trajectories, $T \sim 250 K$, 
and the electronic temperature $T_e = 3,000$ K) using three different approximations of the kernel $K$ .}
\end{figure}

The simulation of the water solvated hydroquinone radical test system in Fig.\ \ref{fg_1} is quite challenging.
Even with an electronic temperature of $T_e = 3,000$ K
the initial self-consistent-field convergence can be slow. Figure \ref{fg_2} shows how a simple linear mixing scheme (with
the fix mixing parameter optimized within a tolerance of 0.1) requires the order of 100 iterations to reach 
an accurate ground state solution. Anderson/Pulay mixing \cite{ASBanerjee16} 
(without restart) accelerates convergence significantly.  A rank-1 update of the kernel $K$, using
the normalized perturbation vector chosen from $v_j = K_{j-1}(\rho[\rho_j]-\rho_j)$ with 
the update as in Eq.\ (\ref{SCF_K}), 
also offers an efficient alternative, though each iteration takes about twice as long to
calculate as in the Anderson/Pulay mixing scheme. Figure \ref{fg_2} shows two cases of rank-1 updates, either using
a single rank-1 based approximation of the kernel (each time updating a scaled delta-function 
estimate of $K_0$ in Eqs.\ (\ref{KCI}) and (\ref{JCI}) with a only single rank-1 update, Eq.\ (\ref{K_up})),
or using multiple accumulated rank-1 updates, Eq.\ (\ref{K_up}), one new update for each new
self-consistent field iteration.  The calculation with the full updates of $K$, which in each self-consistent field iteration
requires as many response calculations as there are atoms in the system, converges rapidly, but is by far the slowest scheme.

\begin{figure}
\includegraphics[scale=0.3]{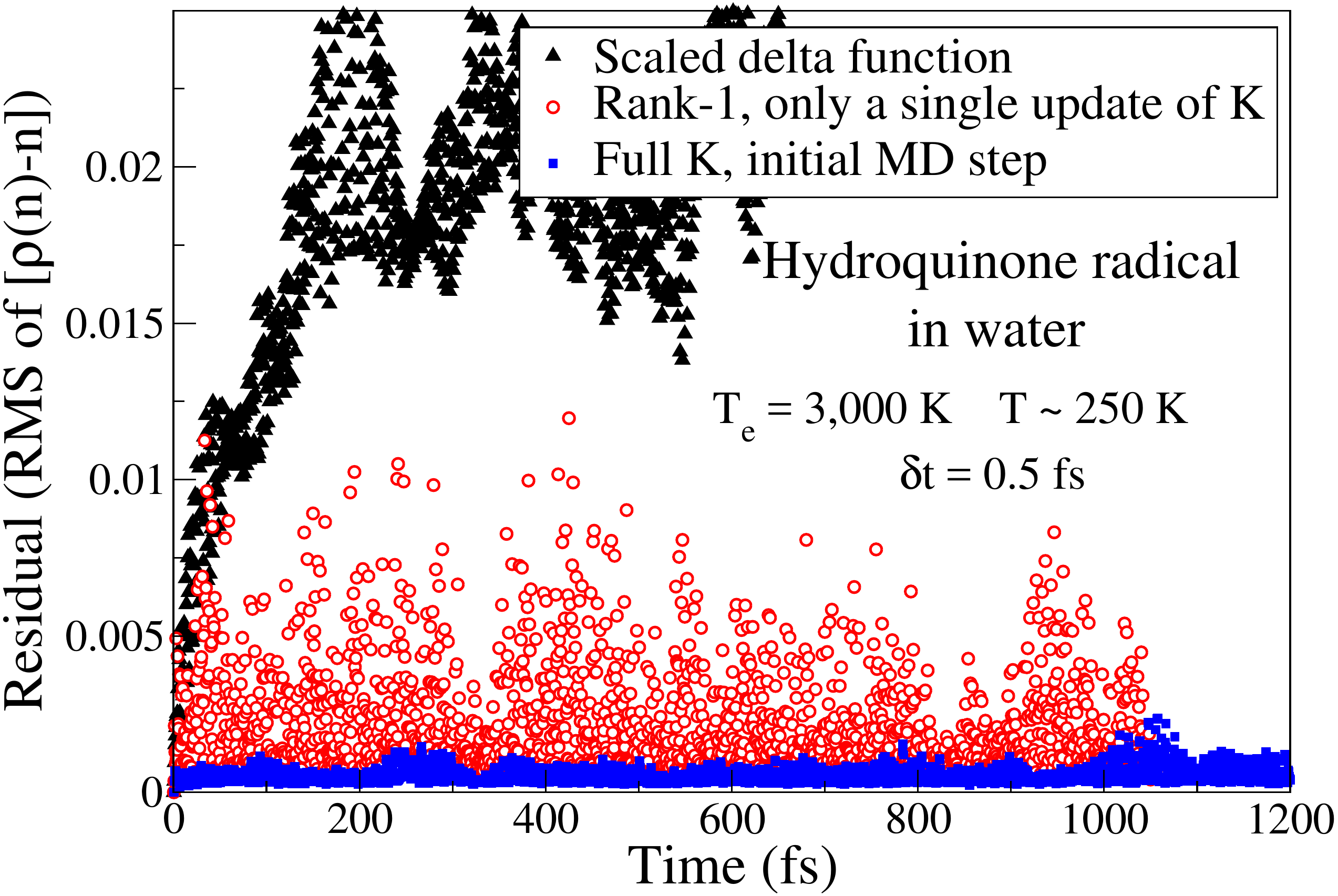}
\caption{\label{fg_4} The the root mean square (RMS) of the residual function ($\rho[n] - n$) as a function
of simulation time for the simulation of the hydroquinone radical in water in Fig.\ \ref{fg_3} .}
\end{figure}

\begin{figure}
\includegraphics[scale=0.3]{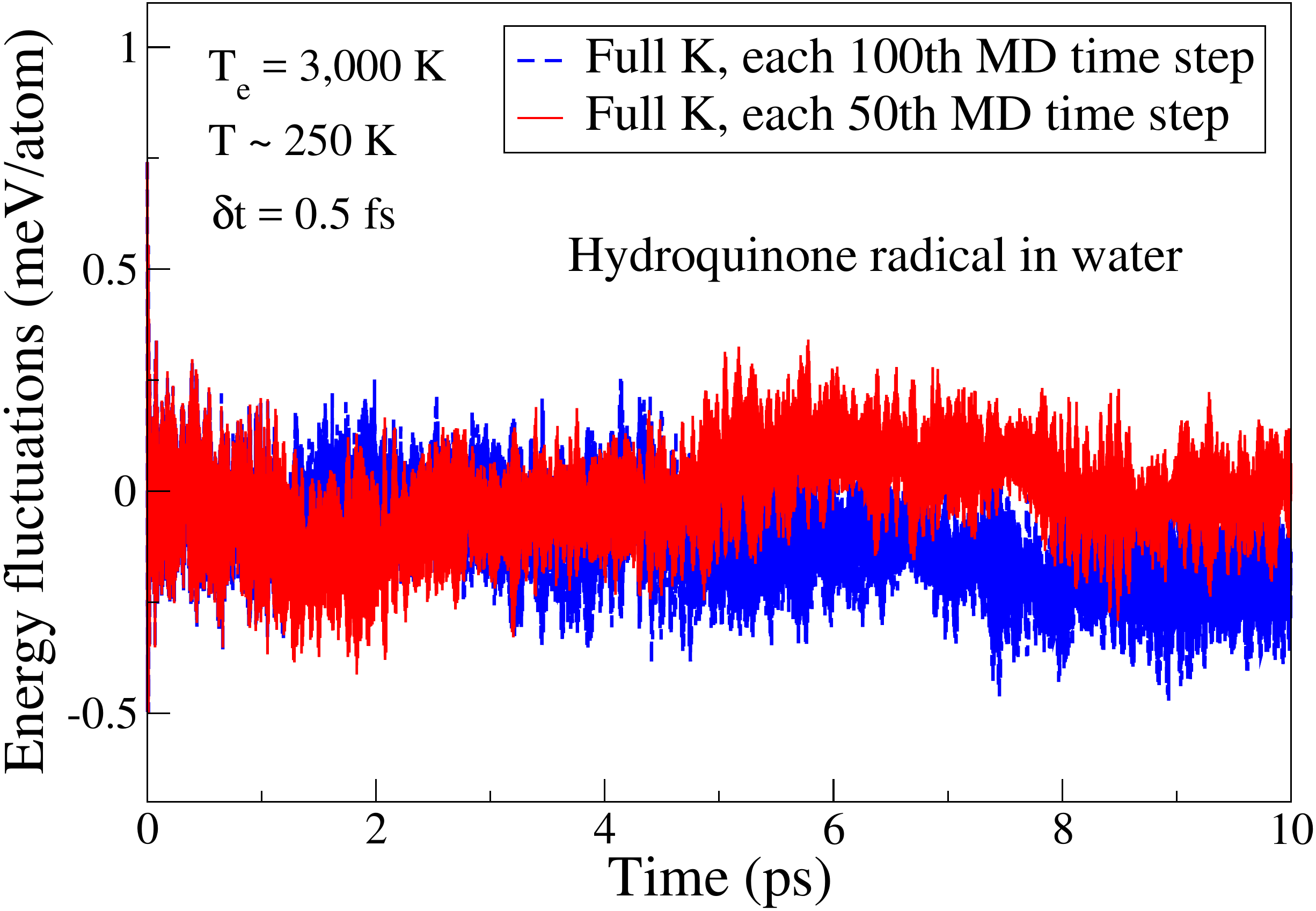}
\caption{\label{fg_5} The fluctuations in the total free energy (potential + entropy + kinetic, corresponding
to the constant of motion in Eq.\ (\ref{ShadowH1})) of 
an extended Lagrangian Born-Oppenheimer molecular dynamics simulation of a hydroquinone radical in water 
(99 atoms).}
\end{figure}

Even if the Anderson/Pulay scheme reaches convergence in the smallest amount of wall-clock time,
it seems hard to apply as approximations of the kernel $K$ 
in the extended Lagrangian Born-Oppenheimer molecular dynamics scheme. In this case the
exact full calculation of the kernel, the rank-1 update approach 
(either using single rank-1 only updates of scaled delta-function approximations or accumulated rank-1 updates), 
or only the scaled delta-function approximation, offer better alternatives. 

Figure \ref{fg_3}
shows the fluctuations of the total free energy for an extended Lagrangian Born-Oppenheimer simulation using either
a scaled delta-function approximation of the kernel, $K = c I$, the single rank-1 updated approximation of $K$,
or a full calculation of the kernel in the initial time step after which $K$ is kept constant.
The rank-1 and full kernel approaches have energy fluctuations that 
are virtually on top of each other for the first 1,000 time steps, whereas the scaled delta-function approximation shows 
fluctuations increasing in size, leading to a significant drift already after about 100 fs of simulation time. 
The behavior can be understood from the ability to keep the dynamical
variable density, $n(t)$, close to the ground state, which can be estimated from the distance to $\rho[n]$,
i.e. the size of the residual function $\rho[n]-n$.
The root means square (RMS) of $\rho[n] - n$
is shown in Fig. \ref{fg_4}. The simulation with the 
scaled delta-function approximation of $K$ shows a significant size of the residual,
whereas the single rank-1 update and
the full kernel approximations keep the residual function small such that $n$ remains close to
the exact ground state. Using the accumulated rank-1 updated kernel (not shown) gives results similar 
or worse compared to the approximation using only single rank-1 updates of scaled delta-functions in each time step.
The advantage with the rank-1 only (or low-rank) updated approximation of the kernel
is that it can be generalized to simulations using a real-space grid, density matrices, or a plane-wave basis functions for which full kernel calculations are not practically feasible. 

Figure \ref{fg_5} shows the total free energy of two longer molecular dynamics simulations of
the hydroquinone radical in water over 10 ps of simulation time. The full kernel, updated once every
100th or 50th time step, was used as an approximation. The simulation remains stable, though it shows some long-term
random-walk-like fluctuations in the total free energy due to small numerical errors in the force evaluations.

\begin{figure}
\includegraphics[scale=0.3]{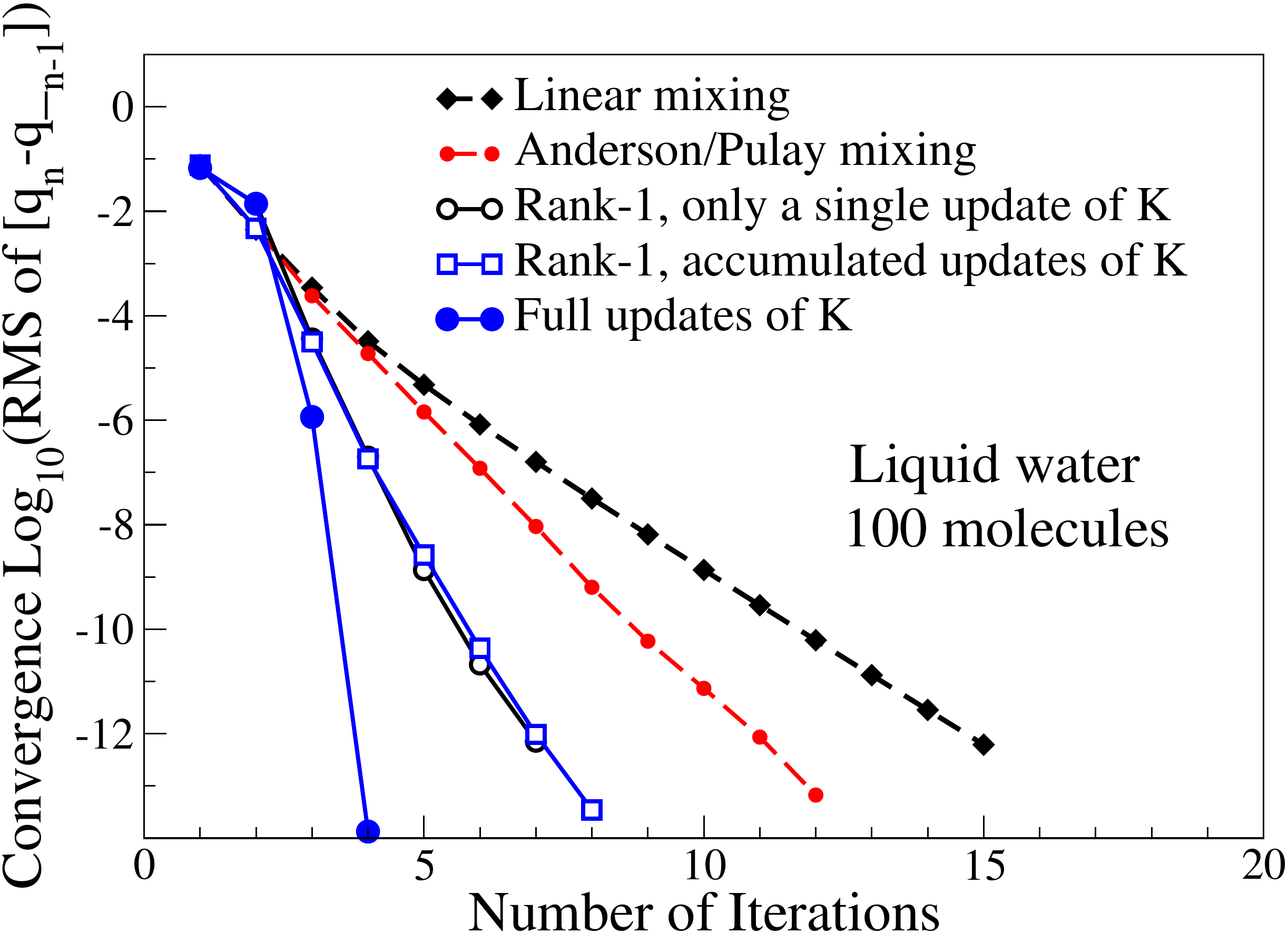}
\caption{\label{fg_6} Convergence or the root mean square (RMS) of the difference in charge between iterations of 
the self-consistent-field optimization
for a water system (with a total of 100 molecules at $T_e = 3,000$ K) using simple linear mixing,
Anderson/Pulay mixing, two different rank-1 updates of the kernel $K$ or full updates of the entire $K$ in each
iterations. In practice the Anderson/Pulay scheme is the fastest.}
\end{figure}

\begin{figure}
\includegraphics[scale=0.3]{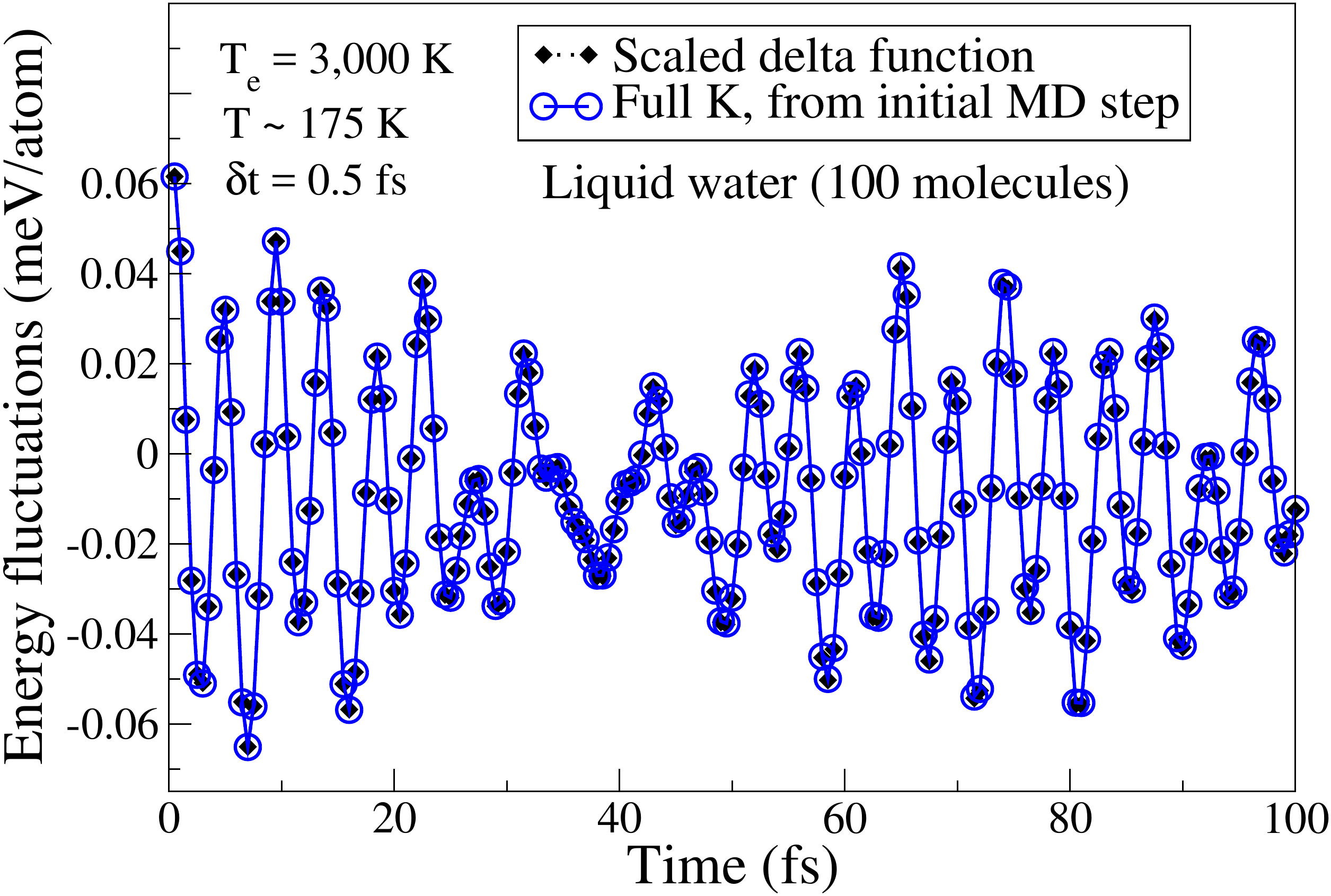}
\caption{\label{fg_7} The fluctuations in the total free energy (potential + entropy + kinetic, 
corresponding to the constant of motion in Eq.\ (\ref{ShadowH1})) of an extended Lagrangian 
Born-Oppenheimer molecular dynamics simulation of a water system
(100 molecules) using either a scaled delta-function approximation of the kernel 
or the full kernel, $K$, calculated at the first time step and then kept constant. 
The scaled delta-function approximation and the full kernel calculation give almost identical results.}
\end{figure}

\begin{figure}
\includegraphics[scale=0.3]{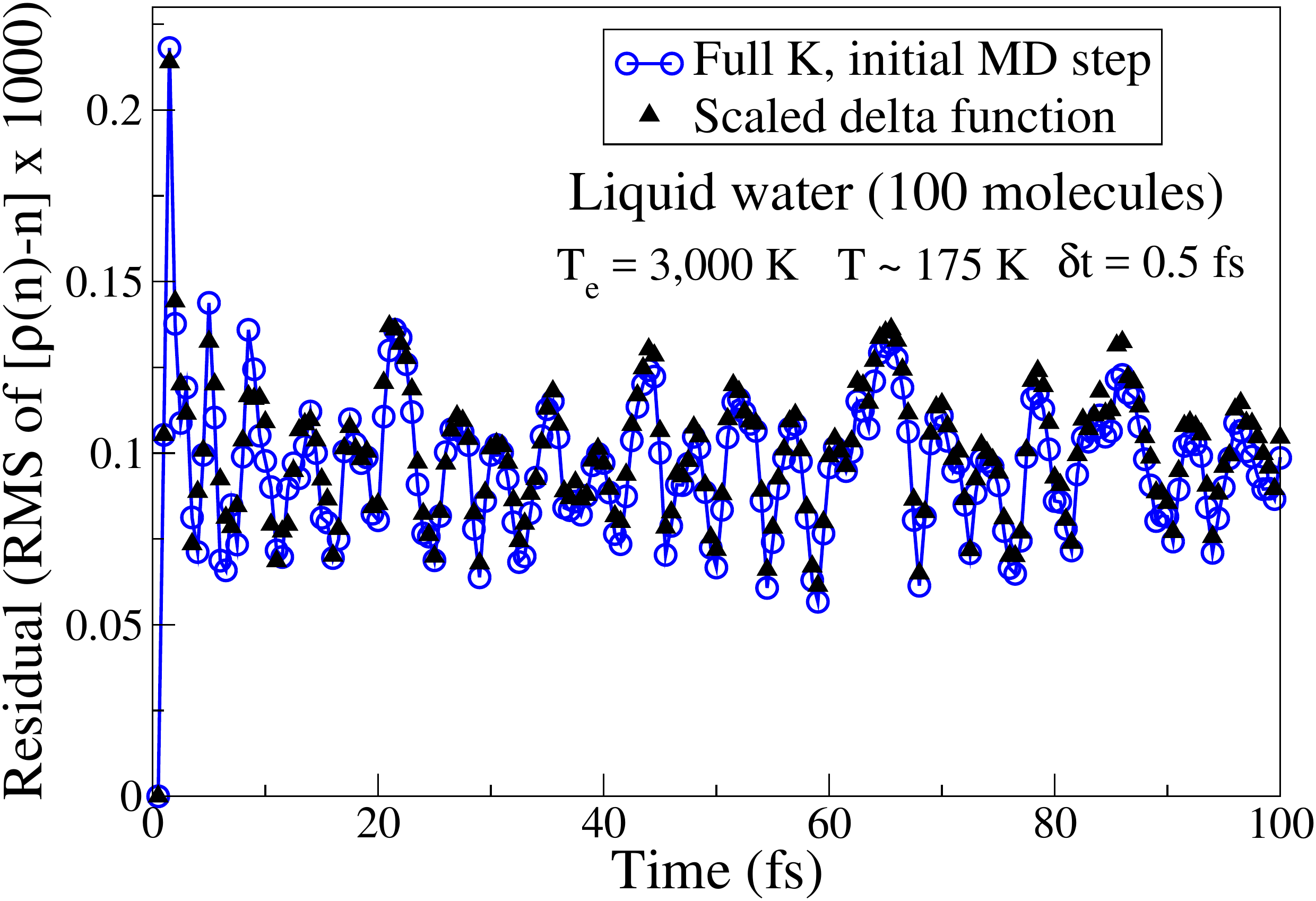}
\caption{\label{fg_8} The root mean square (RMS) of the residual function ($\rho[n] - n$)($\times 1000$) as a function
of simulation time for the water simulation in Fig.\ \ref{fg_7}. The scaled delta-function approximation of the kernel $K$
and the full kernel calculation give almost identical results.}
\end{figure}

\subsection{Water}

A more regular molecular dynamics simulation may be represented by liquid water. Notice, that choosing the same electronic
temperature of $3,000$ K, as in the quinone system, in practice has no influence on the electronic structure of 
liquid water, since the gap is much
larger than for the hydroquinone radical in the example above. In the liquid water
simulation we find a much smaller difference between a fully optimized kernel and a scaled
delta-function approximation of $K$. This is reflected in the relatively fast convergence
for the simple linear mixing approach in Fig. \ref{fg_6}. The energy fluctuation curves 
for the scaled delta-function kernel and the exact full kernel simulations
in Fig. \ref{fg_7} are essentially on top of each other. The reason for this can be understood from
the very small residual function, both for the scaled delta-function approximation and the
exact full kernel (calculated only in the first time step and then kept fix), as shown in Fig. \ref{fg_8}. 

\subsection{Scaling analysis}

\begin{figure}
\includegraphics[scale=0.3]{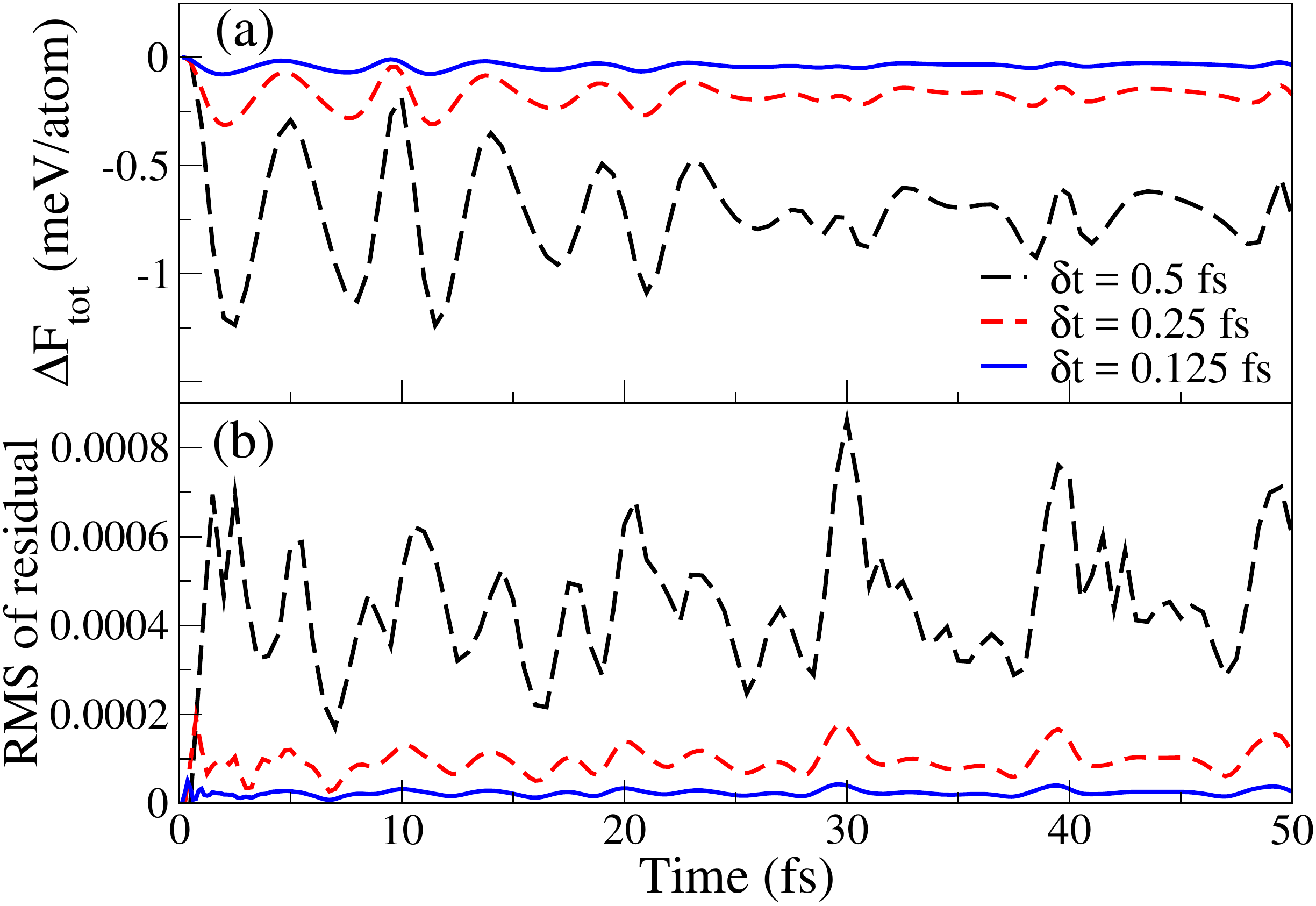}
\caption{\label{fg_9} The upper panel (a) shows fluctuations in the total free energy (corresponding to
the constant of motion in Eq.\ (\ref{ShadowH1})) 
for an extended Lagrangian Born-Oppenheimer molecular dynamics simulation of the 
the hydroqionone radical in water for various sizes of the integration time step ($\delta t =$ 0.5 fs, 0.25 fs, and 0.125 fs). 
The electronic temperature was set to $T_e = 3,000$ K and the statistical temperature of 
the molecular trajectories, $ T\sim 250$ K.
The lower panel (b) shows the corresponding fluctuations of the root mean square (RMS) 
of the residual function ($\rho[n] - n$). The simulations were performed with a full 
calculation of the kernel $K$ that was calculated for the initial configuration and then kept fix.
Both the size of the fluctuations of the total free energy and the RMS of the residual function
scale quadratically with the integration time step.}
\end{figure}

By choosing an appropriate approximation of the kernel $K$, we can stabilize the dynamics
of an extended Lagrangian Born-Oppenheimer molecular dynamics simulation. The kernel
acts like a preconditioner in the equations of motion for the extended electronic degrees
of freedom, which causes the dynamical variable density $n(t)$ to oscillate around 
a closer approximation of the exact Born-Oppenheimer ground state density $\rho_{\rm min}(t)$
and keeps the residual function $\rho[n]-n$ small. For many systems a simple scaled delta-function
approximation of the kernel is sufficiently accurate, but for systems exhibiting 
slow self-consistent field convergence, the approximation based on rank-1 updates or the
infrequent calculation of an exact full kernel using fractional occupation
number, serve as practical alternatives, with 
only a modest computational overhead. A slightly more expensive alternative is to reduce the integration
time step $\delta t$, which rapidly improves accuracy because of the second-order nature of extended
Lagrangian Born-Oppenheimer molecular dynamics for which the residual function scales as $(\rho[n]-n) \sim \delta t^2$.
This quadratic scaling is demonstrated in Fig.\ \ref{fg_9}. 

\begin{figure}
\includegraphics[scale=0.3]{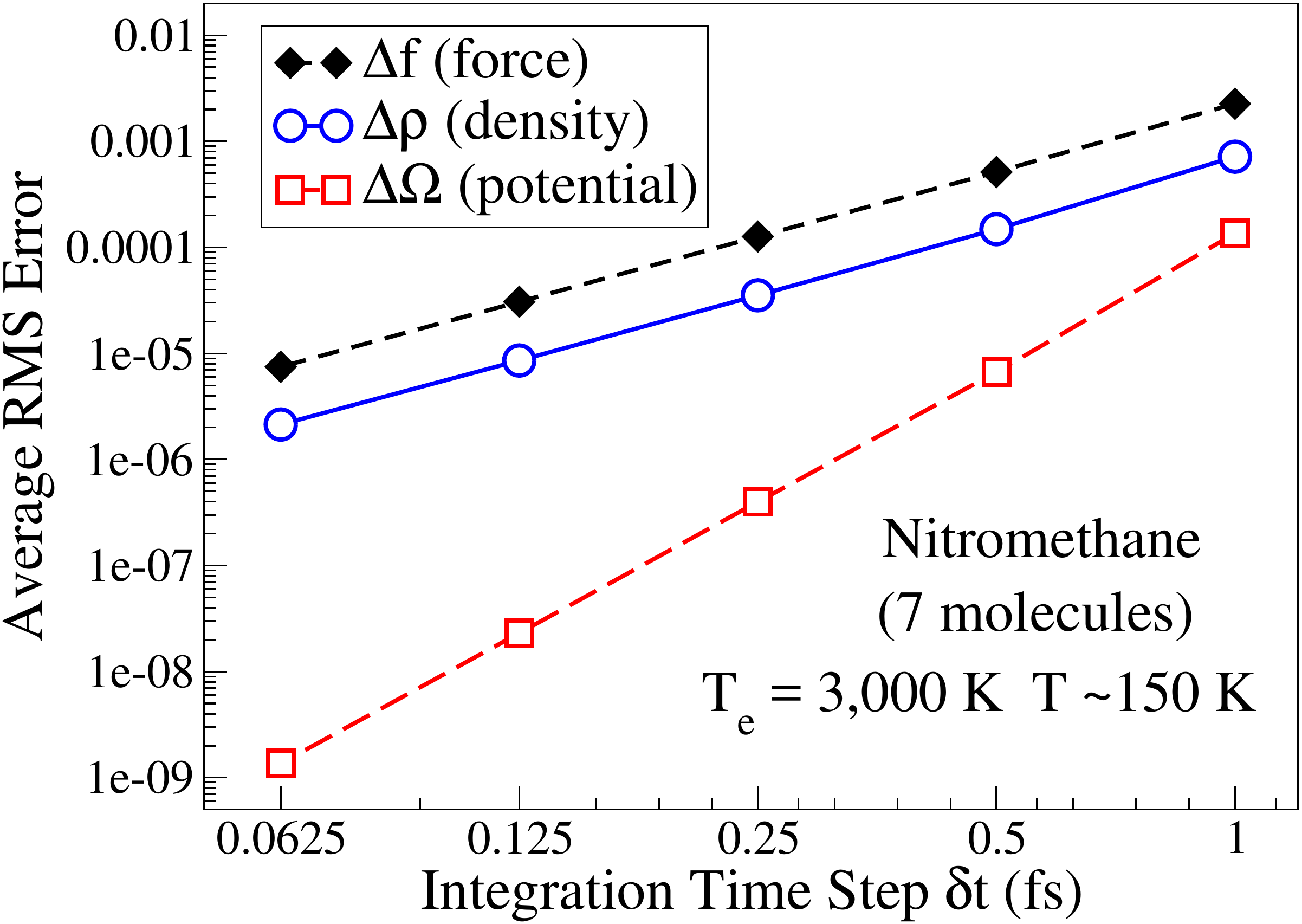}
\caption{\label{fg_10} The scaling of the average (over $\sim 100$ fs of simulation time) 
of the root mean square (RMS) error, as a function of the integration time step $\delta t$,
in the forces $\Delta f$, the electron density $\Delta \rho $ from ($n-\rho_{\rm min}$), and
the free energy potential surface $\Delta \Omega$ from ($\Omega^{(1)}[n]({\bf R})-\Omega_{\rm BO}[\rho_{\rm min}]({\bf R})$), 
during an extended Lagrangian Born-Oppenheimer simulation
with respect to the corresponding "exact" fully converged Born-Oppenheimer values for each configuration, ${\bf R}(t_0+k\delta t)$.
The deviation in the forces and the density scales as $\delta t^2$, whereas the error
in the {\em shadow  potential} scales as $\delta t^4$. The exact full kernel $K$ was recalculated
once every 100th integration time step.}
\end{figure}

The total free energy fluctuation curves in Fig.\ \ref{fg_3} for the full kernel and the rank-1 only approximations
are almost on top of each other for the first 500 fs, even if the residual functions  
in Fig.\ \ref{fg_4} differ in size by almost an order in magnitude. The reason for this is that the error 
in the free energy shadow potential compared to the fully converged regular Born-Oppenheimer free energy potential scales 
as $(\rho[n]-n)^2 \sim \delta t^4$, whereas the amplitude of the total free energy fluctuations are dominated 
by the local truncation error of the Verlet scheme, which scales as $(\rho[n]-n) \sim (\rho_{\rm min}-n) \sim \delta t^2$.
The scaling is demonstrated in Fig.\ \ref{fg_10}. Any error compared to a fully converged regular Born-Oppenheimer
simulation is then small compared to the error due to the finite integration time step. Figure \ref{fg_10}, which
illustrates the results from simulations of liquid nitromethane, also demonstrates the $\delta t^2$ scaling of
the error in the forces acting on the nuclear degrees of freedom.

The residual function, $\rho[n]-n$, plays an important role. It provides a measure of the distance
of $n(t)$ (or $\rho[n](t)$) to the exact ground state density $\rho_{\rm min}(t)$. If the residual function
is too large we may expect that the validity of the linearized shadow potential energy 
surfaces, $U^{(1)}[n]({\bf R})$ or $\Omega^{(1)}[n]({\bf R})$, breaks down. The residual function
is easy to monitor during a simulation and serves as a gauge of the accuracy in a way similar 
to the total free energy fluctuations. The residual function is of particular interest
in simulations of canonical (NVT) ensembles that don't necessarily have a constant of motion \cite{EMartinez15}, 
or in simulations using linear scaling electronic structure theory for systems with fractional occupation numbers, 
where it might be expensive to calculate the electronic entropy.

It is not possible to demonstrate all different aspects of the kernel approximation
techniques. Their relative efficiency may differ significantly depending on the test system.
A wide range of tricks can also be used and combined with the kernel techniques presented here, 
including adaptive back tracking, variable time steps, restarts, and orthonormalized updates.
The complexity, in many ways, is directly related to the challenges of the regular self-consistent 
field optimization for the ground state electronic structure, and we can
most probably not expect a single universal black-box solver that works for all material systems in all situations.

\section{What still remains}

This article presents the development of extended Lagrangian Born-Oppenheimer molecular dynamics.
Even if most parts of the theory seems complete
several important steps are still needed, for example: 1) implementations
in broadly available, state-of-the-art, first principles electronic structure codes 
\cite{RAhlrichs89,MSchmidt93,GKresse96,EHernandez96,ETsuchida98,JSoler02,DBowler01,TOzaki05,JVandevondele05,DBowler06,LKronik06,FreeON,LGenovese08,VBlum09,NHine09,ERudberg_11,JVandevondele12,KAidas13,MArita14,DOseiKuffuor14,XQin15}
that can take full advantage of the approximations of $K$ proposed in this paper, e.g.\ based on low-rank updates calculated from 
quantum perturbation theory; 2) convincing demonstrations of quantum-based simulations, allowing longer 
simulation times and/or larger system sizes than previously possible, for
a broad range of problems across materials science, chemistry and biology;
3) theoretically rigorous (adaptive) integration algorithms that can be used 
for a variety of thermodynamic ensembles; and
4) extensions beyond regular ground state Kohn-Sham Born-Oppenheimer simulations,
such as, orbital-free high-temperature methods, path-integral molecular
dynamics techniques, methods for strong electron correlation, 
accelerated molecular dynamics, polarizable force fields and QM/MM implementations, 
and excited state molecular dynamics. While a few steps
have been taken, a lot of work still remains.

\section{Conclusion and summary}

The unified extended Lagrangian framework introduced by Car and Parrinello made 
first principles molecular dynamics simulations possible 
for a broad range of materials for the first time and set the stage for 
a whole generation of electronic structure modeling. 
In this paper I have reviewed a new approach to first principles 
molecular dynamics. This framework of extended Lagrangian Born-Oppenheimer molecular dynamics was presented
using general Hohenberg-Kohn density functional theory and I discussed some of its similarities
and differences compared to extended Lagrangian Car-Parrinello molecular dynamics.
A key difference is that the accuracy of the electronic degrees of freedom in
extended Lagrangian Born-Oppenheimer molecular dynamics, with respect to the exact Born-Oppenheimer
solution, is of second order in the size of the integration time step and of fourth order
in the potential energy surface, whereas Car-Parrinello molecular dynamics is of first
order in the electronic degrees of freedom and of second order in the potential energy.
This means that the underlying {\em shadow Hamiltonian} of extended Lagrangian Born-Oppenheimer
molecular dynamics closely mimics the exact Born-Oppenheimer Hamiltonian.
An interesting challenge is to design possibly even higher-order formulations.

Improved stability over the most recent formulation of extended Lagrangian Born-Oppenheimer
molecular dynamics was achieved
by generalizing the theory to finite temperature ensembles, using fractional 
occupation numbers in the calculation of the inner-product kernel of the harmonic
oscillator extension, which appears like a preconditioner in the electronic equations of motion.
I showed how this kernel can be estimated on-the-fly using rank-1
updates of approximate kernels with the updates calculated directly
from quantum perturbation theory with the perturbations chosen along optimal directions,
or by the exact full kernel recalculated at some low frequency.
Material systems that normally exhibit slow self-consistent field convergence 
can be simulated using integration time steps of the same order as in direct Born-Oppenheimer
molecular dynamics without the requirement of an expensive, iterative,
non-linear electronic ground state optimization prior to the force evaluations
and without a systematic drift in the total energy.

The formulation of extended Lagrangian Born-Oppenheimer molecular dynamics
presented in this paper represents a framework 
for a next generation extended Lagrangian first principles molecular dynamics
that overcomes shortcomings of regular, direct Born-Oppenheimer simulations,
while maintaining or improving properties of Car-Parrinello molecular dynamics.

\section{Acknowledgements}

This work was supported by the Department of Energy Offices of Basic Energy Sciences (Grant No. LANL2014E8AN)
and performed at Los Alamos National Laboratory (LANL) in New Mexico. 
Discussions with A. Albaugh, B. Aradi, R. Car, E. Chisolm, O. Certik, 
M.J. Cawkwell, Th. Frauenheim, T. Head-Gordon,
S.M. Mniszewski, C.F.A. Negre, and C-K Skylaris are gratefully acknowledged as well 
as generous inspiration from T. Peery and the T-division
international Java group.  LANL, is an affirmative action/equal opportunity employer, 
which is operated by Los Alamos National Security, LLC, for the National Nuclear 
Security Administration of the U.S. DOE under Contract No. DE- AC52-06NA25396.

\bibliography{mndo_new_x}

\begin{thebibliography}{126}%
\makeatletter
\providecommand \@ifxundefined [1]{%
 \@ifx{#1\undefined}
}%
\providecommand \@ifnum [1]{%
 \ifnum #1\expandafter \@firstoftwo
 \else \expandafter \@secondoftwo
 \fi
}%
\providecommand \@ifx [1]{%
 \ifx #1\expandafter \@firstoftwo
 \else \expandafter \@secondoftwo
 \fi
}%
\providecommand \natexlab [1]{#1}%
\providecommand \enquote  [1]{``#1''}%
\providecommand \bibnamefont  [1]{#1}%
\providecommand \bibfnamefont [1]{#1}%
\providecommand \citenamefont [1]{#1}%
\providecommand \href@noop [0]{\@secondoftwo}%
\providecommand \href [0]{\begingroup \@sanitize@url \@href}%
\providecommand \@href[1]{\@@startlink{#1}\@@href}%
\providecommand \@@href[1]{\endgroup#1\@@endlink}%
\providecommand \@sanitize@url [0]{\catcode `\\12\catcode `\$12\catcode
  `\&12\catcode `\#12\catcode `\^12\catcode `\_12\catcode `\%12\relax}%
\providecommand \@@startlink[1]{}%
\providecommand \@@endlink[0]{}%
\providecommand \url  [0]{\begingroup\@sanitize@url \@url }%
\providecommand \@url [1]{\endgroup\@href {#1}{\urlprefix }}%
\providecommand \urlprefix  [0]{URL }%
\providecommand \Eprint [0]{\href }%
\providecommand \doibase [0]{http://dx.doi.org/}%
\providecommand \selectlanguage [0]{\@gobble}%
\providecommand \bibinfo  [0]{\@secondoftwo}%
\providecommand \bibfield  [0]{\@secondoftwo}%
\providecommand \translation [1]{[#1]}%
\providecommand \BibitemOpen [0]{}%
\providecommand \bibitemStop [0]{}%
\providecommand \bibitemNoStop [0]{.\EOS\space}%
\providecommand \EOS [0]{\spacefactor3000\relax}%
\providecommand \BibitemShut  [1]{\csname bibitem#1\endcsname}%
\let\auto@bib@innerbib\@empty
\bibitem [{\citenamefont {Allen}\ and\ \citenamefont
  {Tildesley}(1990)}]{MAllen90}%
  \BibitemOpen
  \bibfield  {author} {\bibinfo {author} {\bibfnamefont {M.}~\bibnamefont
  {Allen}}\ and\ \bibinfo {author} {\bibfnamefont {D.}~\bibnamefont
  {Tildesley}},\ }\href@noop {} {\emph {\bibinfo {title} {Computer Simulation
  of Liquids}}}\ (\bibinfo  {publisher} {Oxford Science},\ \bibinfo {address}
  {London},\ \bibinfo {year} {1990})\BibitemShut {NoStop}%
\bibitem [{\citenamefont {Karplus}\ and\ \citenamefont
  {Petsko}(1990)}]{MKarplus90}%
  \BibitemOpen
  \bibfield  {author} {\bibinfo {author} {\bibfnamefont {M.}~\bibnamefont
  {Karplus}}\ and\ \bibinfo {author} {\bibfnamefont {G.}~\bibnamefont
  {Petsko}},\ }\href@noop {} {\bibfield  {journal} {\bibinfo  {journal}
  {Nature}\ }\textbf {\bibinfo {volume} {347}},\ \bibinfo {pages} {631}
  (\bibinfo {year} {1990})}\BibitemShut {NoStop}%
\bibitem [{\citenamefont {vanGusteren}\ and\ \citenamefont
  {Berendsen}(1990)}]{vanGusteren90}%
  \BibitemOpen
  \bibfield  {author} {\bibinfo {author} {\bibfnamefont {W.~F.}\ \bibnamefont
  {vanGusteren}}\ and\ \bibinfo {author} {\bibfnamefont {H.~J.~C.}\
  \bibnamefont {Berendsen}},\ }\href@noop {} {\bibfield  {journal} {\bibinfo
  {journal} {Angewandte Chemie}\ }\textbf {\bibinfo {volume} {29}},\ \bibinfo
  {pages} {992} (\bibinfo {year} {1990})}\BibitemShut {NoStop}%
\bibitem [{\citenamefont {Tuckerman}\ and\ \citenamefont
  {Martyna}(2000)}]{MTuckerman00}%
  \BibitemOpen
  \bibfield  {author} {\bibinfo {author} {\bibfnamefont {M.~E.}\ \bibnamefont
  {Tuckerman}}\ and\ \bibinfo {author} {\bibfnamefont {G.~J.}\ \bibnamefont
  {Martyna}},\ }\href@noop {} {\bibfield  {journal} {\bibinfo  {journal} {J.
  Phys. Chem. B}\ }\textbf {\bibinfo {volume} {104}},\ \bibinfo {pages} {159}
  (\bibinfo {year} {2000})}\BibitemShut {NoStop}%
\bibitem [{\citenamefont {Frenkel}\ and\ \citenamefont
  {Smit}(2002)}]{DFrenkel02}%
  \BibitemOpen
  \bibfield  {author} {\bibinfo {author} {\bibfnamefont {D.}~\bibnamefont
  {Frenkel}}\ and\ \bibinfo {author} {\bibfnamefont {B.}~\bibnamefont {Smit}},\
  }\enquote {\bibinfo {title} {Understanding molecular simulation},}\ in\
  \href@noop {} {\emph {\bibinfo {booktitle} {Understanding Molecular
  Simulation, Second Edition}}}\ (\bibinfo  {publisher} {Academic Press},\
  \bibinfo {year} {2002})\ \bibinfo {edition} {2nd}\ ed.\BibitemShut {Stop}%
\bibitem [{\citenamefont {Karplus}\ and\ \citenamefont
  {McCammon}(2002)}]{MKarplus02}%
  \BibitemOpen
  \bibfield  {author} {\bibinfo {author} {\bibfnamefont {M.}~\bibnamefont
  {Karplus}}\ and\ \bibinfo {author} {\bibfnamefont {J.}~\bibnamefont
  {McCammon}},\ }\href@noop {} {\bibfield  {journal} {\bibinfo  {journal}
  {Nature Struc. Mol. Bio.}\ }\textbf {\bibinfo {volume} {9}},\ \bibinfo
  {pages} {646} (\bibinfo {year} {2002})}\BibitemShut {NoStop}%
\bibitem [{\citenamefont {Sanbonmatsu}\ and\ \citenamefont
  {Tung}(2007)}]{KYSanbonmatsu07}%
  \BibitemOpen
  \bibfield  {author} {\bibinfo {author} {\bibfnamefont {K.~Y.}\ \bibnamefont
  {Sanbonmatsu}}\ and\ \bibinfo {author} {\bibfnamefont {C.-S.}\ \bibnamefont
  {Tung}},\ }\href@noop {} {\bibfield  {journal} {\bibinfo  {journal} {J.
  Struct. Biol.}\ }\textbf {\bibinfo {volume} {157}},\ \bibinfo {pages} {470}
  (\bibinfo {year} {2007})}\BibitemShut {NoStop}%
\bibitem [{\citenamefont {Orozco}(2014)}]{MOrozco14}%
  \BibitemOpen
  \bibfield  {author} {\bibinfo {author} {\bibfnamefont {M.}~\bibnamefont
  {Orozco}},\ }\href@noop {} {\bibfield  {journal} {\bibinfo  {journal} {Chem.
  Soc. Rev.}\ }\textbf {\bibinfo {volume} {43}},\ \bibinfo {pages} {5051}
  (\bibinfo {year} {2014})}\BibitemShut {NoStop}%
\bibitem [{\citenamefont {Perilla}\ \emph {et~al.}(2015)\citenamefont
  {Perilla}, \citenamefont {Goh}, \citenamefont {Cassidy}, \citenamefont {Lui},
  \citenamefont {Bernardi}, \citenamefont {Rudack}, \citenamefont {Yu},
  \citenamefont {Wu},\ and\ \citenamefont {Schulten}}]{JRPerilla15}%
  \BibitemOpen
  \bibfield  {author} {\bibinfo {author} {\bibfnamefont {J.}~\bibnamefont
  {Perilla}}, \bibinfo {author} {\bibfnamefont {B.}~\bibnamefont {Goh}},
  \bibinfo {author} {\bibfnamefont {C.~K.}\ \bibnamefont {Cassidy}}, \bibinfo
  {author} {\bibfnamefont {B.}~\bibnamefont {Lui}}, \bibinfo {author}
  {\bibfnamefont {R.~C.}\ \bibnamefont {Bernardi}}, \bibinfo {author}
  {\bibfnamefont {T.}~\bibnamefont {Rudack}}, \bibinfo {author} {\bibfnamefont
  {H.}~\bibnamefont {Yu}}, \bibinfo {author} {\bibfnamefont {Z.}~\bibnamefont
  {Wu}}, \ and\ \bibinfo {author} {\bibfnamefont {K.}~\bibnamefont
  {Schulten}},\ }\href@noop {} {\bibfield  {journal} {\bibinfo  {journal}
  {Curr. Opin. Struct. Biol.}\ }\textbf {\bibinfo {volume} {31}},\ \bibinfo
  {pages} {64} (\bibinfo {year} {2015})}\BibitemShut {NoStop}%
\bibitem [{\citenamefont {Marx}\ and\ \citenamefont {Hutter}(2000)}]{DMarx00}%
  \BibitemOpen
  \bibfield  {author} {\bibinfo {author} {\bibfnamefont {D.}~\bibnamefont
  {Marx}}\ and\ \bibinfo {author} {\bibfnamefont {J.}~\bibnamefont {Hutter}},\
  }\enquote {\bibinfo {title} {Modern methods and algorithms of quantum
  chemistry},}\ \ (\bibinfo  {publisher} {ed. J. Grotendorst},\ \bibinfo
  {address} {John von Neumann Institute for Computing, J\"ulich, Germany},\
  \bibinfo {year} {2000})\ \bibinfo {edition} {2nd}\ ed.\BibitemShut {Stop}%
\bibitem [{\citenamefont {Tuckerman}(2002)}]{MTuckerman02}%
  \BibitemOpen
  \bibfield  {author} {\bibinfo {author} {\bibfnamefont {M.~E.}\ \bibnamefont
  {Tuckerman}},\ }\href@noop {} {\bibfield  {journal} {\bibinfo  {journal} {J.
  Phys.: Conden. Matter}\ }\textbf {\bibinfo {volume} {14}},\ \bibinfo {pages}
  {1297} (\bibinfo {year} {2002})}\BibitemShut {NoStop}%
\bibitem [{\citenamefont {Kirchner}\ \emph {et~al.}(2012)\citenamefont
  {Kirchner}, \citenamefont {di~Dio~Philipp},\ and\ \citenamefont
  {Hutter}}]{BKirchner12}%
  \BibitemOpen
  \bibfield  {author} {\bibinfo {author} {\bibfnamefont {B.}~\bibnamefont
  {Kirchner}}, \bibinfo {author} {\bibfnamefont {J.}~\bibnamefont
  {di~Dio~Philipp}}, \ and\ \bibinfo {author} {\bibfnamefont {J.}~\bibnamefont
  {Hutter}},\ }\href@noop {} {\bibfield  {journal} {\bibinfo  {journal} {Top.
  Curr. Chem.}\ }\textbf {\bibinfo {volume} {307}},\ \bibinfo {pages} {109}
  (\bibinfo {year} {2012})}\BibitemShut {NoStop}%
\bibitem [{\citenamefont {Hutter}(2012)}]{JHutter12}%
  \BibitemOpen
  \bibfield  {author} {\bibinfo {author} {\bibfnamefont {J.}~\bibnamefont
  {Hutter}},\ }\href@noop {} {\bibfield  {journal} {\bibinfo  {journal} {WIREs
  Comput. Mol. Sci.}\ }\textbf {\bibinfo {volume} {2}},\ \bibinfo {pages} {604}
  (\bibinfo {year} {2012})}\BibitemShut {NoStop}%
\bibitem [{\citenamefont {Wang}\ and\ \citenamefont
  {Karplus}(1973)}]{MKarplus73}%
  \BibitemOpen
  \bibfield  {author} {\bibinfo {author} {\bibfnamefont {I.~S.~Y.}\
  \bibnamefont {Wang}}\ and\ \bibinfo {author} {\bibfnamefont {M.}~\bibnamefont
  {Karplus}},\ }\href@noop {} {\bibfield  {journal} {\bibinfo  {journal} {J.
  Am. Chem. Soc.}\ }\textbf {\bibinfo {volume} {95}},\ \bibinfo {pages} {8160}
  (\bibinfo {year} {1973})}\BibitemShut {NoStop}%
\bibitem [{\citenamefont {Leforestier}(1978)}]{CLeforestier78}%
  \BibitemOpen
  \bibfield  {author} {\bibinfo {author} {\bibfnamefont {C.}~\bibnamefont
  {Leforestier}},\ }\href@noop {} {\bibfield  {journal} {\bibinfo  {journal}
  {J. Chem. Phys.}\ }\textbf {\bibinfo {volume} {68}},\ \bibinfo {pages} {4406}
  (\bibinfo {year} {1978})}\BibitemShut {NoStop}%
\bibitem [{\citenamefont {Warshel}\ and\ \citenamefont
  {Karplus}(1975)}]{AWarshel75}%
  \BibitemOpen
  \bibfield  {author} {\bibinfo {author} {\bibfnamefont {A.}~\bibnamefont
  {Warshel}}\ and\ \bibinfo {author} {\bibfnamefont {M.}~\bibnamefont
  {Karplus}},\ }\href@noop {} {\bibfield  {journal} {\bibinfo  {journal} {Chem.
  Phys. Lett.}\ }\textbf {\bibinfo {volume} {32}},\ \bibinfo {pages} {11}
  (\bibinfo {year} {1975})}\BibitemShut {NoStop}%
\bibitem [{\citenamefont {Car}\ and\ \citenamefont
  {Parrinello}(1985)}]{RCar85}%
  \BibitemOpen
  \bibfield  {author} {\bibinfo {author} {\bibfnamefont {R.}~\bibnamefont
  {Car}}\ and\ \bibinfo {author} {\bibfnamefont {M.}~\bibnamefont
  {Parrinello}},\ }\href@noop {} {\bibfield  {journal} {\bibinfo  {journal}
  {Phys.\ Rev.\ Lett.}\ }\textbf {\bibinfo {volume} {55}},\ \bibinfo {pages}
  {2471} (\bibinfo {year} {1985})}\BibitemShut {NoStop}%
\bibitem [{\citenamefont {Remler}\ and\ \citenamefont
  {Madden}(1990)}]{DRemler90}%
  \BibitemOpen
  \bibfield  {author} {\bibinfo {author} {\bibfnamefont {D.~K.}\ \bibnamefont
  {Remler}}\ and\ \bibinfo {author} {\bibfnamefont {P.~A.}\ \bibnamefont
  {Madden}},\ }\href@noop {} {\bibfield  {journal} {\bibinfo  {journal} {Mol.\
  Phys.}\ }\textbf {\bibinfo {volume} {70}},\ \bibinfo {pages} {921} (\bibinfo
  {year} {1990})}\BibitemShut {NoStop}%
\bibitem [{\citenamefont {Pastore}\ \emph {et~al.}(1991)\citenamefont
  {Pastore}, \citenamefont {Smargassi},\ and\ \citenamefont
  {Buda}}]{GPastore91}%
  \BibitemOpen
  \bibfield  {author} {\bibinfo {author} {\bibfnamefont {G.}~\bibnamefont
  {Pastore}}, \bibinfo {author} {\bibfnamefont {E.}~\bibnamefont {Smargassi}},
  \ and\ \bibinfo {author} {\bibfnamefont {F.}~\bibnamefont {Buda}},\
  }\href@noop {} {\bibfield  {journal} {\bibinfo  {journal} {Phys. Rev. A}\
  }\textbf {\bibinfo {volume} {44}},\ \bibinfo {pages} {6334} (\bibinfo {year}
  {1991})}\BibitemShut {NoStop}%
\bibitem [{\citenamefont {Bornemann}\ and\ \citenamefont
  {Sch\"{u}tte}(1998)}]{FBornemann98}%
  \BibitemOpen
  \bibfield  {author} {\bibinfo {author} {\bibfnamefont {F.~A.}\ \bibnamefont
  {Bornemann}}\ and\ \bibinfo {author} {\bibfnamefont {C.}~\bibnamefont
  {Sch\"{u}tte}},\ }\href@noop {} {\bibfield  {journal} {\bibinfo  {journal}
  {Numerische Mathematik}\ }\textbf {\bibinfo {volume} {78}},\ \bibinfo {pages}
  {359} (\bibinfo {year} {1998})}\BibitemShut {NoStop}%
\bibitem [{\citenamefont {Carloni}\ \emph {et~al.}(2002)\citenamefont
  {Carloni}, \citenamefont {Rothlisberger},\ and\ \citenamefont
  {Parrinello}}]{PCarloni02}%
  \BibitemOpen
  \bibfield  {author} {\bibinfo {author} {\bibfnamefont {P.}~\bibnamefont
  {Carloni}}, \bibinfo {author} {\bibfnamefont {U.}~\bibnamefont
  {Rothlisberger}}, \ and\ \bibinfo {author} {\bibfnamefont {M.}~\bibnamefont
  {Parrinello}},\ }\href@noop {} {\bibfield  {journal} {\bibinfo  {journal}
  {Acc. Chem. Res.}\ }\textbf {\bibinfo {volume} {35}},\ \bibinfo {pages} {455}
  (\bibinfo {year} {2002})}\BibitemShut {NoStop}%
\bibitem [{\citenamefont {Senn}\ and\ \citenamefont {Thiel}(2009)}]{HSenn09}%
  \BibitemOpen
  \bibfield  {author} {\bibinfo {author} {\bibfnamefont {H.~M.}\ \bibnamefont
  {Senn}}\ and\ \bibinfo {author} {\bibfnamefont {W.}~\bibnamefont {Thiel}},\
  }\href@noop {} {\bibfield  {journal} {\bibinfo  {journal} {Angew. Chem. Int.
  Ed. Engl.}\ }\textbf {\bibinfo {volume} {48}},\ \bibinfo {pages} {1198}
  (\bibinfo {year} {2009})}\BibitemShut {NoStop}%
\bibitem [{\citenamefont {Heitler}\ and\ \citenamefont
  {London}(1927)}]{WHeitler27}%
  \BibitemOpen
  \bibfield  {author} {\bibinfo {author} {\bibfnamefont {W.}~\bibnamefont
  {Heitler}}\ and\ \bibinfo {author} {\bibfnamefont {F.}~\bibnamefont
  {London}},\ }\href@noop {} {\bibfield  {journal} {\bibinfo  {journal} {Z.\
  Phys.}\ }\textbf {\bibinfo {volume} {44}},\ \bibinfo {pages} {455} (\bibinfo
  {year} {1927})}\BibitemShut {NoStop}%
\bibitem [{\citenamefont {Born}\ and\ \citenamefont
  {Oppenheimer}(1927)}]{MBorn27}%
  \BibitemOpen
  \bibfield  {author} {\bibinfo {author} {\bibfnamefont {M.}~\bibnamefont
  {Born}}\ and\ \bibinfo {author} {\bibfnamefont {R.}~\bibnamefont
  {Oppenheimer}},\ }\href@noop {} {\bibfield  {journal} {\bibinfo  {journal}
  {Ann.\ Phys.}\ }\textbf {\bibinfo {volume} {389}},\ \bibinfo {pages} {475}
  (\bibinfo {year} {1927})}\BibitemShut {NoStop}%
\bibitem [{\citenamefont {Barnett}\ and\ \citenamefont
  {Landman}(1993)}]{RBarnett93}%
  \BibitemOpen
  \bibfield  {author} {\bibinfo {author} {\bibfnamefont {R.~N.}\ \bibnamefont
  {Barnett}}\ and\ \bibinfo {author} {\bibfnamefont {U.}~\bibnamefont
  {Landman}},\ }\href@noop {} {\bibfield  {journal} {\bibinfo  {journal}
  {Phys.\ Rev. B}\ }\textbf {\bibinfo {volume} {48}},\ \bibinfo {pages} {2081}
  (\bibinfo {year} {1993})}\BibitemShut {NoStop}%
\bibitem [{\citenamefont {Kresse}\ and\ \citenamefont
  {Hafner}(1993)}]{GKresse93}%
  \BibitemOpen
  \bibfield  {author} {\bibinfo {author} {\bibfnamefont {G.}~\bibnamefont
  {Kresse}}\ and\ \bibinfo {author} {\bibfnamefont {J.}~\bibnamefont
  {Hafner}},\ }\href@noop {} {\bibfield  {journal} {\bibinfo  {journal} {Phys.\
  Rev. B}\ }\textbf {\bibinfo {volume} {47}},\ \bibinfo {pages} {558} (\bibinfo
  {year} {1993})}\BibitemShut {NoStop}%
\bibitem [{\citenamefont {Pulay}\ and\ \citenamefont
  {Fogarasi}(2004)}]{PPulay04}%
  \BibitemOpen
  \bibfield  {author} {\bibinfo {author} {\bibfnamefont {P.}~\bibnamefont
  {Pulay}}\ and\ \bibinfo {author} {\bibfnamefont {G.}~\bibnamefont
  {Fogarasi}},\ }\href@noop {} {\bibfield  {journal} {\bibinfo  {journal}
  {Chem. Phys. Lett.}\ }\textbf {\bibinfo {volume} {386}},\ \bibinfo {pages}
  {272} (\bibinfo {year} {2004})}\BibitemShut {NoStop}%
\bibitem [{\citenamefont {Niklasson}\ \emph {et~al.}(2006)\citenamefont
  {Niklasson}, \citenamefont {Tymczak},\ and\ \citenamefont
  {Challacombe}}]{ANiklasson06}%
  \BibitemOpen
  \bibfield  {author} {\bibinfo {author} {\bibfnamefont {A.~M.~N.}\
  \bibnamefont {Niklasson}}, \bibinfo {author} {\bibfnamefont {C.~J.}\
  \bibnamefont {Tymczak}}, \ and\ \bibinfo {author} {\bibfnamefont
  {M.}~\bibnamefont {Challacombe}},\ }\href@noop {} {\bibfield  {journal}
  {\bibinfo  {journal} {Phys. Rev. Lett.}\ }\textbf {\bibinfo {volume} {97}},\
  \bibinfo {pages} {123001} (\bibinfo {year} {2006})}\BibitemShut {NoStop}%
\bibitem [{\citenamefont {K\"{u}hne}\ \emph {et~al.}(2007)\citenamefont
  {K\"{u}hne}, \citenamefont {Krack}, \citenamefont {Mohamed},\ and\
  \citenamefont {Parrinello}}]{TDKuhne07}%
  \BibitemOpen
  \bibfield  {author} {\bibinfo {author} {\bibfnamefont {T.~D.}\ \bibnamefont
  {K\"{u}hne}}, \bibinfo {author} {\bibfnamefont {M.}~\bibnamefont {Krack}},
  \bibinfo {author} {\bibfnamefont {F.~R.}\ \bibnamefont {Mohamed}}, \ and\
  \bibinfo {author} {\bibfnamefont {M.}~\bibnamefont {Parrinello}},\
  }\href@noop {} {\bibfield  {journal} {\bibinfo  {journal} {Phys. Rev. Lett.}\
  }\textbf {\bibinfo {volume} {98}},\ \bibinfo {pages} {066401} (\bibinfo
  {year} {2007})}\BibitemShut {NoStop}%
\bibitem [{\citenamefont {Hartke}\ and\ \citenamefont
  {Carter}(1992)}]{BHartke92}%
  \BibitemOpen
  \bibfield  {author} {\bibinfo {author} {\bibfnamefont {B.}~\bibnamefont
  {Hartke}}\ and\ \bibinfo {author} {\bibfnamefont {E.}~\bibnamefont
  {Carter}},\ }\href@noop {} {\bibfield  {journal} {\bibinfo  {journal} {Chem.
  Phys. Lett.}\ }\textbf {\bibinfo {volume} {189}},\ \bibinfo {pages} {358}
  (\bibinfo {year} {1992})}\BibitemShut {NoStop}%
\bibitem [{\citenamefont {Lambert}\ \emph {et~al.}(2006)\citenamefont
  {Lambert}, \citenamefont {Clerouin},\ and\ \citenamefont
  {Mazevet}}]{FLambert06}%
  \BibitemOpen
  \bibfield  {author} {\bibinfo {author} {\bibfnamefont {F.}~\bibnamefont
  {Lambert}}, \bibinfo {author} {\bibfnamefont {J.}~\bibnamefont {Clerouin}}, \
  and\ \bibinfo {author} {\bibfnamefont {S.}~\bibnamefont {Mazevet}},\
  }\href@noop {} {\bibfield  {journal} {\bibinfo  {journal} {Eur. Phys. Lett.}\
  }\textbf {\bibinfo {volume} {75}},\ \bibinfo {pages} {681} (\bibinfo {year}
  {2006})}\BibitemShut {NoStop}%
\bibitem [{\citenamefont {Schlegel}\ \emph {et~al.}(2001)\citenamefont
  {Schlegel}, \citenamefont {Millam}, \citenamefont {Iyengar}, \citenamefont
  {Voth}, \citenamefont {Daniels}, \citenamefont {Scusseria},\ and\
  \citenamefont {Frisch}}]{HBSchlegel01}%
  \BibitemOpen
  \bibfield  {author} {\bibinfo {author} {\bibfnamefont {H.~B.}\ \bibnamefont
  {Schlegel}}, \bibinfo {author} {\bibfnamefont {J.~M.}\ \bibnamefont
  {Millam}}, \bibinfo {author} {\bibfnamefont {S.~S.}\ \bibnamefont {Iyengar}},
  \bibinfo {author} {\bibfnamefont {G.~A.}\ \bibnamefont {Voth}}, \bibinfo
  {author} {\bibfnamefont {A.~D.}\ \bibnamefont {Daniels}}, \bibinfo {author}
  {\bibfnamefont {G.}~\bibnamefont {Scusseria}}, \ and\ \bibinfo {author}
  {\bibfnamefont {M.~J.}\ \bibnamefont {Frisch}},\ }\href@noop {} {\bibfield
  {journal} {\bibinfo  {journal} {J. Chem. Phys.}\ }\textbf {\bibinfo {volume}
  {114}},\ \bibinfo {pages} {9758} (\bibinfo {year} {2001})}\BibitemShut
  {NoStop}%
\bibitem [{\citenamefont {Iyengar}\ \emph {et~al.}(2001)\citenamefont
  {Iyengar}, \citenamefont {Schlegel}, \citenamefont {Millam}, \citenamefont
  {Voth}, \citenamefont {Scusseria},\ and\ \citenamefont
  {Frisch}}]{SIyengar01}%
  \BibitemOpen
  \bibfield  {author} {\bibinfo {author} {\bibfnamefont {S.~S.}\ \bibnamefont
  {Iyengar}}, \bibinfo {author} {\bibfnamefont {H.~B.}\ \bibnamefont
  {Schlegel}}, \bibinfo {author} {\bibfnamefont {J.~M.}\ \bibnamefont
  {Millam}}, \bibinfo {author} {\bibfnamefont {G.~A.}\ \bibnamefont {Voth}},
  \bibinfo {author} {\bibfnamefont {G.}~\bibnamefont {Scusseria}}, \ and\
  \bibinfo {author} {\bibfnamefont {M.~J.}\ \bibnamefont {Frisch}},\
  }\href@noop {} {\bibfield  {journal} {\bibinfo  {journal} {J. Chem. Phys.}\
  }\textbf {\bibinfo {volume} {115}},\ \bibinfo {pages} {10291} (\bibinfo
  {year} {2001})}\BibitemShut {NoStop}%
\bibitem [{\citenamefont {Herbert}\ and\ \citenamefont
  {Head-Gordon}(2004)}]{JMHerbert04}%
  \BibitemOpen
  \bibfield  {author} {\bibinfo {author} {\bibfnamefont {J.~M.}\ \bibnamefont
  {Herbert}}\ and\ \bibinfo {author} {\bibfnamefont {M.}~\bibnamefont
  {Head-Gordon}},\ }\href@noop {} {\bibfield  {journal} {\bibinfo  {journal}
  {J. Chem. Phys.}\ }\textbf {\bibinfo {volume} {121}},\ \bibinfo {pages}
  {11542} (\bibinfo {year} {2004})}\BibitemShut {NoStop}%
\bibitem [{\citenamefont {Li}\ \emph {et~al.}(2016)\citenamefont {Li},
  \citenamefont {Haycraft},\ and\ \citenamefont {Iyengar}}]{JLi16}%
  \BibitemOpen
  \bibfield  {author} {\bibinfo {author} {\bibfnamefont {J.}~\bibnamefont
  {Li}}, \bibinfo {author} {\bibfnamefont {C.}~\bibnamefont {Haycraft}}, \ and\
  \bibinfo {author} {\bibfnamefont {S.~S.}\ \bibnamefont {Iyengar}},\
  }\href@noop {} {\bibfield  {journal} {\bibinfo  {journal} {J. Chem. Theory
  Comput.}\ }\textbf {\bibinfo {volume} {12}},\ \bibinfo {pages} {2493}
  (\bibinfo {year} {2016})}\BibitemShut {NoStop}%
\bibitem [{\citenamefont {Goedecker}(1999)}]{SGoedecker99}%
  \BibitemOpen
  \bibfield  {author} {\bibinfo {author} {\bibfnamefont {S.}~\bibnamefont
  {Goedecker}},\ }\href@noop {} {\bibfield  {journal} {\bibinfo  {journal}
  {Rev. Mod. Phys.}\ }\textbf {\bibinfo {volume} {71}},\ \bibinfo {pages}
  {1085} (\bibinfo {year} {1999})}\BibitemShut {NoStop}%
\bibitem [{\citenamefont {Bowler}\ and\ \citenamefont
  {Miyazaki}(2012)}]{DBowler12}%
  \BibitemOpen
  \bibfield  {author} {\bibinfo {author} {\bibfnamefont {D.~R.}\ \bibnamefont
  {Bowler}}\ and\ \bibinfo {author} {\bibfnamefont {T.}~\bibnamefont
  {Miyazaki}},\ }\href@noop {} {\bibfield  {journal} {\bibinfo  {journal} {Rep.
  Prog. Phys.}\ }\textbf {\bibinfo {volume} {75}},\ \bibinfo {pages} {036503}
  (\bibinfo {year} {2012})}\BibitemShut {NoStop}%
\bibitem [{\citenamefont {Andersen}(1980)}]{HCAndersen80}%
  \BibitemOpen
  \bibfield  {author} {\bibinfo {author} {\bibfnamefont {H.~C.}\ \bibnamefont
  {Andersen}},\ }\href@noop {} {\bibfield  {journal} {\bibinfo  {journal} {J.
  Chem. Phys.}\ }\textbf {\bibinfo {volume} {72}},\ \bibinfo {pages} {2384}
  (\bibinfo {year} {1980})}\BibitemShut {NoStop}%
\bibitem [{\citenamefont {Parrinello}\ and\ \citenamefont
  {Rahman}(1980)}]{MParrinello80}%
  \BibitemOpen
  \bibfield  {author} {\bibinfo {author} {\bibfnamefont {M.}~\bibnamefont
  {Parrinello}}\ and\ \bibinfo {author} {\bibfnamefont {A.}~\bibnamefont
  {Rahman}},\ }\href@noop {} {\bibfield  {journal} {\bibinfo  {journal} {Phys.
  Rev. Lett.}\ }\textbf {\bibinfo {volume} {45}},\ \bibinfo {pages} {1196}
  (\bibinfo {year} {1980})}\BibitemShut {NoStop}%
\bibitem [{\citenamefont {Nose}(1984)}]{SNose84}%
  \BibitemOpen
  \bibfield  {author} {\bibinfo {author} {\bibfnamefont {S.}~\bibnamefont
  {Nose}},\ }\href@noop {} {\bibfield  {journal} {\bibinfo  {journal} {J. Chem.
  Phys.}\ }\textbf {\bibinfo {volume} {81}},\ \bibinfo {pages} {511} (\bibinfo
  {year} {1984})}\BibitemShut {NoStop}%
\bibitem [{\citenamefont {Niklasson}(2008{\natexlab{a}})}]{ANiklasson08}%
  \BibitemOpen
  \bibfield  {author} {\bibinfo {author} {\bibfnamefont {A.~M.~N.}\
  \bibnamefont {Niklasson}},\ }\href@noop {} {\bibfield  {journal} {\bibinfo
  {journal} {Phys. Rev. Lett.}\ }\textbf {\bibinfo {volume} {100}},\ \bibinfo
  {pages} {123004} (\bibinfo {year} {2008}{\natexlab{a}})}\BibitemShut
  {NoStop}%
\bibitem [{\citenamefont {Niklasson}\ \emph {et~al.}(2009)\citenamefont
  {Niklasson}, \citenamefont {Steneteg}, \citenamefont {Odell}, \citenamefont
  {Bock}, \citenamefont {Challacombe}, \citenamefont {Tymczak}, \citenamefont
  {Holmstrom}, \citenamefont {Zheng},\ and\ \citenamefont
  {Weber}}]{ANiklasson09}%
  \BibitemOpen
  \bibfield  {author} {\bibinfo {author} {\bibfnamefont {A.~M.~N.}\
  \bibnamefont {Niklasson}}, \bibinfo {author} {\bibfnamefont {P.}~\bibnamefont
  {Steneteg}}, \bibinfo {author} {\bibfnamefont {A.}~\bibnamefont {Odell}},
  \bibinfo {author} {\bibfnamefont {N.}~\bibnamefont {Bock}}, \bibinfo {author}
  {\bibfnamefont {M.}~\bibnamefont {Challacombe}}, \bibinfo {author}
  {\bibfnamefont {C.~J.}\ \bibnamefont {Tymczak}}, \bibinfo {author}
  {\bibfnamefont {E.}~\bibnamefont {Holmstrom}}, \bibinfo {author}
  {\bibfnamefont {G.}~\bibnamefont {Zheng}}, \ and\ \bibinfo {author}
  {\bibfnamefont {V.}~\bibnamefont {Weber}},\ }\href@noop {} {\bibfield
  {journal} {\bibinfo  {journal} {J. Chem. Phys.}\ }\textbf {\bibinfo {volume}
  {130}},\ \bibinfo {pages} {214109} (\bibinfo {year} {2009})}\BibitemShut
  {NoStop}%
\bibitem [{\citenamefont {Odell}\ \emph {et~al.}(2009)\citenamefont {Odell},
  \citenamefont {Delin}, \citenamefont {Johansson}, \citenamefont {Bock},
  \citenamefont {Challacombe},\ and\ \citenamefont {Niklasson}}]{AOdell09}%
  \BibitemOpen
  \bibfield  {author} {\bibinfo {author} {\bibfnamefont {A.}~\bibnamefont
  {Odell}}, \bibinfo {author} {\bibfnamefont {A.}~\bibnamefont {Delin}},
  \bibinfo {author} {\bibfnamefont {B.}~\bibnamefont {Johansson}}, \bibinfo
  {author} {\bibfnamefont {N.}~\bibnamefont {Bock}}, \bibinfo {author}
  {\bibfnamefont {M.}~\bibnamefont {Challacombe}}, \ and\ \bibinfo {author}
  {\bibfnamefont {A.~M.~N.}\ \bibnamefont {Niklasson}},\ }\href@noop {}
  {\bibfield  {journal} {\bibinfo  {journal} {J. Chem. Phys.}\ }\textbf
  {\bibinfo {volume} {131}},\ \bibinfo {pages} {244106} (\bibinfo {year}
  {2009})}\BibitemShut {NoStop}%
\bibitem [{\citenamefont {Steneteg}\ \emph {et~al.}(2010)\citenamefont
  {Steneteg}, \citenamefont {Abrikosov}, \citenamefont {Weber},\ and\
  \citenamefont {Niklasson}}]{PSteneteg10}%
  \BibitemOpen
  \bibfield  {author} {\bibinfo {author} {\bibfnamefont {P.}~\bibnamefont
  {Steneteg}}, \bibinfo {author} {\bibfnamefont {I.~A.}\ \bibnamefont
  {Abrikosov}}, \bibinfo {author} {\bibfnamefont {V.}~\bibnamefont {Weber}}, \
  and\ \bibinfo {author} {\bibfnamefont {A.~M.~N.}\ \bibnamefont {Niklasson}},\
  }\href@noop {} {\bibfield  {journal} {\bibinfo  {journal} {Phys. Rev. B}\
  }\textbf {\bibinfo {volume} {82}},\ \bibinfo {pages} {075110} (\bibinfo
  {year} {2010})}\BibitemShut {NoStop}%
\bibitem [{\citenamefont {Zheng}\ \emph {et~al.}(2011)\citenamefont {Zheng},
  \citenamefont {Niklasson},\ and\ \citenamefont {Karplus}}]{GZheng11}%
  \BibitemOpen
  \bibfield  {author} {\bibinfo {author} {\bibfnamefont {G.}~\bibnamefont
  {Zheng}}, \bibinfo {author} {\bibfnamefont {A.~M.~N.}\ \bibnamefont
  {Niklasson}}, \ and\ \bibinfo {author} {\bibfnamefont {M.}~\bibnamefont
  {Karplus}},\ }\href@noop {} {\bibfield  {journal} {\bibinfo  {journal} {J.
  Chem. Phys.}\ }\textbf {\bibinfo {volume} {135}},\ \bibinfo {pages} {044122}
  (\bibinfo {year} {2011})}\BibitemShut {NoStop}%
\bibitem [{\citenamefont {Cawkwell}\ and\ \citenamefont
  {Niklasson}(2012)}]{MCawkwell12}%
  \BibitemOpen
  \bibfield  {author} {\bibinfo {author} {\bibfnamefont {M.~J.}\ \bibnamefont
  {Cawkwell}}\ and\ \bibinfo {author} {\bibfnamefont {A.~M.~N.}\ \bibnamefont
  {Niklasson}},\ }\href@noop {} {\bibfield  {journal} {\bibinfo  {journal} {J.
  Chem. Phys.}\ }\textbf {\bibinfo {volume} {137}},\ \bibinfo {pages} {134105}
  (\bibinfo {year} {2012})}\BibitemShut {NoStop}%
\bibitem [{\citenamefont {Souvatzis}\ and\ \citenamefont
  {Niklasson}(2014)}]{PSouvatzis14}%
  \BibitemOpen
  \bibfield  {author} {\bibinfo {author} {\bibfnamefont {P.}~\bibnamefont
  {Souvatzis}}\ and\ \bibinfo {author} {\bibfnamefont {A.~M.~N.}\ \bibnamefont
  {Niklasson}},\ }\href@noop {} {\bibfield  {journal} {\bibinfo  {journal} {J.
  Chem. Phys.}\ }\textbf {\bibinfo {volume} {140}},\ \bibinfo {pages} {044117}
  (\bibinfo {year} {2014})}\BibitemShut {NoStop}%
\bibitem [{\citenamefont {Niklasson}\ and\ \citenamefont
  {Cawkwell}(2014)}]{ANiklasson14}%
  \BibitemOpen
  \bibfield  {author} {\bibinfo {author} {\bibfnamefont {A.~M.~N.}\
  \bibnamefont {Niklasson}}\ and\ \bibinfo {author} {\bibfnamefont
  {M.}~\bibnamefont {Cawkwell}},\ }\href@noop {} {\bibfield  {journal}
  {\bibinfo  {journal} {J. Chem. Phys.}\ }\textbf {\bibinfo {volume} {141}},\
  \bibinfo {pages} {164123} (\bibinfo {year} {2014})}\BibitemShut {NoStop}%
\bibitem [{\citenamefont {Aradi}\ \emph {et~al.}(2015)\citenamefont {Aradi},
  \citenamefont {Niklasson},\ and\ \citenamefont {Frauenheim}}]{BAradi15}%
  \BibitemOpen
  \bibfield  {author} {\bibinfo {author} {\bibfnamefont {B.}~\bibnamefont
  {Aradi}}, \bibinfo {author} {\bibfnamefont {A.~M.~N.}\ \bibnamefont
  {Niklasson}}, \ and\ \bibinfo {author} {\bibfnamefont {T.}~\bibnamefont
  {Frauenheim}},\ }\href@noop {} {\bibfield  {journal} {\bibinfo  {journal} {J.
  Chem. Theory Comput.}\ }\textbf {\bibinfo {volume} {11}},\ \bibinfo {pages}
  {3357} (\bibinfo {year} {2015})}\BibitemShut {NoStop}%
\bibitem [{\citenamefont {Martinez}\ \emph {et~al.}(2015)\citenamefont
  {Martinez}, \citenamefont {Cawkwell}, , \citenamefont {Voter},\ and\
  \citenamefont {Niklasson}}]{EMartinez15}%
  \BibitemOpen
  \bibfield  {author} {\bibinfo {author} {\bibfnamefont {E.}~\bibnamefont
  {Martinez}}, \bibinfo {author} {\bibfnamefont {M.~J.}\ \bibnamefont
  {Cawkwell}}, , \bibinfo {author} {\bibfnamefont {A.~F.}\ \bibnamefont
  {Voter}}, \ and\ \bibinfo {author} {\bibfnamefont {A.~M.~N.}\ \bibnamefont
  {Niklasson}},\ }\href@noop {} {\bibfield  {journal} {\bibinfo  {journal} {J.
  Chem. Phys.}\ }\textbf {\bibinfo {volume} {142}},\ \bibinfo {pages} {1770}
  (\bibinfo {year} {2015})}\BibitemShut {NoStop}%
\bibitem [{\citenamefont {Vitale}\ \emph {et~al.}(2017)\citenamefont {Vitale},
  \citenamefont {Dziezic}, \citenamefont {Albaugh}, \citenamefont {Niklasson},
  \citenamefont {Head-Gordon},\ and\ \citenamefont {Skylaris}}]{VVitale17}%
  \BibitemOpen
  \bibfield  {author} {\bibinfo {author} {\bibfnamefont {V.}~\bibnamefont
  {Vitale}}, \bibinfo {author} {\bibfnamefont {J.}~\bibnamefont {Dziezic}},
  \bibinfo {author} {\bibfnamefont {A.}~\bibnamefont {Albaugh}}, \bibinfo
  {author} {\bibfnamefont {A.}~\bibnamefont {Niklasson}}, \bibinfo {author}
  {\bibfnamefont {T.~J.}\ \bibnamefont {Head-Gordon}}, \ and\ \bibinfo {author}
  {\bibfnamefont {C.-K.}\ \bibnamefont {Skylaris}},\ }\href@noop {} {\bibfield
  {journal} {\bibinfo  {journal} {J. Chem. Phys.}\ }\textbf {\bibinfo {volume}
  {12}},\ \bibinfo {pages} {124115} (\bibinfo {year} {2017})}\BibitemShut
  {NoStop}%
\bibitem [{\citenamefont {Albaugh}\ \emph {et~al.}(2017)\citenamefont
  {Albaugh}, \citenamefont {Niklasson},\ and\ \citenamefont
  {Head-Gordon}}]{AAlbaugh17}%
  \BibitemOpen
  \bibfield  {author} {\bibinfo {author} {\bibfnamefont {A.}~\bibnamefont
  {Albaugh}}, \bibinfo {author} {\bibfnamefont {A.}~\bibnamefont {Niklasson}},
  \ and\ \bibinfo {author} {\bibfnamefont {T.~J.}\ \bibnamefont
  {Head-Gordon}},\ }\href@noop {} {\bibfield  {journal} {\bibinfo  {journal}
  {J. Phys. Chem. Lett.}\ }\textbf {\bibinfo {volume} {8}},\ \bibinfo {pages}
  {1714} (\bibinfo {year} {2017})}\BibitemShut {NoStop}%
\bibitem [{\citenamefont {Lin}\ \emph {et~al.}(2014)\citenamefont {Lin},
  \citenamefont {Lu},\ and\ \citenamefont {Shao}}]{LLin14}%
  \BibitemOpen
  \bibfield  {author} {\bibinfo {author} {\bibfnamefont {L.}~\bibnamefont
  {Lin}}, \bibinfo {author} {\bibfnamefont {J.}~\bibnamefont {Lu}}, \ and\
  \bibinfo {author} {\bibfnamefont {S.}~\bibnamefont {Shao}},\ }\href@noop {}
  {\bibfield  {journal} {\bibinfo  {journal} {Entropy}\ }\textbf {\bibinfo
  {volume} {16}},\ \bibinfo {pages} {110} (\bibinfo {year} {2014})}\BibitemShut
  {NoStop}%
\bibitem [{\citenamefont {Herbert}\ and\ \citenamefont
  {Head-Gordon}(2005)}]{JMHerbert05}%
  \BibitemOpen
  \bibfield  {author} {\bibinfo {author} {\bibfnamefont {J.}~\bibnamefont
  {Herbert}}\ and\ \bibinfo {author} {\bibfnamefont {M.}~\bibnamefont
  {Head-Gordon}},\ }\href@noop {} {\bibfield  {journal} {\bibinfo  {journal}
  {Phys. Chem. Chem. Phys.}\ }\textbf {\bibinfo {volume} {7}},\ \bibinfo
  {pages} {3269} (\bibinfo {year} {2005})}\BibitemShut {NoStop}%
\bibitem [{\citenamefont {Kolafa}(2003)}]{JKolafa03}%
  \BibitemOpen
  \bibfield  {author} {\bibinfo {author} {\bibfnamefont {J.}~\bibnamefont
  {Kolafa}},\ }\href@noop {} {\bibfield  {journal} {\bibinfo  {journal} {J.
  Comput. Chem.}\ }\textbf {\bibinfo {volume} {25}},\ \bibinfo {pages} {335}
  (\bibinfo {year} {2003})}\BibitemShut {NoStop}%
\bibitem [{\citenamefont {Niklasson}\ \emph {et~al.}(2011)\citenamefont
  {Niklasson}, \citenamefont {Steneteg},\ and\ \citenamefont
  {Bock}}]{ANiklasson11}%
  \BibitemOpen
  \bibfield  {author} {\bibinfo {author} {\bibfnamefont {A.~M.~N.}\
  \bibnamefont {Niklasson}}, \bibinfo {author} {\bibfnamefont {P.}~\bibnamefont
  {Steneteg}}, \ and\ \bibinfo {author} {\bibfnamefont {N.}~\bibnamefont
  {Bock}},\ }\href@noop {} {\bibfield  {journal} {\bibinfo  {journal} {J. Chem.
  Phys.}\ }\textbf {\bibinfo {volume} {135}},\ \bibinfo {pages} {164111}
  (\bibinfo {year} {2011})}\BibitemShut {NoStop}%
\bibitem [{\citenamefont {Arita}\ \emph {et~al.}(2014)\citenamefont {Arita},
  \citenamefont {Bowler},\ and\ \citenamefont {Miyazaki}}]{MArita14}%
  \BibitemOpen
  \bibfield  {author} {\bibinfo {author} {\bibfnamefont {M.}~\bibnamefont
  {Arita}}, \bibinfo {author} {\bibfnamefont {D.~R.}\ \bibnamefont {Bowler}}, \
  and\ \bibinfo {author} {\bibfnamefont {T.}~\bibnamefont {Miyazaki}},\
  }\href@noop {} {\bibfield  {journal} {\bibinfo  {journal} {J. Chem. Theory
  Comput.}\ }\textbf {\bibinfo {volume} {10}},\ \bibinfo {pages} {5419}
  (\bibinfo {year} {2014})}\BibitemShut {NoStop}%
\bibitem [{\citenamefont {Nomura}\ \emph {et~al.}(2015)\citenamefont {Nomura},
  \citenamefont {Small}, \citenamefont {Kalia}, \citenamefont {Nakano},\ and\
  \citenamefont {Vashista}}]{KNomura15}%
  \BibitemOpen
  \bibfield  {author} {\bibinfo {author} {\bibfnamefont {K.}~\bibnamefont
  {Nomura}}, \bibinfo {author} {\bibfnamefont {P.~E.}\ \bibnamefont {Small}},
  \bibinfo {author} {\bibfnamefont {R.~K.}\ \bibnamefont {Kalia}}, \bibinfo
  {author} {\bibfnamefont {A.}~\bibnamefont {Nakano}}, \ and\ \bibinfo {author}
  {\bibfnamefont {P.}~\bibnamefont {Vashista}},\ }\href@noop {} {\bibfield
  {journal} {\bibinfo  {journal} {Comput. Phys. Comm.}\ }\textbf {\bibinfo
  {volume} {192}},\ \bibinfo {pages} {91} (\bibinfo {year} {2015})}\BibitemShut
  {NoStop}%
\bibitem [{\citenamefont {Albaugh}\ \emph {et~al.}(2015)\citenamefont
  {Albaugh}, \citenamefont {Demardash},\ and\ \citenamefont
  {Head-Gordon}}]{AAlbaugh15}%
  \BibitemOpen
  \bibfield  {author} {\bibinfo {author} {\bibfnamefont {A.}~\bibnamefont
  {Albaugh}}, \bibinfo {author} {\bibfnamefont {O.}~\bibnamefont {Demardash}},
  \ and\ \bibinfo {author} {\bibfnamefont {T.~J.}\ \bibnamefont
  {Head-Gordon}},\ }\href@noop {} {\bibfield  {journal} {\bibinfo  {journal}
  {J. Chem. Phys.}\ }\textbf {\bibinfo {volume} {143}},\ \bibinfo {pages}
  {174104} (\bibinfo {year} {2015})}\BibitemShut {NoStop}%
\bibitem [{\citenamefont {Niklasson}\ \emph {et~al.}(2016)\citenamefont
  {Niklasson}, \citenamefont {Mnizsewski}, \citenamefont {Negre}, \citenamefont
  {Cawkwell}, \citenamefont {Swart}, \citenamefont {Mohd-Yusof}, \citenamefont
  {Germann}, \citenamefont {Wall}, \citenamefont {Bock}, \citenamefont
  {Rubensson},\ and\ \citenamefont {Djidjev}}]{ANiklasson16}%
  \BibitemOpen
  \bibfield  {author} {\bibinfo {author} {\bibfnamefont {A.~M.~N.}\
  \bibnamefont {Niklasson}}, \bibinfo {author} {\bibfnamefont {S.~M.}\
  \bibnamefont {Mnizsewski}}, \bibinfo {author} {\bibfnamefont {C.~F.~A.}\
  \bibnamefont {Negre}}, \bibinfo {author} {\bibfnamefont {M.~J.}\ \bibnamefont
  {Cawkwell}}, \bibinfo {author} {\bibfnamefont {P.~J.}\ \bibnamefont {Swart}},
  \bibinfo {author} {\bibfnamefont {J.}~\bibnamefont {Mohd-Yusof}}, \bibinfo
  {author} {\bibfnamefont {T.~C.}\ \bibnamefont {Germann}}, \bibinfo {author}
  {\bibfnamefont {M.~E.}\ \bibnamefont {Wall}}, \bibinfo {author}
  {\bibfnamefont {N.}~\bibnamefont {Bock}}, \bibinfo {author} {\bibfnamefont
  {E.~H.}\ \bibnamefont {Rubensson}}, \ and\ \bibinfo {author} {\bibfnamefont
  {H.}~\bibnamefont {Djidjev}},\ }\href@noop {} {\bibfield  {journal} {\bibinfo
   {journal} {J. Chem. Phys.}\ }\textbf {\bibinfo {volume} {144}},\ \bibinfo
  {pages} {234101} (\bibinfo {year} {2016})}\BibitemShut {NoStop}%
\bibitem [{\citenamefont {Levy}(1979)}]{MLevy79}%
  \BibitemOpen
  \bibfield  {author} {\bibinfo {author} {\bibfnamefont {M.}~\bibnamefont
  {Levy}},\ }\href@noop {} {\bibfield  {journal} {\bibinfo  {journal} {Proc.
  Nat. Acad. Sci.}\ }\textbf {\bibinfo {volume} {76}},\ \bibinfo {pages} {6062}
  (\bibinfo {year} {1979})}\BibitemShut {NoStop}%
\bibitem [{\citenamefont {Lieb}(1983)}]{ELieb83}%
  \BibitemOpen
  \bibfield  {author} {\bibinfo {author} {\bibfnamefont {E.~H.}\ \bibnamefont
  {Lieb}},\ }\href@noop {} {\bibfield  {journal} {\bibinfo  {journal} {Int. J.
  Quant. Chem.}\ }\textbf {\bibinfo {volume} {24}},\ \bibinfo {pages} {243}
  (\bibinfo {year} {1983})}\BibitemShut {NoStop}%
\bibitem [{\citenamefont {Hohenberg}\ and\ \citenamefont {Kohn}(1964)}]{hohen}%
  \BibitemOpen
  \bibfield  {author} {\bibinfo {author} {\bibfnamefont {P.}~\bibnamefont
  {Hohenberg}}\ and\ \bibinfo {author} {\bibfnamefont {W.}~\bibnamefont
  {Kohn}},\ }\href@noop {} {\bibfield  {journal} {\bibinfo  {journal} {Phys.
  Rev.}\ }\textbf {\bibinfo {volume} {136}},\ \bibinfo {pages} {B:864}
  (\bibinfo {year} {1964})}\BibitemShut {NoStop}%
\bibitem [{\citenamefont {Parr}\ and\ \citenamefont {Yang}(1989)}]{RParr89}%
  \BibitemOpen
  \bibfield  {author} {\bibinfo {author} {\bibfnamefont {R.~G.}\ \bibnamefont
  {Parr}}\ and\ \bibinfo {author} {\bibfnamefont {W.}~\bibnamefont {Yang}},\
  }\href@noop {} {\emph {\bibinfo {title} {Density-functional theory of atoms
  and molecules}}}\ (\bibinfo  {publisher} {Oxford University Press},\ \bibinfo
  {address} {Oxford},\ \bibinfo {year} {1989})\BibitemShut {NoStop}%
\bibitem [{\citenamefont {Dreizler}\ and\ \citenamefont
  {Gross}(1990)}]{RMDreizler90}%
  \BibitemOpen
  \bibfield  {author} {\bibinfo {author} {\bibfnamefont {R.}~\bibnamefont
  {Dreizler}}\ and\ \bibinfo {author} {\bibfnamefont {K.}~\bibnamefont
  {Gross}},\ }\href@noop {} {\emph {\bibinfo {title} {Density-functional
  theory}}}\ (\bibinfo  {publisher} {Springer Verlag},\ \bibinfo {address}
  {Berlin Heidelberg},\ \bibinfo {year} {1990})\BibitemShut {NoStop}%
\bibitem [{\citenamefont {Broyden}(1965)}]{CGBroyden65}%
  \BibitemOpen
  \bibfield  {author} {\bibinfo {author} {\bibfnamefont {C.~G.}\ \bibnamefont
  {Broyden}},\ }\href@noop {} {\bibfield  {journal} {\bibinfo  {journal} {Math.
  Comput.}\ }\textbf {\bibinfo {volume} {19}},\ \bibinfo {pages} {577}
  (\bibinfo {year} {1965})}\BibitemShut {NoStop}%
\bibitem [{\citenamefont {Anderson}(1965)}]{DGAnderson65}%
  \BibitemOpen
  \bibfield  {author} {\bibinfo {author} {\bibfnamefont {D.~G.}\ \bibnamefont
  {Anderson}},\ }\href@noop {} {\bibfield  {journal} {\bibinfo  {journal} {J.
  Assoc. Comput. Mach.}\ }\textbf {\bibinfo {volume} {12}},\ \bibinfo {pages}
  {547} (\bibinfo {year} {1965})}\BibitemShut {NoStop}%
\bibitem [{\citenamefont {Pulay}(1980)}]{PPulay80}%
  \BibitemOpen
  \bibfield  {author} {\bibinfo {author} {\bibfnamefont {P.}~\bibnamefont
  {Pulay}},\ }\href@noop {} {\bibfield  {journal} {\bibinfo  {journal} {Chem.
  Phys. Let.}\ }\textbf {\bibinfo {volume} {73}},\ \bibinfo {pages} {393}
  (\bibinfo {year} {1980})}\BibitemShut {NoStop}%
\bibitem [{\citenamefont {Srivastava}(1984)}]{GPSrivastava84}%
  \BibitemOpen
  \bibfield  {author} {\bibinfo {author} {\bibfnamefont {G.~P.}\ \bibnamefont
  {Srivastava}},\ }\href@noop {} {\bibfield  {journal} {\bibinfo  {journal} {J.
  Phys. A: Math. Gen.}\ }\textbf {\bibinfo {volume} {17}},\ \bibinfo {pages}
  {L317} (\bibinfo {year} {1984})}\BibitemShut {NoStop}%
\bibitem [{\citenamefont {Kerker}(1981)}]{GPKerker81}%
  \BibitemOpen
  \bibfield  {author} {\bibinfo {author} {\bibfnamefont {G.~P.}\ \bibnamefont
  {Kerker}},\ }\href {\doibase 10.1103/PhysRevB.23.3082} {\bibfield  {journal}
  {\bibinfo  {journal} {Phys. Rev. B}\ }\textbf {\bibinfo {volume} {23}},\
  \bibinfo {pages} {3082} (\bibinfo {year} {1981})}\BibitemShut {NoStop}%
\bibitem [{\citenamefont {Johnson}(1988)}]{DDJohnson88}%
  \BibitemOpen
  \bibfield  {author} {\bibinfo {author} {\bibfnamefont {D.~D.}\ \bibnamefont
  {Johnson}},\ }\href {\doibase 10.1103/PhysRevB.38.12807} {\bibfield
  {journal} {\bibinfo  {journal} {Phys. Rev. B}\ }\textbf {\bibinfo {volume}
  {38}},\ \bibinfo {pages} {12807} (\bibinfo {year} {1988})}\BibitemShut
  {NoStop}%
\bibitem [{\citenamefont {Fletcher}(1970)}]{RFletcher70}%
  \BibitemOpen
  \bibfield  {author} {\bibinfo {author} {\bibfnamefont {R.}~\bibnamefont
  {Fletcher}},\ }\href@noop {} {\bibfield  {journal} {\bibinfo  {journal} {Mol.
  Phys.}\ }\textbf {\bibinfo {volume} {19}},\ \bibinfo {pages} {55} (\bibinfo
  {year} {1970})}\BibitemShut {NoStop}%
\bibitem [{\citenamefont {Stich}\ \emph {et~al.}(1989)\citenamefont {Stich},
  \citenamefont {Car}, \citenamefont {Parrinello},\ and\ \citenamefont
  {Baroni}}]{IStich89}%
  \BibitemOpen
  \bibfield  {author} {\bibinfo {author} {\bibfnamefont {I.}~\bibnamefont
  {Stich}}, \bibinfo {author} {\bibfnamefont {R.}~\bibnamefont {Car}}, \bibinfo
  {author} {\bibfnamefont {M.}~\bibnamefont {Parrinello}}, \ and\ \bibinfo
  {author} {\bibfnamefont {S.}~\bibnamefont {Baroni}},\ }\href@noop {}
  {\bibfield  {journal} {\bibinfo  {journal} {Phys. Rev. B}\ }\textbf {\bibinfo
  {volume} {39}},\ \bibinfo {pages} {4997} (\bibinfo {year}
  {1989})}\BibitemShut {NoStop}%
\bibitem [{\citenamefont {Hutter}\ \emph {et~al.}(1994)\citenamefont {Hutter},
  \citenamefont {Luthi},\ and\ \citenamefont {Parrinello}}]{JHutter94}%
  \BibitemOpen
  \bibfield  {author} {\bibinfo {author} {\bibfnamefont {J.}~\bibnamefont
  {Hutter}}, \bibinfo {author} {\bibfnamefont {H.~P.}\ \bibnamefont {Luthi}}, \
  and\ \bibinfo {author} {\bibfnamefont {M.}~\bibnamefont {Parrinello}},\
  }\href@noop {} {\bibfield  {journal} {\bibinfo  {journal} {Comput. Mater.
  Sci}\ }\textbf {\bibinfo {volume} {2}},\ \bibinfo {pages} {244} (\bibinfo
  {year} {1994})}\BibitemShut {NoStop}%
\bibitem [{\citenamefont {Weber}\ \emph {et~al.}(2008)\citenamefont {Weber},
  \citenamefont {VandeVondele}, \citenamefont {Hutter},\ and\ \citenamefont
  {Niklasson}}]{VWeber08a}%
  \BibitemOpen
  \bibfield  {author} {\bibinfo {author} {\bibfnamefont {V.}~\bibnamefont
  {Weber}}, \bibinfo {author} {\bibfnamefont {J.}~\bibnamefont {VandeVondele}},
  \bibinfo {author} {\bibfnamefont {J.}~\bibnamefont {Hutter}}, \ and\ \bibinfo
  {author} {\bibfnamefont {A.~M.~N.}\ \bibnamefont {Niklasson}},\ }\href@noop
  {} {\bibfield  {journal} {\bibinfo  {journal} {J. Chem. Phys.}\ }\textbf
  {\bibinfo {volume} {128}},\ \bibinfo {pages} {084113} (\bibinfo {year}
  {2008})}\BibitemShut {NoStop}%
\bibitem [{\citenamefont {Channel}\ and\ \citenamefont
  {Scovel}(1990)}]{JPChannel90}%
  \BibitemOpen
  \bibfield  {author} {\bibinfo {author} {\bibfnamefont {J.~P.}\ \bibnamefont
  {Channel}}\ and\ \bibinfo {author} {\bibfnamefont {C.}~\bibnamefont
  {Scovel}},\ }\href@noop {} {\bibfield  {journal} {\bibinfo  {journal}
  {Nonlinearity}\ }\textbf {\bibinfo {volume} {3}},\ \bibinfo {pages} {231}
  (\bibinfo {year} {1990})}\BibitemShut {NoStop}%
\bibitem [{\citenamefont {McLachlan}\ and\ \citenamefont
  {Atela}(1992)}]{McLachlan92}%
  \BibitemOpen
  \bibfield  {author} {\bibinfo {author} {\bibfnamefont {R.}~\bibnamefont
  {McLachlan}}\ and\ \bibinfo {author} {\bibfnamefont {P.}~\bibnamefont
  {Atela}},\ }\href@noop {} {\bibfield  {journal} {\bibinfo  {journal}
  {Nonlinearity}\ }\textbf {\bibinfo {volume} {5}},\ \bibinfo {pages} {541}
  (\bibinfo {year} {1992})}\BibitemShut {NoStop}%
\bibitem [{\citenamefont {Leimkuhler}\ and\ \citenamefont
  {Reich}(2004)}]{BJLeimkuhler04}%
  \BibitemOpen
  \bibfield  {author} {\bibinfo {author} {\bibfnamefont {B.}~\bibnamefont
  {Leimkuhler}}\ and\ \bibinfo {author} {\bibfnamefont {S.}~\bibnamefont
  {Reich}},\ }\href@noop {} {\emph {\bibinfo {title} {Simulating Hamiltonian
  Dynamics}}}\ (\bibinfo  {publisher} {Cambridge University Press},\ \bibinfo
  {address} {Cambridge},\ \bibinfo {year} {2004})\BibitemShut {NoStop}%
\bibitem [{\citenamefont {Engel}\ \emph {et~al.}(2005)\citenamefont {Engel},
  \citenamefont {Skeel},\ and\ \citenamefont {Drees}}]{RDEngle05}%
  \BibitemOpen
  \bibfield  {author} {\bibinfo {author} {\bibfnamefont {R.~D.}\ \bibnamefont
  {Engel}}, \bibinfo {author} {\bibfnamefont {R.~D.}\ \bibnamefont {Skeel}}, \
  and\ \bibinfo {author} {\bibfnamefont {M.}~\bibnamefont {Drees}},\
  }\href@noop {} {\bibfield  {journal} {\bibinfo  {journal} {J. Comput. Phys.}\
  }\textbf {\bibinfo {volume} {206}},\ \bibinfo {pages} {432} (\bibinfo {year}
  {2005})}\BibitemShut {NoStop}%
\bibitem [{\citenamefont {Harris}(1985)}]{JHarris85}%
  \BibitemOpen
  \bibfield  {author} {\bibinfo {author} {\bibfnamefont {J.}~\bibnamefont
  {Harris}},\ }\href@noop {} {\bibfield  {journal} {\bibinfo  {journal} {Phys.
  Rev. B}\ }\textbf {\bibinfo {volume} {31}},\ \bibinfo {pages} {1770}
  (\bibinfo {year} {1985})}\BibitemShut {NoStop}%
\bibitem [{\citenamefont {Foulkes}\ and\ \citenamefont
  {Haydock}(1989)}]{MFoulkes89}%
  \BibitemOpen
  \bibfield  {author} {\bibinfo {author} {\bibfnamefont {W.~M.~C.}\
  \bibnamefont {Foulkes}}\ and\ \bibinfo {author} {\bibfnamefont
  {R.}~\bibnamefont {Haydock}},\ }\href@noop {} {\bibfield  {journal} {\bibinfo
   {journal} {Phys. Rev. B}\ }\textbf {\bibinfo {volume} {39}},\ \bibinfo
  {pages} {12520} (\bibinfo {year} {1989})}\BibitemShut {NoStop}%
\bibitem [{\citenamefont {Negre}\ \emph {et~al.}(2016)\citenamefont {Negre},
  \citenamefont {Mnizsewski}, \citenamefont {Niklasson}, \citenamefont
  {Mnizsewski}, \citenamefont {Cawkwell}, \citenamefont {Bock}, \citenamefont
  {Wall},\ and\ \citenamefont {Niklasson}}]{CNegre16}%
  \BibitemOpen
  \bibfield  {author} {\bibinfo {author} {\bibfnamefont {C.~F.~A.}\
  \bibnamefont {Negre}}, \bibinfo {author} {\bibfnamefont {S.~M.}\ \bibnamefont
  {Mnizsewski}}, \bibinfo {author} {\bibfnamefont {A.~M.~N.}\ \bibnamefont
  {Niklasson}}, \bibinfo {author} {\bibfnamefont {S.~M.}\ \bibnamefont
  {Mnizsewski}}, \bibinfo {author} {\bibfnamefont {M.~J.}\ \bibnamefont
  {Cawkwell}}, \bibinfo {author} {\bibfnamefont {N.}~\bibnamefont {Bock}},
  \bibinfo {author} {\bibfnamefont {M.~E.}\ \bibnamefont {Wall}}, \ and\
  \bibinfo {author} {\bibfnamefont {A.~M.~N.}\ \bibnamefont {Niklasson}},\
  }\href@noop {} {\bibfield  {journal} {\bibinfo  {journal} {J. Chem. Theory
  Comput.}\ }\textbf {\bibinfo {volume} {12}},\ \bibinfo {pages} {3063}
  (\bibinfo {year} {2016})}\BibitemShut {NoStop}%
\bibitem [{\citenamefont {Hirakawa}\ \emph {et~al.}(2017)\citenamefont
  {Hirakawa}, \citenamefont {suzuki}, \citenamefont {Bowler},\ and\
  \citenamefont {Myazaki}}]{THirakawa17}%
  \BibitemOpen
  \bibfield  {author} {\bibinfo {author} {\bibfnamefont {T.}~\bibnamefont
  {Hirakawa}}, \bibinfo {author} {\bibfnamefont {T.}~\bibnamefont {suzuki}},
  \bibinfo {author} {\bibfnamefont {D.~R.}\ \bibnamefont {Bowler}}, \ and\
  \bibinfo {author} {\bibfnamefont {T.}~\bibnamefont {Myazaki}},\ }\href@noop
  {} {\bibfield  {journal} {\bibinfo  {journal} {arXiv
  http://arxiv.org/abs/1705.01448}\ } (\bibinfo {year} {2017})}\BibitemShut
  {NoStop}%
\bibitem [{\citenamefont {Pecora}\ and\ \citenamefont
  {Carrol}(1990)}]{LPecora90}%
  \BibitemOpen
  \bibfield  {author} {\bibinfo {author} {\bibfnamefont {L.~M.}\ \bibnamefont
  {Pecora}}\ and\ \bibinfo {author} {\bibfnamefont {T.~L.}\ \bibnamefont
  {Carrol}},\ }\href@noop {} {\bibfield  {journal} {\bibinfo  {journal} {Phys.
  Rev. Lett.}\ }\textbf {\bibinfo {volume} {64}},\ \bibinfo {pages} {821}
  (\bibinfo {year} {1990})}\BibitemShut {NoStop}%
\bibitem [{\citenamefont {Pecora}\ and\ \citenamefont
  {Carrol}(2015)}]{LPecora15}%
  \BibitemOpen
  \bibfield  {author} {\bibinfo {author} {\bibfnamefont {L.~M.}\ \bibnamefont
  {Pecora}}\ and\ \bibinfo {author} {\bibfnamefont {T.~L.}\ \bibnamefont
  {Carrol}},\ }\href@noop {} {\bibfield  {journal} {\bibinfo  {journal}
  {Chaos}\ }\textbf {\bibinfo {volume} {25}},\ \bibinfo {pages} {097611}
  (\bibinfo {year} {2015})}\BibitemShut {NoStop}%
\bibitem [{\citenamefont {Odell}\ \emph {et~al.}(2011)\citenamefont {Odell},
  \citenamefont {Delin}, \citenamefont {Johansson}, \citenamefont {Cawkwell},\
  and\ \citenamefont {Niklasson}}]{AOdell11}%
  \BibitemOpen
  \bibfield  {author} {\bibinfo {author} {\bibfnamefont {A.}~\bibnamefont
  {Odell}}, \bibinfo {author} {\bibfnamefont {A.}~\bibnamefont {Delin}},
  \bibinfo {author} {\bibfnamefont {B.}~\bibnamefont {Johansson}}, \bibinfo
  {author} {\bibfnamefont {M.~J.}\ \bibnamefont {Cawkwell}}, \ and\ \bibinfo
  {author} {\bibfnamefont {A.~M.~N.}\ \bibnamefont {Niklasson}},\ }\href@noop
  {} {\bibfield  {journal} {\bibinfo  {journal} {J. Chem. Phys.}\ }\textbf
  {\bibinfo {volume} {135}},\ \bibinfo {pages} {224105} (\bibinfo {year}
  {2011})}\BibitemShut {NoStop}%
\bibitem [{\citenamefont {Tangney}\ and\ \citenamefont
  {Scandolo}(2002)}]{PTangney02}%
  \BibitemOpen
  \bibfield  {author} {\bibinfo {author} {\bibfnamefont {P.}~\bibnamefont
  {Tangney}}\ and\ \bibinfo {author} {\bibfnamefont {S.}~\bibnamefont
  {Scandolo}},\ }\href@noop {} {\bibfield  {journal} {\bibinfo  {journal} {J.
  Chem. Phys.}\ }\textbf {\bibinfo {volume} {116}},\ \bibinfo {pages} {14}
  (\bibinfo {year} {2002})}\BibitemShut {NoStop}%
\bibitem [{\citenamefont {Tangney}(2006)}]{PTangney06}%
  \BibitemOpen
  \bibfield  {author} {\bibinfo {author} {\bibfnamefont {P.}~\bibnamefont
  {Tangney}},\ }\href@noop {} {\bibfield  {journal} {\bibinfo  {journal} {J.
  Chem. Phys.}\ }\textbf {\bibinfo {volume} {124}},\ \bibinfo {pages} {44111}
  (\bibinfo {year} {2006})}\BibitemShut {NoStop}%
\bibitem [{\citenamefont {Souvatzis}\ and\ \citenamefont
  {Niklasson}(2013)}]{PSouvatzis13}%
  \BibitemOpen
  \bibfield  {author} {\bibinfo {author} {\bibfnamefont {P.}~\bibnamefont
  {Souvatzis}}\ and\ \bibinfo {author} {\bibfnamefont {A.~M.~N.}\ \bibnamefont
  {Niklasson}},\ }\href@noop {} {\bibfield  {journal} {\bibinfo  {journal} {J.
  Chem. Phys.}\ }\textbf {\bibinfo {volume} {139}},\ \bibinfo {pages} {214102}
  (\bibinfo {year} {2013})}\BibitemShut {NoStop}%
\bibitem [{\citenamefont {Mermin}(1965)}]{NMermin65}%
  \BibitemOpen
  \bibfield  {author} {\bibinfo {author} {\bibfnamefont {N.~D.}\ \bibnamefont
  {Mermin}},\ }\href@noop {} {\bibfield  {journal} {\bibinfo  {journal} {Phys.
  Rev. B}\ }\textbf {\bibinfo {volume} {137}},\ \bibinfo {pages} {A1441}
  (\bibinfo {year} {1965})}\BibitemShut {NoStop}%
\bibitem [{\citenamefont {Weinert}\ and\ \citenamefont
  {Davenport}(1992)}]{MWeinert92}%
  \BibitemOpen
  \bibfield  {author} {\bibinfo {author} {\bibfnamefont {M.}~\bibnamefont
  {Weinert}}\ and\ \bibinfo {author} {\bibfnamefont {J.~W.}\ \bibnamefont
  {Davenport}},\ }\href@noop {} {\bibfield  {journal} {\bibinfo  {journal}
  {Phys. Rev. B}\ }\textbf {\bibinfo {volume} {45}},\ \bibinfo {pages} {R13709}
  (\bibinfo {year} {1992})}\BibitemShut {NoStop}%
\bibitem [{\citenamefont {Wentzcovitch}\ \emph {et~al.}(1992)\citenamefont
  {Wentzcovitch}, \citenamefont {Martins},\ and\ \citenamefont
  {Allen}}]{RWentzcovitch92}%
  \BibitemOpen
  \bibfield  {author} {\bibinfo {author} {\bibfnamefont {R.~M.}\ \bibnamefont
  {Wentzcovitch}}, \bibinfo {author} {\bibfnamefont {J.~L.}\ \bibnamefont
  {Martins}}, \ and\ \bibinfo {author} {\bibfnamefont {P.~B.}\ \bibnamefont
  {Allen}},\ }\href@noop {} {\bibfield  {journal} {\bibinfo  {journal} {Phys.
  Rev. B}\ }\textbf {\bibinfo {volume} {45}},\ \bibinfo {pages} {R11372}
  (\bibinfo {year} {1992})}\BibitemShut {NoStop}%
\bibitem [{\citenamefont {Niklasson}(2008{\natexlab{b}})}]{ANiklasson08b}%
  \BibitemOpen
  \bibfield  {author} {\bibinfo {author} {\bibfnamefont {A.~M.~N.}\
  \bibnamefont {Niklasson}},\ }\href@noop {} {\bibfield  {journal} {\bibinfo
  {journal} {J. Chem. Phys.}\ }\textbf {\bibinfo {volume} {129}},\ \bibinfo
  {pages} {244107} (\bibinfo {year} {2008}{\natexlab{b}})}\BibitemShut
  {NoStop}%
\bibitem [{\citenamefont {Nishimoto}(2017)}]{YNishimoto17}%
  \BibitemOpen
  \bibfield  {author} {\bibinfo {author} {\bibfnamefont {Y.}~\bibnamefont
  {Nishimoto}},\ }\href@noop {} {\bibfield  {journal} {\bibinfo  {journal} {J.
  Chem. Phys.}\ }\textbf {\bibinfo {volume} {146}},\ \bibinfo {pages} {084101}
  (\bibinfo {year} {2017})}\BibitemShut {NoStop}%
\bibitem [{\citenamefont {Niklasson}\ and\ \citenamefont
  {Challacombe}(2004)}]{ANiklasson04}%
  \BibitemOpen
  \bibfield  {author} {\bibinfo {author} {\bibfnamefont {A.~M.~N.}\
  \bibnamefont {Niklasson}}\ and\ \bibinfo {author} {\bibfnamefont
  {M.}~\bibnamefont {Challacombe}},\ }\href@noop {} {\bibfield  {journal}
  {\bibinfo  {journal} {Phys. Rev. Lett.}\ }\textbf {\bibinfo {volume} {92}},\
  \bibinfo {pages} {193001} (\bibinfo {year} {2004})}\BibitemShut {NoStop}%
\bibitem [{\citenamefont {Weber}\ \emph {et~al.}(2004)\citenamefont {Weber},
  \citenamefont {Niklasson},\ and\ \citenamefont {Challacombe}}]{VWeber04}%
  \BibitemOpen
  \bibfield  {author} {\bibinfo {author} {\bibfnamefont {V.}~\bibnamefont
  {Weber}}, \bibinfo {author} {\bibfnamefont {A.~M.~N.}\ \bibnamefont
  {Niklasson}}, \ and\ \bibinfo {author} {\bibfnamefont {M.}~\bibnamefont
  {Challacombe}},\ }\href@noop {} {\bibfield  {journal} {\bibinfo  {journal}
  {Phys. Rev. Lett.}\ }\textbf {\bibinfo {volume} {92}},\ \bibinfo {pages}
  {193002} (\bibinfo {year} {2004})}\BibitemShut {NoStop}%
\bibitem [{\citenamefont {Weber}\ \emph {et~al.}(2005)\citenamefont {Weber},
  \citenamefont {Niklasson},\ and\ \citenamefont {Challacombe}}]{VWeber05}%
  \BibitemOpen
  \bibfield  {author} {\bibinfo {author} {\bibfnamefont {V.}~\bibnamefont
  {Weber}}, \bibinfo {author} {\bibfnamefont {A.~M.~N.}\ \bibnamefont
  {Niklasson}}, \ and\ \bibinfo {author} {\bibfnamefont {M.}~\bibnamefont
  {Challacombe}},\ }\href@noop {} {\bibfield  {journal} {\bibinfo  {journal}
  {J. Chem. Phys.}\ }\textbf {\bibinfo {volume} {123}},\ \bibinfo {pages}
  {44106} (\bibinfo {year} {2005})}\BibitemShut {NoStop}%
\bibitem [{\citenamefont {Niklasson}\ \emph {et~al.}(2015)\citenamefont
  {Niklasson}, \citenamefont {Cawkwell}, \citenamefont {Rubensson},\ and\
  \citenamefont {Rudberg}}]{ANiklasson15}%
  \BibitemOpen
  \bibfield  {author} {\bibinfo {author} {\bibfnamefont {A.~M.~N.}\
  \bibnamefont {Niklasson}}, \bibinfo {author} {\bibfnamefont {M.~J.}\
  \bibnamefont {Cawkwell}}, \bibinfo {author} {\bibfnamefont {E.~H.}\
  \bibnamefont {Rubensson}}, \ and\ \bibinfo {author} {\bibfnamefont
  {E.}~\bibnamefont {Rudberg}},\ }\href@noop {} {\bibfield  {journal} {\bibinfo
   {journal} {Phys. Rev. E}\ }\textbf {\bibinfo {volume} {92}},\ \bibinfo
  {pages} {063301} (\bibinfo {year} {2015})}\BibitemShut {NoStop}%
\bibitem [{\citenamefont {Sherman}\ and\ \citenamefont
  {Morrison}(1950)}]{JSherman50}%
  \BibitemOpen
  \bibfield  {author} {\bibinfo {author} {\bibfnamefont {J.}~\bibnamefont
  {Sherman}}\ and\ \bibinfo {author} {\bibfnamefont {W.~J.}\ \bibnamefont
  {Morrison}},\ }\href@noop {} {\bibfield  {journal} {\bibinfo  {journal} {Ann.
  Math. Statist.}\ }\textbf {\bibinfo {volume} {21}},\ \bibinfo {pages} {124}
  (\bibinfo {year} {1950})}\BibitemShut {NoStop}%
\bibitem [{\citenamefont {Elstner}\ \emph {et~al.}(1998)\citenamefont
  {Elstner}, \citenamefont {Poresag}, \citenamefont {Jungnickel}, \citenamefont
  {Elsner}, \citenamefont {Haugk}, \citenamefont {Frauenheim}, \citenamefont
  {Suhai},\ and\ \citenamefont {Seifert}}]{MElstner98}%
  \BibitemOpen
  \bibfield  {author} {\bibinfo {author} {\bibfnamefont {M.}~\bibnamefont
  {Elstner}}, \bibinfo {author} {\bibfnamefont {D.}~\bibnamefont {Poresag}},
  \bibinfo {author} {\bibfnamefont {G.}~\bibnamefont {Jungnickel}}, \bibinfo
  {author} {\bibfnamefont {J.}~\bibnamefont {Elsner}}, \bibinfo {author}
  {\bibfnamefont {M.}~\bibnamefont {Haugk}}, \bibinfo {author} {\bibfnamefont
  {T.}~\bibnamefont {Frauenheim}}, \bibinfo {author} {\bibfnamefont
  {S.}~\bibnamefont {Suhai}}, \ and\ \bibinfo {author} {\bibfnamefont
  {G.}~\bibnamefont {Seifert}},\ }\href@noop {} {\bibfield  {journal} {\bibinfo
   {journal} {Phys. Rev. B}\ }\textbf {\bibinfo {volume} {58}},\ \bibinfo
  {pages} {7260} (\bibinfo {year} {1998})}\BibitemShut {NoStop}%
\bibitem [{\citenamefont {Finnis}\ \emph {et~al.}(1998)\citenamefont {Finnis},
  \citenamefont {Paxton}, \citenamefont {Methfessel},\ and\ \citenamefont {van
  Schilfgarde}}]{MFinnis98}%
  \BibitemOpen
  \bibfield  {author} {\bibinfo {author} {\bibfnamefont {M.~W.}\ \bibnamefont
  {Finnis}}, \bibinfo {author} {\bibfnamefont {A.~T.}\ \bibnamefont {Paxton}},
  \bibinfo {author} {\bibfnamefont {M.}~\bibnamefont {Methfessel}}, \ and\
  \bibinfo {author} {\bibfnamefont {M.}~\bibnamefont {van Schilfgarde}},\
  }\href@noop {} {\bibfield  {journal} {\bibinfo  {journal} {Phys. Rev. Lett.}\
  }\textbf {\bibinfo {volume} {81}},\ \bibinfo {pages} {5149} (\bibinfo {year}
  {1998})}\BibitemShut {NoStop}%
\bibitem [{\citenamefont {Frauenheim}\ \emph {et~al.}(2000)\citenamefont
  {Frauenheim}, \citenamefont {Seifert}, \citenamefont {aand Z.~Hajnal},
  \citenamefont {Jungnickel}, \citenamefont {Poresag}, \citenamefont {Suhai},\
  and\ \citenamefont {Scholz}}]{TFrauenheim00}%
  \BibitemOpen
  \bibfield  {author} {\bibinfo {author} {\bibfnamefont {T.}~\bibnamefont
  {Frauenheim}}, \bibinfo {author} {\bibfnamefont {G.}~\bibnamefont {Seifert}},
  \bibinfo {author} {\bibfnamefont {M.~E.}\ \bibnamefont {aand Z.~Hajnal}},
  \bibinfo {author} {\bibfnamefont {G.}~\bibnamefont {Jungnickel}}, \bibinfo
  {author} {\bibfnamefont {D.}~\bibnamefont {Poresag}}, \bibinfo {author}
  {\bibfnamefont {S.}~\bibnamefont {Suhai}}, \ and\ \bibinfo {author}
  {\bibfnamefont {R.}~\bibnamefont {Scholz}},\ }\href@noop {} {\bibfield
  {journal} {\bibinfo  {journal} {Phys. Stat. sol.}\ }\textbf {\bibinfo
  {volume} {217}},\ \bibinfo {pages} {41} (\bibinfo {year} {2000})}\BibitemShut
  {NoStop}%
\bibitem [{\citenamefont {Cawkwell}\ and\ \citenamefont
  {et~al.}(2010)}]{LATTE}%
  \BibitemOpen
  \bibfield  {author} {\bibinfo {author} {\bibfnamefont {M.~J.}\ \bibnamefont
  {Cawkwell}}\ and\ \bibinfo {author} {\bibnamefont {et~al.}},\ }\href@noop {}
  {\enquote {\bibinfo {title} {{\sc LATTE}},}\ } (\bibinfo {year} {2010}),\
  \bibinfo {note} {\mbox{L}os Alamos National Laboratory (LA- CC-10004),
  http://www.github.com/lanl/latte}\BibitemShut {NoStop}%
\bibitem [{\citenamefont {Sanville}\ \emph {et~al.}(2010)\citenamefont
  {Sanville}, \citenamefont {Bock}, \citenamefont {Challacombe}, \citenamefont
  {Niklasson}, \citenamefont {Cawkwell}, \citenamefont {Dattelbaum},\ and\
  \citenamefont {Sheffield}}]{ESanville10}%
  \BibitemOpen
  \bibfield  {author} {\bibinfo {author} {\bibfnamefont {E.}~\bibnamefont
  {Sanville}}, \bibinfo {author} {\bibfnamefont {N.}~\bibnamefont {Bock}},
  \bibinfo {author} {\bibfnamefont {W.~M.}\ \bibnamefont {Challacombe}},
  \bibinfo {author} {\bibfnamefont {A.~M.~N.}\ \bibnamefont {Niklasson}},
  \bibinfo {author} {\bibfnamefont {M.~J.}\ \bibnamefont {Cawkwell}}, \bibinfo
  {author} {\bibfnamefont {D.~M.}\ \bibnamefont {Dattelbaum}}, \ and\ \bibinfo
  {author} {\bibfnamefont {S.}~\bibnamefont {Sheffield}},\ }\enquote {\bibinfo
  {title} {Proceedings of the fourteenth international detonation symposium},}\
  \ (\bibinfo  {publisher} {Office of Naval Research, Arlington VA,
  ONR-351-10-185},\ \bibinfo {year} {2010})\ pp.\ \bibinfo {pages}
  {91--101}\BibitemShut {NoStop}%
\bibitem [{\citenamefont {Pulay}(1982)}]{PPulay82}%
  \BibitemOpen
  \bibfield  {author} {\bibinfo {author} {\bibfnamefont {P.}~\bibnamefont
  {Pulay}},\ }\href@noop {} {\bibfield  {journal} {\bibinfo  {journal} {J.
  Comput. Chem.}\ }\textbf {\bibinfo {volume} {3}},\ \bibinfo {pages} {556}
  (\bibinfo {year} {1982})}\BibitemShut {NoStop}%
\bibitem [{\citenamefont {Banerjee}\ \emph {et~al.}(2016)\citenamefont
  {Banerjee}, \citenamefont {Suryanaryana},\ and\ \citenamefont
  {Pask}}]{ASBanerjee16}%
  \BibitemOpen
  \bibfield  {author} {\bibinfo {author} {\bibfnamefont {A.~S.}\ \bibnamefont
  {Banerjee}}, \bibinfo {author} {\bibfnamefont {P.}~\bibnamefont
  {Suryanaryana}}, \ and\ \bibinfo {author} {\bibfnamefont {J.~E.}\
  \bibnamefont {Pask}},\ }\href@noop {} {\bibfield  {journal} {\bibinfo
  {journal} {Chem. Phys. Lett.}\ }\textbf {\bibinfo {volume} {647}},\ \bibinfo
  {pages} {31} (\bibinfo {year} {2016})}\BibitemShut {NoStop}%
\bibitem [{\citenamefont {Ahlrichs}\ \emph {et~al.}(1989)\citenamefont
  {Ahlrichs}, \citenamefont {B\"{a}r}, \citenamefont {H\"{a}ser}, \citenamefont
  {Hom},\ and\ \citenamefont {K\"{o}lmel}}]{RAhlrichs89}%
  \BibitemOpen
  \bibfield  {author} {\bibinfo {author} {\bibfnamefont {R.}~\bibnamefont
  {Ahlrichs}}, \bibinfo {author} {\bibfnamefont {M.}~\bibnamefont {B\"{a}r}},
  \bibinfo {author} {\bibfnamefont {M.}~\bibnamefont {H\"{a}ser}}, \bibinfo
  {author} {\bibfnamefont {H.}~\bibnamefont {Hom}}, \ and\ \bibinfo {author}
  {\bibfnamefont {C.}~\bibnamefont {K\"{o}lmel}},\ }\href@noop {} {\bibfield
  {journal} {\bibinfo  {journal} {Chem. Phys. Lett.}\ }\textbf {\bibinfo
  {volume} {162}},\ \bibinfo {pages} {165} (\bibinfo {year}
  {1989})}\BibitemShut {NoStop}%
\bibitem [{\citenamefont {Schmidt}\ \emph {et~al.}(1993)\citenamefont
  {Schmidt}, \citenamefont {Baldridge}, \citenamefont {Boatz}, \citenamefont
  {Elbert}, \citenamefont {Gordon}, \citenamefont {Jensen}, \citenamefont
  {Koseki}, \citenamefont {Matsunaga}, \citenamefont {Nguyen}, \citenamefont
  {Su}, \citenamefont {Windus}, \citenamefont {Dupuis},\ and\ \citenamefont
  {Montgomery}}]{MSchmidt93}%
  \BibitemOpen
  \bibfield  {author} {\bibinfo {author} {\bibfnamefont {M.~W.}\ \bibnamefont
  {Schmidt}}, \bibinfo {author} {\bibfnamefont {K.~K.}\ \bibnamefont
  {Baldridge}}, \bibinfo {author} {\bibfnamefont {J.~A.}\ \bibnamefont
  {Boatz}}, \bibinfo {author} {\bibfnamefont {S.~T.}\ \bibnamefont {Elbert}},
  \bibinfo {author} {\bibfnamefont {M.~S.}\ \bibnamefont {Gordon}}, \bibinfo
  {author} {\bibfnamefont {J.~H.}\ \bibnamefont {Jensen}}, \bibinfo {author}
  {\bibfnamefont {S.}~\bibnamefont {Koseki}}, \bibinfo {author} {\bibfnamefont
  {N.}~\bibnamefont {Matsunaga}}, \bibinfo {author} {\bibfnamefont {K.~A.}\
  \bibnamefont {Nguyen}}, \bibinfo {author} {\bibfnamefont {S.~J.}\
  \bibnamefont {Su}}, \bibinfo {author} {\bibfnamefont {T.~L.}\ \bibnamefont
  {Windus}}, \bibinfo {author} {\bibfnamefont {M.}~\bibnamefont {Dupuis}}, \
  and\ \bibinfo {author} {\bibfnamefont {J.~A.}\ \bibnamefont {Montgomery}},\
  }\href@noop {} {\bibfield  {journal} {\bibinfo  {journal} {J. Comput. Chem.}\
  }\textbf {\bibinfo {volume} {14}},\ \bibinfo {pages} {1347} (\bibinfo {year}
  {1993})}\BibitemShut {NoStop}%
\bibitem [{\citenamefont {Kresse}\ and\ \citenamefont
  {Furthmuller}(1996)}]{GKresse96}%
  \BibitemOpen
  \bibfield  {author} {\bibinfo {author} {\bibfnamefont {G.}~\bibnamefont
  {Kresse}}\ and\ \bibinfo {author} {\bibfnamefont {J.}~\bibnamefont
  {Furthmuller}},\ }\href@noop {} {\bibfield  {journal} {\bibinfo  {journal}
  {Phys. Rev. B}\ }\textbf {\bibinfo {volume} {54}},\ \bibinfo {pages} {11169}
  (\bibinfo {year} {1996})}\BibitemShut {NoStop}%
\bibitem [{\citenamefont {Hern{\'a}ndez}\ \emph {et~al.}(1996)\citenamefont
  {Hern{\'a}ndez}, \citenamefont {Gillan},\ and\ \citenamefont
  {Goringe}}]{EHernandez96}%
  \BibitemOpen
  \bibfield  {author} {\bibinfo {author} {\bibfnamefont {E.}~\bibnamefont
  {Hern{\'a}ndez}}, \bibinfo {author} {\bibfnamefont {M.~J.}\ \bibnamefont
  {Gillan}}, \ and\ \bibinfo {author} {\bibfnamefont {C.}~\bibnamefont
  {Goringe}},\ }\href@noop {} {\bibfield  {journal} {\bibinfo  {journal} {Phys.
  Rev. B}\ }\textbf {\bibinfo {volume} {53}},\ \bibinfo {pages} {7147}
  (\bibinfo {year} {1996})}\BibitemShut {NoStop}%
\bibitem [{\citenamefont {Tsuchida}\ and\ \citenamefont
  {Tsukada}(1998)}]{ETsuchida98}%
  \BibitemOpen
  \bibfield  {author} {\bibinfo {author} {\bibfnamefont {E.}~\bibnamefont
  {Tsuchida}}\ and\ \bibinfo {author} {\bibfnamefont {M.}~\bibnamefont
  {Tsukada}},\ }\href@noop {} {\bibfield  {journal} {\bibinfo  {journal} {J.
  Phys. Soc. Jpn.}\ }\textbf {\bibinfo {volume} {67}},\ \bibinfo {pages} {340}
  (\bibinfo {year} {1998})}\BibitemShut {NoStop}%
\bibitem [{\citenamefont {Soler}\ \emph {et~al.}(2002)\citenamefont {Soler},
  \citenamefont {Artacho}, \citenamefont {Gale}, \citenamefont {Garcia},
  \citenamefont {Junquera}, \citenamefont {Ordejon},\ and\ \citenamefont
  {Sanchez-Portal}}]{JSoler02}%
  \BibitemOpen
  \bibfield  {author} {\bibinfo {author} {\bibfnamefont {J.~M.}\ \bibnamefont
  {Soler}}, \bibinfo {author} {\bibfnamefont {E.}~\bibnamefont {Artacho}},
  \bibinfo {author} {\bibfnamefont {J.~D.}\ \bibnamefont {Gale}}, \bibinfo
  {author} {\bibfnamefont {A.}~\bibnamefont {Garcia}}, \bibinfo {author}
  {\bibfnamefont {J.}~\bibnamefont {Junquera}}, \bibinfo {author}
  {\bibfnamefont {P.}~\bibnamefont {Ordejon}}, \ and\ \bibinfo {author}
  {\bibfnamefont {D.}~\bibnamefont {Sanchez-Portal}},\ }\href@noop {}
  {\bibfield  {journal} {\bibinfo  {journal} {J. Phys.: Condens. Matter}\
  }\textbf {\bibinfo {volume} {14}},\ \bibinfo {pages} {2745} (\bibinfo {year}
  {2002})}\BibitemShut {NoStop}%
\bibitem [{\citenamefont {Bowler}\ \emph {et~al.}(2001)\citenamefont {Bowler},
  \citenamefont {Miyazaki},\ and\ \citenamefont {Gillan}}]{DBowler01}%
  \BibitemOpen
  \bibfield  {author} {\bibinfo {author} {\bibfnamefont {D.~R.}\ \bibnamefont
  {Bowler}}, \bibinfo {author} {\bibfnamefont {T.}~\bibnamefont {Miyazaki}}, \
  and\ \bibinfo {author} {\bibfnamefont {M.}~\bibnamefont {Gillan}},\
  }\href@noop {} {\bibfield  {journal} {\bibinfo  {journal} {Comput. Phys.
  Comm.}\ }\textbf {\bibinfo {volume} {137}},\ \bibinfo {pages} {255} (\bibinfo
  {year} {2001})}\BibitemShut {NoStop}%
\bibitem [{\citenamefont {Ozaki}\ and\ \citenamefont {Kino}(2005)}]{TOzaki05}%
  \BibitemOpen
  \bibfield  {author} {\bibinfo {author} {\bibfnamefont {T.}~\bibnamefont
  {Ozaki}}\ and\ \bibinfo {author} {\bibfnamefont {H.}~\bibnamefont {Kino}},\
  }\href@noop {} {\bibfield  {journal} {\bibinfo  {journal} {Phys.\ Rev.\ B}\
  }\textbf {\bibinfo {volume} {72}},\ \bibinfo {pages} {045121} (\bibinfo
  {year} {2005})}\BibitemShut {NoStop}%
\bibitem [{\citenamefont {VandeVondele}\ \emph {et~al.}(2005)\citenamefont
  {VandeVondele}, \citenamefont {Krack}, \citenamefont {Mohammed},
  \citenamefont {Parrinello}, \citenamefont {Chassing},\ and\ \citenamefont
  {Hutter}}]{JVandevondele05}%
  \BibitemOpen
  \bibfield  {author} {\bibinfo {author} {\bibfnamefont {J.}~\bibnamefont
  {VandeVondele}}, \bibinfo {author} {\bibfnamefont {M.}~\bibnamefont {Krack}},
  \bibinfo {author} {\bibfnamefont {F.}~\bibnamefont {Mohammed}}, \bibinfo
  {author} {\bibfnamefont {M.}~\bibnamefont {Parrinello}}, \bibinfo {author}
  {\bibfnamefont {T.}~\bibnamefont {Chassing}}, \ and\ \bibinfo {author}
  {\bibfnamefont {J.}~\bibnamefont {Hutter}},\ }\href@noop {} {\bibfield
  {journal} {\bibinfo  {journal} {Comput. Phys. Commun.}\ }\textbf {\bibinfo
  {volume} {167}},\ \bibinfo {pages} {103} (\bibinfo {year}
  {2005})}\BibitemShut {NoStop}%
\bibitem [{\citenamefont {Bowler}\ \emph {et~al.}(2006)\citenamefont {Bowler},
  \citenamefont {Choudhury}, \citenamefont {Gillan},\ and\ \citenamefont
  {Miyazaki}}]{DBowler06}%
  \BibitemOpen
  \bibfield  {author} {\bibinfo {author} {\bibfnamefont {D.~R.}\ \bibnamefont
  {Bowler}}, \bibinfo {author} {\bibfnamefont {R.}~\bibnamefont {Choudhury}},
  \bibinfo {author} {\bibfnamefont {M.~J.}\ \bibnamefont {Gillan}}, \ and\
  \bibinfo {author} {\bibfnamefont {T.}~\bibnamefont {Miyazaki}},\ }\href@noop
  {} {\bibfield  {journal} {\bibinfo  {journal} {Phys. Stat. Sol. B}\ }\textbf
  {\bibinfo {volume} {243}},\ \bibinfo {pages} {898} (\bibinfo {year}
  {2006})}\BibitemShut {NoStop}%
\bibitem [{\citenamefont {Kronik}\ \emph {et~al.}(2006)\citenamefont {Kronik},
  \citenamefont {Makmal}, \citenamefont {Tiago}, \citenamefont {Alemany},
  \citenamefont {Huang},\ and\ \citenamefont {Saad}}]{LKronik06}%
  \BibitemOpen
  \bibfield  {author} {\bibinfo {author} {\bibfnamefont {L.}~\bibnamefont
  {Kronik}}, \bibinfo {author} {\bibfnamefont {A.}~\bibnamefont {Makmal}},
  \bibinfo {author} {\bibfnamefont {M.}~\bibnamefont {Tiago}}, \bibinfo
  {author} {\bibfnamefont {M.~M.}\ \bibnamefont {Alemany}}, \bibinfo {author}
  {\bibfnamefont {X.}~\bibnamefont {Huang}}, \ and\ \bibinfo {author}
  {\bibfnamefont {Y.}~\bibnamefont {Saad}},\ }\href@noop {} {\bibfield
  {journal} {\bibinfo  {journal} {Phys. Stat. Solidi.}\ }\textbf {\bibinfo
  {volume} {243}},\ \bibinfo {pages} {1063} (\bibinfo {year}
  {2006})}\BibitemShut {NoStop}%
\bibitem [{\citenamefont {Bock}\ \emph {et~al.}(2008)\citenamefont {Bock},
  \citenamefont {Challacombe}, \citenamefont {Gan}, \citenamefont {Henkelman},
  \citenamefont {Nemeth}, \citenamefont {Niklasson}, \citenamefont {Odell},
  \citenamefont {Schwegler}, \citenamefont {Tymczak},\ and\ \citenamefont
  {Weber}}]{FreeON}%
  \BibitemOpen
  \bibfield  {author} {\bibinfo {author} {\bibfnamefont {N.}~\bibnamefont
  {Bock}}, \bibinfo {author} {\bibfnamefont {M.}~\bibnamefont {Challacombe}},
  \bibinfo {author} {\bibfnamefont {C.~K.}\ \bibnamefont {Gan}}, \bibinfo
  {author} {\bibfnamefont {G.}~\bibnamefont {Henkelman}}, \bibinfo {author}
  {\bibfnamefont {K.}~\bibnamefont {Nemeth}}, \bibinfo {author} {\bibfnamefont
  {A.~M.~N.}\ \bibnamefont {Niklasson}}, \bibinfo {author} {\bibfnamefont
  {A.}~\bibnamefont {Odell}}, \bibinfo {author} {\bibfnamefont
  {E.}~\bibnamefont {Schwegler}}, \bibinfo {author} {\bibfnamefont {C.~J.}\
  \bibnamefont {Tymczak}}, \ and\ \bibinfo {author} {\bibfnamefont
  {V.}~\bibnamefont {Weber}},\ }\href {http://www.nongnu.org/freeon/} {\enquote
  {\bibinfo {title} {{\sc FreeON}},}\ } (\bibinfo {year} {2008}),\ \bibinfo
  {note} {\mbox{L}os Alamos National Laboratory (LA-CC 01-2; LA- CC-04-086),
  Copyright University of California.}\BibitemShut {Stop}%
\bibitem [{\citenamefont {Genovese}\ \emph {et~al.}(2008)\citenamefont
  {Genovese}, \citenamefont {Neelov}, \citenamefont {Goedecker}, \citenamefont
  {Deutsch}, \citenamefont {Ghasemi}, \citenamefont {Willand}, \citenamefont
  {Caliste}, \citenamefont {Zilerberg}, \citenamefont {Rayson}, \citenamefont
  {Bergman},\ and\ \citenamefont {Schneider}}]{LGenovese08}%
  \BibitemOpen
  \bibfield  {author} {\bibinfo {author} {\bibfnamefont {L.}~\bibnamefont
  {Genovese}}, \bibinfo {author} {\bibfnamefont {A.}~\bibnamefont {Neelov}},
  \bibinfo {author} {\bibfnamefont {S.}~\bibnamefont {Goedecker}}, \bibinfo
  {author} {\bibfnamefont {T.}~\bibnamefont {Deutsch}}, \bibinfo {author}
  {\bibfnamefont {S.~A.}\ \bibnamefont {Ghasemi}}, \bibinfo {author}
  {\bibfnamefont {A.}~\bibnamefont {Willand}}, \bibinfo {author} {\bibfnamefont
  {D.}~\bibnamefont {Caliste}}, \bibinfo {author} {\bibfnamefont
  {O.}~\bibnamefont {Zilerberg}}, \bibinfo {author} {\bibfnamefont
  {M.}~\bibnamefont {Rayson}}, \bibinfo {author} {\bibfnamefont
  {A.}~\bibnamefont {Bergman}}, \ and\ \bibinfo {author} {\bibfnamefont
  {R.}~\bibnamefont {Schneider}},\ }\href@noop {} {\bibfield  {journal}
  {\bibinfo  {journal} {J. Chem. Phys.}\ }\textbf {\bibinfo {volume} {129}},\
  \bibinfo {pages} {014109} (\bibinfo {year} {2008})}\BibitemShut {NoStop}%
\bibitem [{\citenamefont {Blum}\ \emph {et~al.}(2009)\citenamefont {Blum},
  \citenamefont {Gehrke}, \citenamefont {Hanke}, \citenamefont {Havu},
  \citenamefont {Havu}, \citenamefont {Ren}, \citenamefont {Reuter},\ and\
  \citenamefont {Sheffler}}]{VBlum09}%
  \BibitemOpen
  \bibfield  {author} {\bibinfo {author} {\bibfnamefont {V.}~\bibnamefont
  {Blum}}, \bibinfo {author} {\bibfnamefont {R.}~\bibnamefont {Gehrke}},
  \bibinfo {author} {\bibfnamefont {F.}~\bibnamefont {Hanke}}, \bibinfo
  {author} {\bibfnamefont {P.}~\bibnamefont {Havu}}, \bibinfo {author}
  {\bibfnamefont {V.}~\bibnamefont {Havu}}, \bibinfo {author} {\bibfnamefont
  {X.}~\bibnamefont {Ren}}, \bibinfo {author} {\bibfnamefont {K.}~\bibnamefont
  {Reuter}}, \ and\ \bibinfo {author} {\bibfnamefont {M.}~\bibnamefont
  {Sheffler}},\ }\href@noop {} {\bibfield  {journal} {\bibinfo  {journal}
  {Comput. Phys. Commun.}\ }\textbf {\bibinfo {volume} {180}},\ \bibinfo
  {pages} {2175} (\bibinfo {year} {2009})}\BibitemShut {NoStop}%
\bibitem [{\citenamefont {Hine}\ \emph {et~al.}(2009)\citenamefont {Hine},
  \citenamefont {Haynes}, \citenamefont {Mostofi}, \citenamefont {Skylaris},\
  and\ \citenamefont {Payne}}]{NHine09}%
  \BibitemOpen
  \bibfield  {author} {\bibinfo {author} {\bibfnamefont {N.~D.}\ \bibnamefont
  {Hine}}, \bibinfo {author} {\bibfnamefont {P.~D.}\ \bibnamefont {Haynes}},
  \bibinfo {author} {\bibfnamefont {A.~A.}\ \bibnamefont {Mostofi}}, \bibinfo
  {author} {\bibfnamefont {C.-K.}\ \bibnamefont {Skylaris}}, \ and\ \bibinfo
  {author} {\bibfnamefont {M.~C.}\ \bibnamefont {Payne}},\ }\href@noop {}
  {\bibfield  {journal} {\bibinfo  {journal} {Comput. Phys. Comm.}\ }\textbf
  {\bibinfo {volume} {180}},\ \bibinfo {pages} {1041} (\bibinfo {year}
  {2009})}\BibitemShut {NoStop}%
\bibitem [{\citenamefont {Rudberg}\ \emph {et~al.}(2011)\citenamefont
  {Rudberg}, \citenamefont {Rubensson},\ and\ \citenamefont
  {Salek}}]{ERudberg_11}%
  \BibitemOpen
  \bibfield  {author} {\bibinfo {author} {\bibfnamefont {E.}~\bibnamefont
  {Rudberg}}, \bibinfo {author} {\bibfnamefont {E.~H.}\ \bibnamefont
  {Rubensson}}, \ and\ \bibinfo {author} {\bibfnamefont {P.}~\bibnamefont
  {Salek}},\ }\href@noop {} {\bibfield  {journal} {\bibinfo  {journal} {J.
  Chem. Theory Comput.}\ }\textbf {\bibinfo {volume} {7}},\ \bibinfo {pages}
  {340} (\bibinfo {year} {2011})}\BibitemShut {NoStop}%
\bibitem [{\citenamefont {VandeVondele}\ \emph {et~al.}(2012)\citenamefont
  {VandeVondele}, \citenamefont {Borstnik},\ and\ \citenamefont
  {Hutter}}]{JVandevondele12}%
  \BibitemOpen
  \bibfield  {author} {\bibinfo {author} {\bibfnamefont {J.}~\bibnamefont
  {VandeVondele}}, \bibinfo {author} {\bibfnamefont {U.}~\bibnamefont
  {Borstnik}}, \ and\ \bibinfo {author} {\bibfnamefont {J.}~\bibnamefont
  {Hutter}},\ }\href@noop {} {\bibfield  {journal} {\bibinfo  {journal} {J.
  Chem. Theory Comput.}\ }\textbf {\bibinfo {volume} {8}},\ \bibinfo {pages}
  {3565} (\bibinfo {year} {2012})}\BibitemShut {NoStop}%
\bibitem [{\citenamefont {Aidas}\ and\ \citenamefont
  {et~al.}(2013)}]{KAidas13}%
  \BibitemOpen
  \bibfield  {author} {\bibinfo {author} {\bibfnamefont {K.}~\bibnamefont
  {Aidas}}\ and\ \bibinfo {author} {\bibnamefont {et~al.}},\ }\href@noop {}
  {\bibfield  {journal} {\bibinfo  {journal} {WIREs Comput. Mol. Sci.}\
  }\textbf {\bibinfo {volume} {4}},\ \bibinfo {pages} {269} (\bibinfo {year}
  {2013})}\BibitemShut {NoStop}%
\bibitem [{\citenamefont {Osei-Kuffuor}\ \emph {et~al.}(2014)\citenamefont
  {Osei-Kuffuor}, \citenamefont {Fattebert},\ and\ \citenamefont
  {Gygi}}]{DOseiKuffuor14}%
  \BibitemOpen
  \bibfield  {author} {\bibinfo {author} {\bibfnamefont {D.}~\bibnamefont
  {Osei-Kuffuor}}, \bibinfo {author} {\bibfnamefont {J.~L.}\ \bibnamefont
  {Fattebert}}, \ and\ \bibinfo {author} {\bibfnamefont {F.}~\bibnamefont
  {Gygi}},\ }\href@noop {} {\bibfield  {journal} {\bibinfo  {journal} {Phys.
  Rev. Lett.}\ }\textbf {\bibinfo {volume} {112}},\ \bibinfo {pages} {046401}
  (\bibinfo {year} {2014})}\BibitemShut {NoStop}%
\bibitem [{\citenamefont {Qin}\ \emph {et~al.}(2015)\citenamefont {Qin},
  \citenamefont {Shang}, \citenamefont {Xiang}, \citenamefont {Li},\ and\
  \citenamefont {Yang}}]{XQin15}%
  \BibitemOpen
  \bibfield  {author} {\bibinfo {author} {\bibfnamefont {X.~M.}\ \bibnamefont
  {Qin}}, \bibinfo {author} {\bibfnamefont {H.~H.}\ \bibnamefont {Shang}},
  \bibinfo {author} {\bibfnamefont {H.~J.}\ \bibnamefont {Xiang}}, \bibinfo
  {author} {\bibfnamefont {Z.~Y.}\ \bibnamefont {Li}}, \ and\ \bibinfo {author}
  {\bibfnamefont {J.~L.}\ \bibnamefont {Yang}},\ }\href@noop {} {\bibfield
  {journal} {\bibinfo  {journal} {Int. J. Quantum Chem.}\ }\textbf {\bibinfo
  {volume} {115}},\ \bibinfo {pages} {647} (\bibinfo {year}
  {2015})}\BibitemShut {NoStop}%
\end{thebibliography}

\end{document}